\documentclass[a4paper,11pt]{article}
\usepackage{jheppub}
\usepackage{xcolor}
\usepackage{subcaption}
\usepackage{tikz}
\usepackage{multirow}
\usepackage{needspace}
\usepackage[normalem]{ulem}
\newif\ifmarkup     \markuptrue
\newif\ifshownotes  \shownotestrue
\definecolor{revnote}{rgb}{0,0.5,0}
\definecolor{revlink}{rgb}{0.3,0.3,0.3}

\newcommand{\del}[1]{\ifmarkup{\leavevmode\color{red}\sout{#1}}\fi}

\ifmarkup\hypersetup{linkcolor=revlink,citecolor=revlink,urlcolor=revlink}\fi

\newcommand{\CJTGraph}[3]{%
  \begingroup
  \def\CJTDrawingType{#2}\def\CJTDoubleType{dressed}%
  \ifx\CJTDrawingType\CJTDoubleType
    \def\CJTOuterShift{.95pt}\def\CJTTangentShift{1.175pt}%
  \else
    \def\CJTOuterShift{.225pt}\def\CJTTangentShift{.45pt}%
  \fi
  \mathord{\vcenter{\hbox{%
    \begin{tikzpicture}[
      x=1cm,y=1cm,
      cjt bare/.style={draw=#1,line width=.45pt,line cap=round,line join=round},
      cjt dressed/.style={cjt bare,double=white,double distance=1.0pt},
      cjt line/.style={cjt #2},
      cjt external/.style={draw=black,line width=.45pt,line cap=round,line join=round},
    ]
      #3%
    \end{tikzpicture}%
  }}}%
  \endgroup
}

\newcommand{\CJTTadpole}[2]{%
  \CJTGraph{#1}{#2}{%
    \draw[cjt external] (0,-.34)--(0,-.14);
    \draw[cjt line] (0,.06) circle[radius=.20];
  }%
}

\newcommand{\CJTSelfTadpole}[2]{%
  \CJTGraph{#1}{#2}{%
    \draw[cjt external,yshift=-\CJTTangentShift] (-.46,-.20)--(.46,-.20);
    \draw[cjt line] (0,0) circle[radius=.20];
  }%
}

\newcommand{\CJTBubble}[2]{%
  \CJTGraph{#1}{#2}{%
    \draw[cjt external] (-.50,0)--([xshift=-\CJTOuterShift]-.20,0);
    \draw[cjt external] ([xshift=\CJTOuterShift].20,0)--(.50,0);
    \draw[cjt line] (0,0) circle[radius=.20];
  }%
}

\newcommand{\CJTBubbleZero}[2]{\CJTBubble{#1}{#2}^{(0)}}

\newcommand{\CJTDressedVertexTadpole}[2]{%
  \CJTGraph{#1}{#2}{%
    \draw[cjt external] (0,-.34)--(0,-.14);
    \draw[cjt line] (0,.06) circle[radius=.20];
    \fill[black] (0,-.14) circle[radius=.055];
  }%
}

\arxivnumber{} 

\title{\boldmath Thermal Resummation Effects on Electroweak Phase Transitions and Gravitational Waves in Conformal and Non-Conformal 2HDMs}

\author{
Yu-Bo Zhao$^a$, Ji-Wei Li$^a$, Zhi-Wei Wang$^a$
}
\affiliation[a]{School of Physics, University of Electronic Science and Technology of China, No.~2006, Xiyuan Avenue, West Hi-Tech Zone, Chengdu, China}

\emailAdd{zhiwei.wang@uestc.edu.cn}
\emailAdd{jiweili@std.uestc.edu.cn}
\emailAdd{yubo.zhao@std.uestc.edu.cn}

\abstract
{We quantify how the choice of thermal resummation prescription affects electroweak phase transitions and the associated gravitational-wave predictions in classically conformal and non-conformal two-Higgs-doublet models. We treat the scalar sector with multi-field partial dressing (PD), which retains the full thermal functions and determines the field-dependent scalar masses and mixing from gap equations, and compare it with the Parwani and Arnold--Espinosa prescriptions at the same zero-temperature input points, so that all differences originate in the finite-temperature treatment. Completed first-order transitions dominate the conformal scan under all three prescriptions, with PD typically giving the strongest and most supercooled transitions (down to \(T_p/T_c\simeq0.5\), compared with \(T_p/T_c\gtrsim0.8\) for Parwani and Arnold--Espinosa). This trend carries over to the gravitational-wave predictions: at every conformal point where all three predict completion, PD gives the largest peak amplitude, followed by Parwani and Arnold--Espinosa. The non-conformal model has no such universal ordering, and over extended regions of its parameter space the prescription changes the type of thermal history itself. For representative benchmarks, the PD and Arnold--Espinosa predictions differ by up to about nine orders of magnitude in peak amplitude and by up to a factor of about fifty in peak frequency. The PD calculation also reveals a theoretical problem: contrary to the path independence assumed in earlier multi-field studies, the dressed tadpole is not a gradient, so the reconstructed multi-field potential depends on the integration path. In our benchmark tests, changing the path shifts the transition temperatures by at most about 7\% and the peak amplitude by up to a factor of about three; this is well below the largest prescription differences but constitutes an additional ambiguity intrinsic to multi-field PD. Thermal resummation is therefore a major source of uncertainty in these phase-transition and gravitational-wave predictions.}

\begin{document}
\maketitle
\flushbottom

\section{Introduction}
\label{sec:intro}
The thermal history of electroweak symmetry breaking is sensitive to physics beyond the Standard Model and has direct implications for gravitational wave (GW) phenomenology. In the Standard Model with the observed Higgs mass, the electroweak phase transition (EWPT) is a smooth crossover rather than a first-order transition~\cite{Kajantie:1996mn,Kajantie:1998yc,Laine:2015kra,DOnofrio:2016Crossover}. A first-order electroweak phase transition (FOEWPT) therefore requires new dynamics beyond the minimal Higgs sector, most commonly in the form of additional scalar degrees of freedom that modify the finite temperature effective potential~\cite{quiros1999finite,Morrissey:2012db}. Such a transition can provide the required departure from thermal equilibrium for electroweak baryogenesis and source a stochastic gravitational wave background~\cite{Caprini_2016,Caprini:2018mtu,Caprini:2024LISA,Caprini:2025WeakToStrong,Caprini:2020LISAUpdate}. The resulting gravitational wave signal can lie in the frequency range targeted by future space-based interferometers such as LISA, Taiji, TianQin, DECIGO and BBO~\cite{2017arXiv170200786A,Corbin:2005ny,Kawamura:2011zz,TianQin:2015yph,Hu:2017mde,Liu:2023ymk,Thrane:2013oya,Schmitz:2020syl}. The EWPT thus links collider physics, cosmology and gravitational-wave astronomy.

A reliable prediction of these gravitational-wave observables requires accurate values of the transition temperature, the released vacuum energy and the inverse time scale of bubble nucleation, all of which are fixed by the finite-temperature effective potential over the field region relevant to the transition. 
Perturbative evaluations of this potential are infrared sensitive: loop corrections from the static Matsubara modes of bosons are enhanced by the Bose factor $T/m$. Near the transition, the field-dependent masses of light scalars and longitudinal gauge bosons become comparable to or smaller than their thermal masses, so a fixed-order loop expansion converges poorly and thermal resummation is required.
In electroweak phase-transition studies it is usually implemented through the Parwani or Arnold--Espinosa prescriptions~\cite{Dolan:1973qd,Arnold:1992rz,Parwani:1991gq,Carrington1992Effective}, which supplement the field-dependent masses with the leading high-temperature thermal masses. These keep only the leading term of the expansion in $m/T$, which is adequate for modes light compared with $T$. In strong first-order transitions, however, the scalar background and with it some field-dependent masses become comparable to $T$. The full thermal functions then suppress the contributions of these heavy modes by Boltzmann factors, an effect that truncated thermal masses miss~\cite{Navarrete:2026NoHighT}. The resummation prescription is therefore a significant source of theoretical uncertainty in phase-transition and gravitational-wave predictions~\cite{Croon:2020,Kainulainen:2019Validity,Cho:2026Resummation,Lofgren:2023Resummation}, alongside gauge, renormalization-scheme and scale dependence~\cite{Qin:2025GaugeInvariant,Zhu:2025Uncertainties}.

To go beyond this high-temperature mass approximation, we use partial dressing (PD)~\cite{Boyd:1993,Curtin:2022,Bahl2024Improved}, which determines the scalar masses from gap equations with the full thermal functions and therefore includes subleading thermal corrections. These masses are inserted into the one-loop tadpole after differentiation, rather than into the potential itself as in full dressing; this avoids overcounting higher-loop contributions. The recently proposed hybrid partial dressing (HPD) scheme further improves the sunset treatment by solving a Hartree gap equation and adding the full two-loop vacuum sunset to the potential~\cite{Berthiaume2026Hybrid}, although its original treatment covers only one scalar and two scalars without mixing. Ref.~\cite{Curtin:2016} pointed out that PD, formulated along a single field direction, has no unique extension to excursions along several directions. Ref.~\cite{Bahl2024Improved} proposed such an extension, a line integral of the dressed tadpole in field space, and argued that it is path independent because the curl of a gradient vanishes. Ref.~\cite{Bittar:2025lcr} applied this construction, along a straight line, to the non-conformal two-Higgs-doublet model and found that the prescription can change the nature of the transition and its gravitational-wave signal, with the symmetry non-restoration of Arnold--Espinosa absent in PD. We depart from these studies in two respects. First, we show that the dressed tadpole is not a gradient: the multi-field PD potential depends on the integration path, and the straight line is only one of many inequivalent choices. Second, we use the classically conformal version of the model as a laboratory for supercooled transitions, which proceed at temperatures well below the heavy scalar masses, where the high-temperature approximation behind Parwani and Arnold--Espinosa becomes unreliable; there PD typically yields the strongest and most supercooled transitions of the three prescriptions.

The two-Higgs-doublet model (2HDM)~\cite{Branco_2012,Cline:1996mga,Dorsch:2016nrg,Basler_2017} is a natural testing ground for multi-field PD: its first-order transitions are typically driven by heavy scalars with large quartic couplings, for which truncated thermal masses are inadequate, and these scalars mix in a field-dependent way. In its classically conformal version, electroweak symmetry breaking and the SM-like Higgs mass arise radiatively along a Gildener--Weinberg flat direction~\cite{Coleman:1973jx,Gildener:1976ih,Lee:2012jn,Fuyuto:2015ScaleInvariant}; with no tree-level potential along this direction, supercooling is expected. Yet our earlier comparison with the non-conformal 2HDM, which has explicit mass terms and a softly broken \(\mathbb Z_2\) symmetry, found only limited supercooling in the conformal model, using the Parwani prescription alone~\cite{Benincasa:2026c2hdm}. We therefore apply PD, Parwani and Arnold--Espinosa to the same accepted zero-temperature points in each model, so that any difference at a given point is due to the prescription alone. Across the two scans, these differences test whether model-level conclusions, such as this limited supercooling, survive a change of prescription. Conversely, since the two models share the same field content and differ only by tree-level mass terms, comparing them isolates the effect of classical conformal symmetry on the structure of the prescription dependence.

The path dependence arises because PD reconstructs the potential by integrating the dressed tadpole through field space, which yields a unique potential only if the tadpole is a gradient. The undressed one-loop tadpole is a gradient, but the dressed tadpole generally has a nonzero curl (Eq.~\eqref{eq:pd-curl}), so multi-field PD does not define a unique potential. We define the PD potential by integrating along the straight path from the origin and estimate the resulting ambiguity with two alternative contours for 2HDM benchmarks (Appendix~\ref{app:path-dependence}). Our subsequent work~\cite{LiWangZhao:MultifieldCJT} traces the curl to the zero-momentum sunset self-energy in the gap equation and derives a path-independent completion of PD.

Numerically, the path ambiguity is modest: for our benchmarks it shifts the transition temperatures by at most about \(7\%\) and the gravitational-wave peak amplitude by up to a factor of about three. The prescription dependence can be far larger: for individual benchmarks, the peak amplitudes of PD and Arnold--Espinosa differ by up to about nine orders of magnitude and the peak frequencies by up to a factor of about fifty. In the conformal model, completed first-order transitions dominate under all three prescriptions, but PD typically gives the strongest and most supercooled ones, reaching \(\alpha=\mathcal O(1)\) and \(T_p/T_c\simeq0.5\), whereas Parwani and Arnold--Espinosa stay at \(T_p/T_c\gtrsim0.8\); the limited conformal supercooling is thus specific to the truncated prescriptions. At every conformal point where all three prescriptions predict a completed transition, the peak amplitudes follow PD\,\(>\)\,Parwani\,\(>\)\,Arnold--Espinosa. The non-conformal model, with its explicit mass scales, has no such ordering, although PD gives the largest peak amplitude at most such points. Over extended regions of its parameter space, the prescription even decides whether a point shows a completed first-order transition, a second-order transition, or electroweak symmetry non-restoration. Throughout, PD dresses only the scalars, the longitudinal gauge bosons receive leading Debye masses, the tunneling path is taken along the SM-like direction, as checked at a representative benchmark, and the gravitational-wave spectra assume a non-runaway plasma energy budget~\cite{Bodeker:2009qy,Espinosa:2010hh,Bodeker:2017cim,Hindmarsh:2015qta}.

The paper is organized as follows. Section~\ref{sec:PD} formulates the multi-field scalar gap equation with field-dependent mixing, contrasts full and partial dressing, shows that the multi-field PD potential is path dependent, and casts the Parwani and Arnold--Espinosa prescriptions as high-temperature truncations of full dressing. Section~\ref{sec:2HDM} specifies the conformal and non-conformal 2HDMs, their zero-temperature matching conditions and the scan constraints, and Section~\ref{sec:thermal-potential} constructs the three finite-temperature potentials, including the 2HDM gap-equation solver. Section~\ref{sec:phase_transition_dynamics} compares the prescriptions point by point, from the type of thermal history to the transition strength, supercooling and duration, and Section~\ref{sec:gw_implications} carries this comparison over to the gravitational-wave spectra. Section~\ref{sec:conclusion} concludes. Appendices~\ref{app:identities} and~\ref{app:counterterm} collect thermal-function identities and counterterm formulae, and Appendix~\ref{app:path-dependence} quantifies the path dependence of PD for representative benchmarks.

\section[Thermal Resummation Prescriptions]{Thermal Resummation Prescriptions}
\label{sec:PD}

Starting from the multi-field scalar gap equation with field-dependent mixing, we compare full dressing (FD) with partial dressing (PD). We find that the multi-field PD tadpole generally has a nonzero curl, so the potential obtained by integrating it depends on the path. We therefore define the PD potential used here by a straight-path integral. Finally, we formulate the Parwani and Arnold--Espinosa prescriptions, against which PD is compared, as high-temperature truncations of FD.

\subsection{Gap Equation for Multiple Fields} 
Consider a theory containing multiple real scalar fields \(\varphi_a\) with a general tree level scalar potential \(V_0\) that allows for mixing. The field dependent tree level mass matrix is defined by
\begin{equation}
m^2_{ab}(\varphi)
\equiv
\frac{\partial^2V_0}{\partial\varphi_a\partial\varphi_b}.
\label{eq:tree_mass_matrix}
\end{equation}
Let \(\mathbf u(\varphi)\) be the orthogonal matrix that diagonalizes \(\mathbf{m}^2(\varphi)\),
\begin{equation}
\big(\mathbf u^T(\varphi)\mathbf{m}^2(\varphi)\mathbf u(\varphi)\big)_{AB}
=
\delta_{AB}m_A^2(\varphi),
\label{eq:zero_temp_mass_rotation}
\end{equation}
where capital Latin indices label the mass basis. In terms of the mass eigenvalues \(m_A^2\), the finite temperature one-loop effective potential takes the form~\cite{Laine:2016}
\begin{equation}
V
=
V_0+V_1,\qquad V_1=\frac12\sum_A\mathcal{J}(m_A^2).
\label{eq:oneloop}
\end{equation}
The scalar loop function is

\begin{equation*}
\mathcal J(m^2)=\int_K\ln(K^2+m^2),\qquad
\int_K\equiv T\sum_{n\in\mathbb Z}\int\frac{d^3k}{(2\pi)^3}.
\end{equation*}
Here $K=(2\pi nT,\mathbf k)$ is a bosonic Euclidean momentum. The one-loop term $V_1$ includes the Coleman--Weinberg and thermal contributions\footnote{Since our discussion of resummation focuses on how the mass argument is modified, the explicit temperature dependence will be suppressed in the notation below.}. The explicit loop functions are given in Appendix~\ref{app:identities}.

Following Ref.~\cite{Bahl2024Improved}, we define the multi-field PD gap equation as
\begin{equation}
M^2_{ab}(\varphi,T)
=
\left.
\frac{\partial^2V}{\partial\varphi_a\partial\varphi_b}
\right|_{(m_A^2,\mathbf u)\to(M_A^2,\mathbf U)},
\label{eq:gap-compact}
\end{equation}
where the corrected mass matrix is diagonalized according to
\begin{equation}
\big(\mathbf U^T(\varphi,T)\mathbf{M}^2(\varphi,T)\mathbf U(\varphi,T)\big)_{AB}
=
\delta_{AB}M_A^2(\varphi,T).
\label{eq:dressed_mass_rotation}
\end{equation}
Using Eq.~\eqref{eq:oneloop}, Eq.~\eqref{eq:gap-compact} becomes
\begin{equation}
\begin{aligned}
M^2_{ab}
&=m^2_{ab}
+\frac12\sum_A
\left.
\left[
\mathcal J'(m_A^2)\partial_a\partial_bm_A^2
+\mathcal J''(m_A^2)\partial_am_A^2\partial_bm_A^2
\right]
\right|_{(m_A^2,\mathbf u)\to(M_A^2,\mathbf U)}
\\[3pt]
&=m^2_{ab}
+\frac12\,\CJTSelfTadpole{black}{dressed}_{ab}
+\frac12\,\CJTBubbleZero{black}{dressed}_{ab}.
\end{aligned}
\label{eq:gap-expanded}
\end{equation}
where $\partial_a\equiv\partial/\partial\varphi_a$. The second line contains the tadpole self-energy from a quartic vertex and the bubble self-energy from two cubic vertices, both evaluated at zero external momentum. Single and double internal lines denote tree-level and dressed propagators, respectively. 

It should be emphasized that the replacement in Eq.~\eqref{eq:gap-expanded} acts not only on the mass arguments of the loop functions, but also on the derivatives of the mass eigenvalues. This can be done by applying the Feynman--Hellmann theorem, which expresses these mass derivatives in terms of the mass eigenvalues and the mixing matrix~\cite{Bittar:2025lcr}. 
For the tree-level mass matrix \(\mathbf{m}^2(\varphi)\), one has
\begin{equation}
    \partial_am^2_A=\big(\mathbf u^T\partial_a\mathbf{m}^2\mathbf u\big)_{AA}.
    \label{eq:feynhell}
\end{equation}
The second derivative of the mass eigenvalue is more involved. Nevertheless, it can be derived in close analogy with the second-order perturbative correction to energy levels in quantum mechanics, yielding
\begin{equation}
\begin{aligned}
\partial_a\partial_b m_A^2
={}&
\left(\mathbf u^T\partial_a\partial_b \mathbf{m}^2\mathbf u\right)_{AA}
\\
&+
\sum_{B\neq A}
\frac{
\left(\mathbf u^T\partial_a\mathbf{m}^2\mathbf u\right)_{AB}
\left(\mathbf u^T\partial_b\mathbf{m}^2\mathbf u\right)_{BA}
+
\left(\mathbf u^T\partial_b\mathbf{m}^2\mathbf u\right)_{AB}
\left(\mathbf u^T\partial_a\mathbf{m}^2\mathbf u\right)_{BA}
}{
m_A^2-m_B^2
}.
\end{aligned}
\label{eq:feynhell-second}
\end{equation}
For degenerate eigenvalues, the corresponding term in the second line is taken to be zero. These relations are essential for solving the gap equation numerically.

Equation~\eqref{eq:gap-expanded} should be understood as a set of nonlinear algebraic equations for the dressed mass matrix \(\mathbf M^2(\varphi,T)\) at each point in field space and at each temperature. In practice, the equation is solved numerically by iteration: starting from an initial spectrum, the right-hand side is evaluated, the resulting matrix is diagonalized to obtain updated eigenvalues \(M_A^2\) and mixing matrix \(\mathbf U\), and the procedure is iterated until convergence. During this iteration, the derivatives of the tree level mass matrix \(\partial_a\mathbf{m}^2\) and \(\partial_a\partial_b \mathbf{m}^2\) are fixed once and for all and are not updated. 

\subsection{Full Dressing and Partial Dressing}
Partial dressing (PD) was originally introduced under the name of tadpole resummation~\cite{Boyd:1993,Curtin:2016}.
It is useful to compare the ordinary one-loop result, the full dressing prescription, and the partial dressing prescription directly at the level of the tadpoles. At one loop, before any resummation is performed, the tadpoles are given by
\begin{equation}
\begin{aligned}
\partial_a V
&=\partial_a V_0
+\frac12\sum_A\mathcal J^{\prime}(m_A^2)\,\partial_a m_A^2
\\[3pt]
&=\partial_a V_0+\frac12\,\CJTTadpole{black}{bare}_a.
\end{aligned}
\end{equation}
Here \(\mathcal J^{\prime}(m_A^2)\) represents the loop contribution with the tree-level mass eigenvalue \(m_A^2\), while \(\partial_a m_A^2\) gives the corresponding cubic interaction vertex.

The full dressing (FD) prescription is obtained by substituting the solution of the gap equation directly into the one-loop effective potential~\cite{Espinosa:1992,Espinosa:1993},
\begin{equation}
V_{\mathrm{FD}}
=
V\big|_{(m_A^2,\mathbf u)\to(M_A^2,\mathbf U)}
=
V_0
+
\frac12\sum_A \mathcal J(M_A^2).
\end{equation}
Differentiating this expression with respect to $\varphi_a$ gives the FD tadpole, with a filled dot marking the dressed vertex,
\begin{equation}
\begin{aligned}
\partial_a V_{\mathrm{FD}}
&=\partial_a V_0
+\frac12\sum_A\mathcal J^{\prime}(M_A^2)\,\partial_a M_A^2
\\[3pt]
&=\partial_a V_0+\frac12\,\CJTDressedVertexTadpole{black}{dressed}_a.
\end{aligned}
\end{equation}
As a result, both the propagator and the cubic vertex are dressed: the loop function depends on \(M_A^2\), and the vertex is given by \(\partial_a M_A^2\).

PD reverses this order: it first differentiates the one-loop potential and then replaces the masses and mixing matrix, giving the loop tadpole $\mathcal T_a^{\mathrm{PD}}$:
\begin{equation}
\begin{aligned}
\mathcal T_a^{\mathrm{PD}}
&\equiv\left.\partial_aV_1\right|_{(m_A^2,\mathbf u)\to(M_A^2,\mathbf U)}
=\frac12\sum_A\mathcal J^{\prime}(M_A^2)\,
\left.\partial_a m_A^2\right|_{\mathbf u\to\mathbf U}
\\[3pt]
&=\frac12\,\CJTTadpole{black}{dressed}_a.
\end{aligned}
\label{eq:PDtadpole}
\end{equation}
The full scalar tadpole is $\partial_aV_0+\mathcal T_a^{\mathrm{PD}}$. As Eq.~\eqref{eq:PDtadpole} shows, this construction dresses the propagators while retaining the tree-level vertices, rotated into the dressed mass basis, without differentiating the dressed masses at the vertices.

Because the undressed one-loop tadpole is a gradient, replacing its masses and mixing matrix may seem to preserve this property, as assumed in the path-independence argument of Ref.~\cite{Bahl2024Improved}. However, we find that the gradient property does not generally survive the PD replacement. To check this directly, we use Eq.~\eqref{eq:feynhell} to write the dressed tadpole as
\begin{equation}
\mathcal T_a^{\mathrm{PD}}
=\frac12\sum_A\mathcal J^{\prime}\big(M_A^2\big)\,
\big(\mathbf U^T\partial_a\mathbf m^2\mathbf U\big)_{AA}.
\label{eq:PDloop-tadpole}
\end{equation}
In this expression, the dressed masses and mixing matrix depend on the background through the gap equation, whereas the vertex contains only the derivative of the tree-level mass matrix. Differentiating the replaced tadpole, we find that its curl is generally nonzero:
\begin{equation}
\begin{aligned}
\partial_a\mathcal T_b^{\mathrm{PD}}-\partial_b\mathcal T_a^{\mathrm{PD}}
&=\frac12\sum_A\Big[
\mathcal J''(M_A^2)\,(\partial_a M_A^2)\,
\big(\mathbf U^T\partial_b\mathbf m^2\mathbf U\big)_{AA}
\\
&\qquad+2\mathcal J'(M_A^2)\,
\big(\mathbf U^T\partial_b\mathbf m^2\,\partial_a\mathbf U\big)_{AA}
-(a\leftrightarrow b)\Big]\not\equiv0,
\end{aligned}
\label{eq:pd-curl}
\end{equation}
which means that different paths between the same endpoints can give different potentials: the direct multi-field extension of PD does not define a unique effective potential. This is a theoretical inconsistency of the construction, not a numerical artifact. We quantify its numerical impact for 2HDM benchmarks in Appendix~\ref{app:path-dependence}, where changing the integration path shifts the transition temperatures by at most about \(7\%\) and the gravitational-wave peak amplitude by up to a factor of about three. Our subsequent work~\cite{LiWangZhao:MultifieldCJT} shows, for general scalar theories, that the curl comes entirely from the zero-momentum sunset self-energy in the gap equation and is not removed by adding the lollipop. PD completed with both the lollipop and the exact momentum-overlap factors is instead the gradient of the variational Cornwall--Jackiw--Tomboulis (CJT) potential~\cite{Cornwall1974Effective}, and hence path independent for any number of mixing fields.

 In this work, we define the PD potential using the straight path $\gamma_a(t)=\varphi_at$, with $t\in[0,1]$. With the field-independent integration constant set to zero, this gives 
\begin{equation}
V_{\mathrm{PD}}(\varphi,T)
=
V_0(\varphi)
+
\int_0^1 dt \,\varphi_a \mathcal T_a^{\mathrm{PD}}(\varphi t,T),
\label{eq:VPD}
\end{equation}
where the index $a$ is summed and all quantities in the integrand are evaluated on the path. Below, $V_{\mathrm{PD}}$ always denotes this straight-path result.

Besides this difference in integrability, FD and PD also differ in how they count resummed diagrams. In FD, inserting the dressed masses directly into the one-loop potential dresses both the internal propagator and the interaction vertex; differentiating \(V_{\mathrm{FD}}\) therefore produces vertex-dressing terms such as \(\partial_a M_A^2\). In a loop expansion, these terms do not simply reproduce the daisy-chain resummation and lead to a miscounting of daisy and superdaisy contributions beyond one loop~\cite{Boyd:1993,Curtin:2016,Bahl2024Improved}. PD avoids this overcounting by applying the replacement at the tadpole level: the internal line is dressed through \(M_A^2\), while the vertex remains at its tree-level form, up to the rotation to the corrected mass basis.

However, the PD prescription should not be interpreted as a complete reconstruction of the full higher-loop effective potential. Certain subleading diagrams, such as lollipop contributions and sunset-type terms with their exact combinatorial factors, are not fully reproduced by the minimal PD construction. For the sunset contribution, a factor $2/3$ corrects the gap equation when the internal masses are equal and the temperature is high, whereas mixed-mass channels require different factors~\cite{Bahl2024Improved}. Lollipop terms must instead be added separately. Although the sunset corrections were small in the examples of Ref.~\cite{Bahl2024Improved}, their size has not been tested across our 2HDM scan.
Hybrid partial dressing (HPD) improves this sunset treatment by using Hartree masses and adding the full two-loop sunset directly to the potential~\cite{Berthiaume2026Hybrid}. Its original treatment, however, covers one scalar and two scalars without mixing. For the present numerical comparison, we retain minimal PD and omit the additional corrections described above.

\Needspace{5\baselineskip}
\subsection{Truncated Full Dressing: Parwani and Arnold--Espinosa}
The FD prescription uses the full gap-equation masses \(M_A^2\). A commonly used approximation is obtained by truncating the gap equation to leading order and evaluating the resulting self-energies in the high-temperature expansion. In the presence of scalar mixing, this gives a truncated thermal mass matrix of the form
\begin{equation}
\bar{\mathbf m}^2(\varphi,T)
=
\mathbf m^2(\varphi)+\boldsymbol{\Pi}(T),
\label{eq:truncatedmass}
\end{equation}
whose eigenvalues are denoted by \(\bar m_A^2\). Here \(\boldsymbol{\Pi}(T)\) is the leading high-temperature self-energy matrix, which gives a simple \(T^2\)-type correction. At the one-loop level, the Parwani~\cite{Parwani:1991gq} and Arnold--Espinosa~\cite{Arnold:1992rz} prescriptions can be understood as specific implementations of this truncated full dressing approximation. We compare their predictions with PD.

The Parwani prescription inserts these truncated thermal masses directly into the one-loop potential,
\begin{equation}
V_{\mathrm{Parwani}}
=
V_0
+
\frac12\sum_A \mathcal J(\bar m_A^2).
\end{equation}
The Arnold--Espinosa prescription instead separates the bosonic one-loop function into a soft zero-mode contribution and a hard contribution from the non-zero Matsubara modes\cite{Laine:2016,Bahl2024Improved},
\begin{equation}
\mathcal J(m_A^2)
=
\mathcal J_{\mathrm{hard}}(m_A^2)
+
\mathcal J_{\mathrm{soft}}(m_A^2).
\end{equation}
Here \(\mathcal J_{\mathrm{soft}}\) is obtained by isolating the zero mode from the Matsubara sum,
\begin{equation}
\mathcal J_{\mathrm{soft}}(m_A^2)
=
T
\int\frac{d^3k}{(2\pi)^3}
\ln\left(k^2+m_A^2\right),
\end{equation}
while \(\mathcal J_{\mathrm{hard}}\) contains the remaining non-zero Matsubara modes. For details of this decomposition, see Appendix~\ref{app:identities}. In dimensional regularization, the soft contribution evaluates to
\begin{equation}
\mathcal J_{\mathrm{soft}}(m_A^2)
=
-\frac{T}{6\pi}
\left(m_A^2\right)^{3/2}.
\end{equation}
The Arnold--Espinosa prescription implements the thermal mass replacement only in this soft zero mode contribution,
\begin{equation}
V_{\mathrm{AE}}
=
V_0
+
\frac12\sum_A
\left[
\mathcal J_{\mathrm{hard}}(m_A^2)
+
\mathcal J_{\mathrm{soft}}(\bar m_A^2)
\right].
\end{equation}
Equivalently,
\begin{equation}
V_{\mathrm{AE}}
=
V_0
+
\frac12\sum_A\mathcal J(m_A^2)
-
\frac{T}{12\pi}
\sum_A 
\left[
\left(\bar m_A^2\right)^{3/2}
-
\left(m_A^2\right)^{3/2}
\right],
\label{eq:AE}
\end{equation}
Thus, at one loop, both prescriptions can be viewed as truncated full dressing schemes based on the high temperature solution \(\bar m_A^2\), but they implement the replacement in different parts of the potential~\cite{Curtin:2016}. Parwani applies the replacement throughout the one-loop function, whereas Arnold--Espinosa applies it only to the infrared-sensitive Matsubara zero mode, i.e.\ to the daisy contribution. Both schemes share the same limitation: their thermal masses rely on the high-temperature expansion and can therefore miss the Boltzmann suppression of heavy field-dependent modes, which PD retains through the full thermal functions in its gap equation.

\section{Two-Higgs-doublet Model}
\label{sec:2HDM}
The two-Higgs-doublet model (2HDM) is a minimal extension of the Standard Model scalar sector, obtained by adding a second electroweak scalar doublet while leaving the gauge structure unchanged~\cite{Branco_2012}. Its enlarged Higgs spectrum, including additional CP-even, CP-odd, and charged scalars, can significantly affect the finite temperature effective potential and allow for a first-order electroweak phase transition~\cite{Cline:1996mga,Dorsch:2016nrg}. To compare the thermal prescriptions at fixed inputs, we first specify the zero-temperature spectra, matching conditions, and scan constraints for the classically conformal and non-conformal 2HDMs. We begin with the conformal model, in which electroweak symmetry breaking and the SM-like Higgs mass arise at one loop, then turn to the non-conformal model with a softly broken \(\mathbb Z_2\) symmetry, and close with the constraints and scan ranges (Section~\ref{sec:constraints}). Each prescription is then applied to the same accepted points.

\subsection{Conformal 2HDM and Tree Level Spectrum}
We start with the classically conformal, CP conserving two-Higgs-doublet model (C2HDM). The scalar sector contains two complex \(SU(2)_L\) doublets \(\Phi_1\) and \(\Phi_2\) with hypercharge \(1/2\). Classical conformal invariance forbids all explicit quadratic terms in the tree-level potential, so electroweak symmetry breaking is generated radiatively, making the vacuum structure particularly sensitive to loop effects~\cite{Coleman:1973jx,Gildener:1976ih}. We impose the usual \(\mathbb{Z}_2\) symmetry,
\begin{equation}
\Phi_1\to \Phi_1,
\qquad
\Phi_2\to -\Phi_2 ,
\label{eq:Z2}
\end{equation}
which, together with the appropriate Yukawa assignments, avoids tree-level flavor-changing neutral currents~\cite{Glashow1977Natural,Paschos1977Diagonal}. Since the parameter region studied below is dominated by the top quark contribution, we do not distinguish between the different \(\mathbb{Z}_2\) symmetric Yukawa types and neglect lighter fermions.

The tree-level scalar potential is then
\begin{align}
V_0 ={}& \frac{\lambda_1}{2}\left(\Phi_1^\dagger\Phi_1\right)^2
+ \frac{\lambda_2}{2}\left(\Phi_2^\dagger\Phi_2\right)^2
+ \lambda_3\left(\Phi_1^\dagger\Phi_1\right)\left(\Phi_2^\dagger\Phi_2\right)
\nonumber\\
&+ \lambda_4\left(\Phi_1^\dagger\Phi_2\right)\left(\Phi_2^\dagger\Phi_1\right)
+ \frac{\lambda_5}{2}\left[
\left(\Phi_1^\dagger\Phi_2\right)^2+\text{h.c.}
\right].
\label{eq:V0_conformal_2HDM}
\end{align}
Here all quartic couplings are real and dimensionless. The absence of the usual mass terms follows from classical conformal invariance. The two Higgs doublets \(\Phi_{1,2}\) can be parametrized as
\begin{equation}
\Phi_1=
\begin{pmatrix}
\phi_1^+\\[2pt]
\dfrac{1}{\sqrt2}(h_1+i a_1)
\end{pmatrix},
\qquad
\Phi_2=
\begin{pmatrix}
\phi_2^+\\[2pt]
\dfrac{1}{\sqrt2}(h_2+i a_2)
\end{pmatrix},
\label{eq:doublets_conformal_2HDM}
\end{equation}
The electroweak vacuum is located at
\[
h_i=v_i,\qquad a_i=\phi_i^\pm=0 .
\]
where \(v_1\equiv v\cos\beta \equiv vc_\beta\) and \(v_2\equiv v\sin\beta \equiv vs_\beta\). Thus \(h_{1,2}\) denote the full neutral CP-even field directions, while \(a_{1,2}\) and \(\phi_{1,2}^\pm\) denote the CP-odd and charged fields. 

The tree-level tadpole conditions at the electroweak vacuum are
\begin{equation}
\begin{aligned}
\left.\frac{\partial V_0}{\partial h_1}\right|_{\rm vac}\,=\frac{1}{2}v_1\big(\lambda_1 v_1^2 + \lambda_{345}v_2^2\big) = 0,
\\
\left.\frac{\partial V_0}{\partial h_2}\right|_{\rm vac}\,=\frac{1}{2}v_2\big(\lambda_2 v_2^2 + \lambda_{345}v_1^2\big) = 0,
\label{eq:tadpole_tree}
\end{aligned}
\end{equation}
with \(\lambda_{345}\equiv\lambda_3+\lambda_4+\lambda_5\). It follows that
\begin{equation}
\frac{\lambda_1}{\lambda_2}=\tan^4\beta,\qquad
\lambda_{345}=-\sqrt{\lambda_1\lambda_2}.
\end{equation}
Because the classically conformal potential in Eq.~\eqref{eq:V0_conformal_2HDM} is homogeneous in the background fields, these conditions do not fix the radial scale at tree level. Instead they select a Gildener--Weinberg ray in the CP-even field space,
\begin{equation}
h_1=c_\beta h,\qquad h_2=s_\beta h ,
\label{eq:c2hdm_flat_ray}
\end{equation}
along which the tree-level potential is flat. The radial field and the orthogonal CP-even direction are
\begin{equation}
h = c_\beta\,h_1 + s_\beta\,h_2,
\qquad
H=-s_\beta\,h_1+c_\beta\,h_2 .
\label{eq:flat_dir}
\end{equation}
The electroweak scale and the mass of the SM-like scalar are therefore fixed only after the one loop potential is included. The scalar fluctuation along the \(h\) direction is the scalon-like state and is identified with the SM-like Higgs boson in the phenomenologically relevant region.

The tree-level vacuum mass matrix is defined as the Hessian of \(V_0\) evaluated at the electroweak vacuum:
\begin{equation}
\big(\mathcal{M}^2_0\big)_{ab}
=
\left.
\frac{\partial^2 V_0}{\partial \varphi_a\partial \varphi_b}
\right|_{\rm vac},
\qquad
a,b=1,\ldots,8 .
\end{equation}
with \(\varphi_a\in\{h_1,h_2,a_1,a_2,\phi_1^\pm,\phi_2^\pm\}\). At the electrically neutral CP-conserving vacuum, this \(8\times 8\) Hessian is automatically block diagonal: electric charge conservation and CP conservation forbid quadratic mixing between the neutral CP-even, neutral CP-odd, and charged sectors. Therefore, one may write~\cite{Lee:2012jn}
\begin{equation}
\mathcal{M}^2_0
=
\operatorname{diag}
\left(
\mathcal M_{{\mathcal H},0}^2\,,\,
\mathcal M_{{\mathcal A}}^2\,,\,
\mathcal M_{{\mathcal H}^\pm}^2\,,\,
\mathcal M_{{\mathcal H}^\pm}^2
\right).
\label{eq:full_hessian_block_diagonal}
\end{equation}
It is useful to introduce the common angular matrix
\begin{equation}
P_\beta \equiv
\begin{pmatrix}
s_\beta^2 & -s_\beta c_\beta\\[4pt]
-s_\beta c_\beta & c_\beta^2
\end{pmatrix}.
\label{eq:Pbeta_def}
\end{equation}
Then the nontrivial tree-level vacuum mass blocks are
\begin{equation}
\mathcal{M}_{{\mathcal H},0}^2=m_{{\mathrm H},0}^2 P_\beta,
\qquad
\mathcal{M}_{{\mathcal A}}^2=m_{{\mathrm A}^0}^2 P_\beta,
\qquad
\mathcal{M}_{{\mathcal H}^\pm}^2=m_{\mathrm H^\pm}^2 P_\beta,
\label{eq:tree_level_mass_blocks}
\end{equation}
with
\begin{equation}
m_{{\mathrm H},0}^2\equiv-\lambda_{345}v^2,
\qquad
m_{{\mathrm A}^0}^2\equiv-\lambda_5 v^2,
\qquad
m_{\mathrm H^\pm}^2\equiv-\frac{\lambda_4+\lambda_5}{2}v^2 .
\label{eq:tree_level_mass_parameters}
\end{equation}
Each block has one zero eigenvalue along the vacuum direction \((c_\beta,s_\beta)^T\). In the CP-even sector this zero mode is the tree-level flat direction, while in the CP-odd and charged sectors it corresponds to the Goldstone modes. Equivalently, diagonalizing these blocks amounts to separating the vacuum direction from the orthogonal massive direction in each sector. The resulting scalar spectrum is therefore organized into the CP-even states \(\mathrm h,\mathrm H\), the neutral Goldstone and CP-odd states \(\mathrm G^0,\mathrm A^0\), and the charged states \(\mathrm G^\pm,\mathrm H^\pm\).

\subsection{Coleman--Weinberg Potential and Parameter Matching}
The vacuum mass matrices introduced above are sufficient for defining the tree-level spectrum at the electroweak minimum. However, both the Coleman--Weinberg potential and the finite temperature effective potential are functions of a generic scalar background. Therefore, in the loop corrections we must use the full field dependent spectrum rather than only the vacuum mass matrices.  Let \(\varphi\) denote a generic scalar background in the full scalar field space. The scalar field-dependent mass matrix is defined by
\begin{equation}
\bigl[\mathbf m_{\rm S}^2(\varphi)\bigr]_{ab}
\equiv
\frac{\partial^2 V_0}{\partial\varphi_a\,\partial\varphi_b},
\qquad
\mathbf u^T(\varphi)\mathbf m_{\rm S}^2(\varphi)\mathbf u(\varphi)
=
\operatorname{diag}\{m_{A}^2(\varphi)\},
\label{eq:c2hdm_field_mass_general}
\end{equation}
where the capital Latin index \(A\) labels the scalar spectrum,
\begin{equation}
A\in S=\{\mathrm h,\mathrm H,\mathrm G^0,\mathrm A^0,\mathrm G^\pm,\mathrm H^\pm\}.
\end{equation}
The non scalar field dependent masses, denoted collectively by \(m_{V,F}^2(\varphi)\), are obtained from the gauge and Yukawa interactions. Together these eigenvalues form the set \(m_i^2(\varphi)\) entering the Coleman--Weinberg potential and the perturbative finite temperature potential. Evaluating Eq.~\eqref{eq:c2hdm_field_mass_general} at the electroweak vacuum reproduces the vacuum matrices in Eq.~\eqref{eq:tree_level_mass_blocks}.

Using these field dependent masses, and working in the \(\overline{\rm MS}\) scheme and Landau gauge, the Coleman--Weinberg potential of the \(i\) th mode is
\begin{equation}
V_{{\rm CW},i}(m_i^2(\varphi))
=
\frac{1}{64\pi^2}
(-1)^{2s_i} n_i
m_i^4(\varphi)
\left(
\ln\frac{m_i^2(\varphi)}{\bar\mu^2}
-c_i
\right).
\label{eq:c2hdm_vcw}
\end{equation}
Here \(s_i\), \(n_i\), and \(m_i^2(\varphi)\) are respectively the spin, multiplicity factor, and field-dependent squared mass of the \(i\) th mode. The species index \(i=S,V,F\) includes the scalar (S) eigenvalues defined in Eq.~\eqref{eq:c2hdm_field_mass_general}, the electroweak gauge bosons (V), and the top quark (F). The multiplicity factors and renormalization constants used in Eq.~\eqref{eq:c2hdm_vcw} are
\begin{equation}
\begin{aligned}
&n_\varphi=1,\quad n_t=12,\quad
n_{W_T}=4,\quad n_{W_L}=2,\quad
n_{Z_T}=2,\quad n_{Z_L}=1,\\
&c_\varphi=c_t=c_{W_L}=c_{Z_L}=\frac{3}{2},
\qquad
c_{W_T}=c_{Z_T}=\frac{1}{2}.
\end{aligned}
\label{eq:CW_multiplicities_constants}
\end{equation}
Here \(\varphi\) denotes each scalar eigenmode. The transverse and longitudinal gauge polarizations are treated separately. 

The Coleman--Weinberg correction by itself would in general shift the location of the electroweak minimum and modify the mass matrices in all scalar sectors. We therefore adopt a finite counterterm prescription in which the vacuum position and the CP-odd and charged mass blocks are kept at their tree level forms, while the CP-even flat direction curvature is left to be generated radiatively. The zero temperature one loop effective potential is thus written as
\begin{equation}
V^{T=0}
=
V_0+V_{\rm CT}+V_{\rm CW}.
\label{eq:c2hdm_zeroT_eff_potential}
\end{equation}
The finite counterterms are fixed by requiring that the one loop correction does not shift the vacuum and does not modify the CP-odd and charged mass blocks,
\begin{alignat}{2} 
\text{(Tadpoles):}\quad &\left.\frac{\partial\bigl(V_{\rm CW}+V_{\rm CT}\bigr)}{\partial h_i}\right|_{\rm vac} &\,{}= 0,
\label{eq:CTtadpole}
\\[2pt] 
\text{(CP-odd):}\quad &\left.\frac{\partial^2\bigl(V_{\rm CW}+V_{\rm CT}\bigr)}{\partial a_i\,\partial a_j}\right|_{\rm vac} &\,{}= 0, 
\\[2pt] 
\text{(Charged):}\quad &\left.\frac{\partial^2\bigl(V_{\rm CW}+V_{\rm CT}\bigr)}{\partial\phi_i^-\,\partial\phi_j^+}\right|_{\rm vac} &\,{}= 0. 
\label{eq:charged}
\end{alignat}
The explicit counterterm potential and the solutions for the coefficients are given in Appendix~\ref{app:counterterm}. 

The first condition Eq.~\eqref{eq:CTtadpole} ensures that the electroweak vacuum remains located at \((v_1,v_2)\) after including the Coleman--Weinberg correction. The second and third conditions restore the CP-odd and charged vacuum mass matrices to their tree level forms in Eq.~\eqref{eq:tree_level_mass_blocks}. Therefore the input masses in these sectors, defined as zero-momentum curvature masses of the one-loop potential, can be matched using the same tree-level relations as follows
\begin{equation}
\lambda_5 = -\frac{m_{{\mathrm A}^0}^2}{v^2},
\qquad
\lambda_4 = \frac{m_{{\mathrm A}^0}^2-2m_{{\mathrm H}^\pm}^2}{v^2}.
\label{eq:l45}
\end{equation}
Together with the flat direction conditions in Eq.~\eqref{eq:tadpole_tree}, one obtains
\begin{equation}
\lambda_1=-\lambda_{345}\tan^2\!\beta,
\qquad
\lambda_2=-\lambda_{345}\cot^2\!\beta,
\qquad
\lambda_3=\lambda_{345}-\lambda_4-\lambda_5 .
\label{eq:l123_from_l345}
\end{equation}
Thus, once \(m_{{\mathrm A}^0}\), \(m_{{\mathrm H}^\pm}\), and \(\tan\beta\) are specified, the five quartic couplings are determined up to the single remaining parameter \(\lambda_{345}\). This remaining freedom is fixed by the neutral CP-even sector, where the treatment differs from the CP-odd and charged sectors.

In contrast to the CP-odd and charged blocks, no renormalization condition is imposed to restore the CP-even mass matrix to its tree level form. In particular, the one loop curvature along the tree level flat direction is not subtracted away. This is essential in the classically conformal model: the observed SM-like Higgs mass is generated radiatively from the Coleman--Weinberg correction, rather than inserted as an independent tree level mass parameter. 

The one loop corrected CP-even mass matrix at the electroweak vacuum reads
\begin{equation}
\mathcal{M}_{\mathcal H}^2
=
\mathcal{M}_{{\mathcal H},0}^2+\Delta,
\label{eq:MH2}
\end{equation}
where the correction matrix is
\begin{equation}
\Delta_{ij}
=
\left.
\frac{\partial^2\bigl(V_{\rm CW}+V_{\rm CT}\bigr)}
{\partial h_i\,\partial h_j}
\right|_{\rm vac}.
\label{eq:Delta_CP_even}
\end{equation}
Identifying the radiatively generated CP-even eigenvalue on the SM-like branch (Section~\ref{sec:constraints}) with the observed Higgs mass \(m_\mathrm h=125.09~\mathrm{GeV}\) fixes \(\lambda_{345}\) through
\begin{equation}
\det\!\left(\mathcal{M}_{\mathcal H}^2-m_\mathrm h^2\mathbf{1}\right)=0 .
\label{eq:det_cond}
\end{equation}
Using Eq.~\eqref{eq:tree_level_mass_blocks}, this condition can be written as the implicit one-dimensional equation
\begin{equation}
\lambda_{345}
=
\frac{
(\Delta_{11}-m_{\mathrm h}^2)(\Delta_{22}-m_{\mathrm h}^2)-\Delta_{12}^2
}{
v_2^2(\Delta_{22}-m_{\mathrm h}^2)
+v_1^2(\Delta_{11}-m_{\mathrm h}^2)
+2v_1v_2\Delta_{12}
}.
\label{eq:l345_exact}
\end{equation}
Since \(\Delta_{ij}\) depends on the field-dependent spectrum and therefore on \(\lambda_{345}\) itself, we solve Eq.~\eqref{eq:l345_exact} as a nonlinear equation for $\lambda_{345}$. 

In summary, the conformal parametrization is specified by
\begin{equation}
\left\{m_{{\mathrm A}^0},\ m_{{\mathrm H}^\pm},\ \tan\beta \right\},
\label{eq:conformal_parametrization}
\end{equation}
with \(m_{\mathrm h}\) fixed to its observed value through the CP-even matching condition above.  To complete the numerical setup, we also specify two technical ingredients that enter the evaluation of the one loop potential and the counterterm conditions: the Goldstone infrared regulator and the choice of the renormalization scale.

A numerical complication arises from the Goldstone catastrophe. The counterterm conditions Eq.~\eqref{eq:CTtadpole}--\eqref{eq:charged} require first and second derivatives of the CW potential at the electroweak vacuum. In Landau gauge, however, the CP-odd and charged Goldstone modes are massless at this point, which leads to the familiar Goldstone infrared problem~\cite{Martin2014Taming}.
We regularize the corresponding derivatives by shifting the Goldstone masses as \(m_G^2\to m_G^2+\mu_{\rm IR}^2\). In the numerical analysis we take \(\mu_{\rm IR}^2=1~{\rm GeV}^2\), and have checked that moderate variations of this regulator do not affect the main results within the numerical accuracy of the computation. With this prescription, the counterterms keep the CP-odd and charged Goldstone massless in \(V_0+V_{\rm CW}+V_{\rm CT}\) at the electroweak minimum.

The counterterm conditions are imposed at the electroweak vacuum, so we also set the renormalization scale in the Coleman--Weinberg potential to $\bar\mu=v$. Along the flat direction, the tree-level potential vanishes, while the finite counterterm contribution cancels the explicit scale dependence of the one loop Coleman-Weinberg potential, as shown in Appendix~\ref{app:counterterm}. 

\subsection{Non-conformal 2HDM}
We now turn to the non-conformal CP-conserving two-Higgs-doublet model (NC2HDM). In contrast to the classically conformal case, the scalar potential already contains dimensionful quadratic terms at tree level and allows a soft breaking of the \(\mathbb Z_2\) symmetry in Eq.~\eqref{eq:Z2} through the bilinear parameter \(m_{12}^2\)~\cite{Branco_2012}.
The tree-level scalar potential is therefore given by
\begin{align}
V_0 ={}& m_{11}^2\,\Phi_1^\dagger\Phi_1
+ m_{22}^2\,\Phi_2^\dagger\Phi_2
- m_{12}^2\left(\Phi_1^\dagger\Phi_2+\text{h.c.}\right)
\nonumber\\
&+ \frac{\lambda_1}{2}\left(\Phi_1^\dagger\Phi_1\right)^2
+ \frac{\lambda_2}{2}\left(\Phi_2^\dagger\Phi_2\right)^2
+ \lambda_3\left(\Phi_1^\dagger\Phi_1\right)\left(\Phi_2^\dagger\Phi_2\right)
\nonumber\\
&+ \lambda_4\left(\Phi_1^\dagger\Phi_2\right)\left(\Phi_2^\dagger\Phi_1\right)
+ \frac{\lambda_5}{2}\left[\left(\Phi_1^\dagger\Phi_2\right)^2+\text{h.c.}\right].
\label{eq:V0_nonconformal_2HDM}
\end{align}
All parameters in Eq.~\eqref{eq:V0_nonconformal_2HDM} are taken to be real. As a result, CP-even/odd bilinear mixing is absent, and the neutral mass eigenstates are CP eigenstates.

The mass spectrum and physical parametrization can be obtained in complete analogy with the conformal case~\cite{Benincasa:2026c2hdm}. The only difference is that, in the non-conformal model, the counterterm conditions can be chosen such that the tadpoles and all blocks of the scalar mass matrix retain their tree-level forms. The explicit counterterm conditions are given in Appendix~\ref{app:counterterm}.

As a result, the physical parametrization can be determined directly from the tree-level relations. For completeness, we list the relevant relations below. The tadpole conditions at the electroweak vacuum are
\begin{equation}
\begin{aligned}
\left.\frac{\partial V_0}{\partial h_1}\right|_{\rm vac}
&=
m^2_{11}v_1-m^2_{12}v_2
+\frac{1}{2}v_1\left(\lambda_1 v_1^2+\lambda_{345}v_2^2\right)
=0,
\\
\left.\frac{\partial V_0}{\partial h_2}\right|_{\rm vac}
&=
m^2_{22}v_2-m^2_{12}v_1
+\frac{1}{2}v_2\left(\lambda_2 v_2^2+\lambda_{345}v_1^2\right)
=0.
\label{eq:tadpole_tree_NC}
\end{aligned}
\end{equation}
The vacuum mass matrix is block diagonal, as in the conformal case in Eq.~\eqref{eq:full_hessian_block_diagonal}. The CP-odd and charged blocks take the same form,
\begin{equation}
\mathcal{M}_{{\mathcal A}}^2=m_{{\mathrm A}^0}^2 P_\beta,
\qquad
\mathcal{M}_{{\mathcal H}^\pm}^2=m_{\mathrm H^\pm}^2 P_\beta,
\label{eq:tree_level_mass_blocks_NC}
\end{equation}
with
\begin{equation}
m_{{\mathrm A}^0}^2\equiv -\lambda_5 v^2+M^2,
\qquad
m_{\mathrm H^\pm}^2\equiv -\frac{\lambda_4+\lambda_5}{2}v^2+M^2,
\label{eq:CPoddcharged_NC}
\end{equation}
where
\begin{equation}
M^2\equiv \frac{m_{12}^2}{s_\beta c_\beta}.
\end{equation}
Unlike in the conformal case, the presence of dimensionful mass parameters makes the CP-even spectrum already massive at tree level. The corresponding CP-even mass matrix is given by
\begin{equation}
\mathcal{M}_{{\mathcal H}}^2 =
\begin{pmatrix}
\lambda_1 v^2c_\beta^2+ M^2 s_\beta^2
&
\left(\lambda_{345}v^2- M^2\right)s_\beta c_\beta
\\
\left(\lambda_{345}v^2- M^2\right)  s_\beta c_\beta
&
\lambda_2 v^2s_\beta^2+M^2 c_\beta^2
\end{pmatrix}.
\label{eq:CPeven_NC}
\end{equation}
To facilitate comparison with the C2HDM analysis, we impose the alignment limit in the NC2HDM, such that the CP-even mass eigenstates follow the same directions as the flat and orthogonal directions in the conformal case in Eq.~\eqref{eq:flat_dir}:
\begin{equation}
h = c_\beta\,h_1 + s_\beta\,h_2,
\qquad
H=-s_\beta\,h_1+c_\beta\,h_2 .
\label{eq:NC_alignment_basis}
\end{equation}
Diagonalising Eq.~\eqref{eq:CPeven_NC} in this basis gives
\begin{equation}
\begin{aligned}
\lambda_1 v^2c_\beta^2+M^2 s_\beta^2
&=
m_{{\mathrm H}}^2s_\beta^2+m_{\mathrm h}^2c_\beta^2,
\\
\lambda_2 v^2s_\beta^2+M^2 c_\beta^2
&=
m_{{\mathrm H}}^2c_\beta^2+m_{\mathrm h}^2s_\beta^2,
\\
\lambda_{345}v^2- M^2
&=
m_{\mathrm h}^2-m_{{\mathrm H}}^2.
\label{eq:CPeven_eigenstates_NC}
\end{aligned}
\end{equation}
Solving Eqs.~\eqref{eq:tadpole_tree_NC}, \eqref{eq:CPoddcharged_NC} and \eqref{eq:CPeven_eigenstates_NC}, one obtains
\begin{equation}
\begin{aligned}
m_{11}^2
&=
-\tfrac{1}{2}m_{\mathrm h}^2+M^2\sin^2\beta,
\\
m_{22}^2
&=
-\tfrac{1}{2}m_{\mathrm h}^2+M^2\cos^2\beta,
\\
\lambda_1 v^2
&=
m_{\mathrm h}^2+\left(m_{{\mathrm H}}^2-M^2\right)\tan^2\beta,
\\
\lambda_2 v^2
&=
m_{\mathrm h}^2+\left(m_{{\mathrm H}}^2-M^2\right)\cot^2\beta,
\\
\lambda_3 v^2
&=
m_{\mathrm h}^2-m_{{\mathrm H}}^2+2m_{\mathrm H^\pm}^2-M^2,
\\
\lambda_4 v^2
&=
m_{{\mathrm A}^0}^2-2m_{\mathrm H^\pm}^2+M^2,
\\
\lambda_5 v^2
&=
-m_{{\mathrm A}^0}^2+M^2.
\end{aligned}
\end{equation}
Therefore, the physical parametrization is specified by
\begin{equation}
\left\{
m_{\mathrm h},\ m_{{\mathrm H}},\ m_{{\mathrm A}^0},\ m_{{\mathrm H}^\pm},\ \tan\beta,\ M
\right\}.
\label{eq:nonconformal_parametrization}
\end{equation}
In the numerical scan, \(m_{\mathrm h}=125.09~\mathrm{GeV}\) is fixed. The actual scan variables are therefore \(m_{{\mathrm H}}\), \(m_{{\mathrm A}^0}\), \(m_{{\mathrm H}^\pm}\), \(\tan\beta\), and \(M\).

\subsection{Theoretical and Experimental Constraints}
\label{sec:constraints}
Before performing the finite temperature analysis, we impose the zero-temperature constraints following Ref.~\cite{Benincasa:2026c2hdm}, summarized below for completeness. We require the quartic couplings to remain in the perturbative regime,
\begin{equation}
    |\lambda_i| < 4\pi,
    \qquad i=1,\ldots,5 .
    \label{eq:perturbativity}
\end{equation}
Tree-level scalar unitarity requires \(|e_i|<8\pi\) for all eigenvalues of the \(2\to2\) scalar scattering matrix. For the CP-conserving \(\mathbb Z_2\)-symmetric 2HDM, these eigenvalues are~\cite{Becirevic:2015fmu,Akeroyd:2000Unitarity}
\begin{equation}
\begin{split}
e_{1,2} &= \lambda_3 \pm \lambda_4,\qquad
e_{3,4} = \lambda_3 \pm \lambda_5,\qquad
e_{5,6} = \lambda_3+2\lambda_4 \pm 3\lambda_5,\\
e_{7,8} &= \frac{1}{2}\left[
\lambda_1+\lambda_2
\pm \sqrt{(\lambda_1-\lambda_2)^2+4\lambda_4^2}
\right],\\
e_{9,10} &= \frac{1}{2}\left[
\lambda_1+\lambda_2
\pm \sqrt{(\lambda_1-\lambda_2)^2+4\lambda_5^2}
\right],\\
e_{11,12} &= \frac{3}{2}(\lambda_1+\lambda_2)
\pm \sqrt{
\frac{9}{4}(\lambda_1-\lambda_2)^2+(2\lambda_3+\lambda_4)^2
}.
\end{split}
\label{eq:unitarity_eigenvalues}
\end{equation}

The tree-level scalar potential is also required to be bounded from below.  For the non-conformal model this gives the standard conditions \cite{Deshpande:1977rw,Branco_2012}
\begin{equation}
    \lambda_1>0,
    \qquad
    \lambda_2>0,
    \qquad
    \sqrt{\lambda_1\lambda_2}
    +\lambda_3
    +\min\{0,\lambda_4-|\lambda_5|\}
    >0 .
    \label{eq:bfb_conditions}
\end{equation}
For the conformal model, the Gildener--Weinberg direction lies on the boundary of tree-level stability~\cite{Gildener:1976ih,Lee:2012jn}.
We therefore require the quartic potential to be non negative, allowing equality along the flat direction, and impose that the one loop Coleman--Weinberg potential generates a stable electroweak vacuum.

The collider and electroweak precision constraints include direct LEP limits and the correlated oblique-parameter fit. We implement the LEP constraints by imposing~\cite{ParticleDataGroup:2020ssz}
\begin{equation}
    m_{{\mathrm A}^0} \gtrsim 90 \text{~GeV}\quad (e^+e^- \rightarrow hA),\quad m_{{\mathrm H}^\pm} \gtrsim 80 \text{~GeV}\quad (e^+e^- \rightarrow H^+H^-).
\label{eq:lep_bounds}
\end{equation}

Electroweak precision constraints are imposed through the oblique parameters \(S\), \(T\) and \(U\)~\cite{Peskin:1991sw}. In extended Higgs sectors these parameters mainly constrain custodial symmetry breaking effects induced by mass splittings among the additional scalar states. We use the correlated global fit of Ref.~\cite{Lu:2022bgw},
\begin{equation}
\begin{aligned}
    \widehat S &= 0.06\pm0.10,
    &
    \widehat T &= 0.11\pm0.12,
    &
    \widehat U &= -0.02\pm0.09,
    \\
    \rho_{ST} &= 0.90,
    &
    \rho_{SU} &= -0.57,
    &
    \rho_{TU} &= -0.82 .
\end{aligned}
\label{eq:stu_global_fit}
\end{equation}
For each parameter point, we compute the scalar-sector contributions to \((S,T,U)\) and require them to lie inside the correlated 95\% confidence ellipsoid,
\begin{equation}
    \Delta\chi^2_{STU}
    =
    \Delta X_i
    \left(C^{-1}\right)_{ij}
    \Delta X_j
    \leq
    \chi^2_{3,\,0.95},
    \qquad
    \Delta X=(S-\widehat S,\ T-\widehat T,\ U-\widehat U),
    \label{eq:stu_chi2}
\end{equation}
where \(i,j=S,T,U\),
\begin{equation}
    C_{ij}
    =
    \rho_{ij}\sigma_i\sigma_j,
    \qquad
    \rho_{ii}=1,
    \qquad
    (\sigma_{\rm S},\sigma_T,\sigma_U)=(0.10,0.12,0.09),
    \label{eq:stu_covariance}
\end{equation}
and \(\chi^2_{3,\,0.95}=7.815\). The explicit one loop functions entering the scalar contributions to \(S\), \(T\) and \(U\) are not repeated here~\cite{Grimus:2008Oblique}. Their main effect is to disfavor large mass splittings among the heavy scalar states, especially splittings involving the charged Higgs boson.

\paragraph{Scan ranges and strategy.}
In both the conformal and non-conformal scans we use the common input ranges
\begin{equation}
     m_{{\mathrm A}^0}\in[90,1000]~{\rm GeV},\qquad
     m_{{\mathrm H}^\pm}\in[80,1000]~{\rm GeV},\qquad
    \tan\beta\in[0.1,5].
\end{equation}
The two scans differ only in the CP-even sector and in the presence of a soft scale. In the conformal model, the CP-even masses arise radiatively through the Coleman--Weinberg mechanism, with the Gildener--Weinberg condition fixing the scale along the flat direction. In the non-conformal model, we additionally scan
\begin{equation}
    m_{{\mathrm H}}\in[m_{\mathrm h}  ,1000~{\rm GeV}],\qquad
    M\in[100,1000]~{\rm GeV},
    \qquad
    m_{12}^2=M^2s_\beta c_\beta .
\end{equation}
Points passing the zero temperature constraints are then passed unchanged to the PD, AE and Parwani finite temperature analyses.

\paragraph{Branching.}
The conformal scan requires an additional zero-temperature branch identification.  For fixed \((m_{{\mathrm A}^0},m_{{\mathrm H}^\pm},\tan\beta)\), the one loop matching condition in Eq.~\eqref{eq:det_cond} can admit more than one solution for $\lambda_{345}$, corresponding to different CP-even spectra and mixings at the electroweak vacuum. We therefore identify the observed Higgs with the mass eigenstate that is sufficiently aligned with the electroweak vacuum direction, following the C2HDM criterion~\cite{Lee:2012jn,Ahriche:2021frb}. 

Denoting by \(O\) the orthogonal matrix that diagonalizes the CP-even mass matrix in Eq.~\eqref{eq:MH2} at the electroweak vacuum, the corresponding mass eigen directions are \(H_i=O_{1i}h_1+O_{2i}h_2\), with \(m_{H_1}<m_{H_2}\). Since the gauge boson masses are generated along the vacuum direction \((c_\beta,s_\beta)\), the tree-level reduced coupling of \(H_i\) to vector bosons is given by its projection onto this direction, 
\begin{equation}
g_{H_iVV}=c_\beta O_{1i}+s_\beta O_{2i}.
\label{eq:c2hdm_gHVV}
\end{equation}
For each parameter point, we retain the zero temperature branches in which either \(H_1\) or \(H_2\) has mass \(125.09~\mathrm{GeV}\) and satisfies
\begin{equation}
g_{125,VV}^2>0.8 .
\label{eq:c2hdm_sm_like_branch_condition}
\end{equation}
This selects the branch in which the 125 GeV scalar has a sufficiently SM-like electroweak gauge coupling.

\begin{figure}[t!]
\centering
        \includegraphics[
        width=0.65\textwidth,
        height=0.35\textheight
    ]{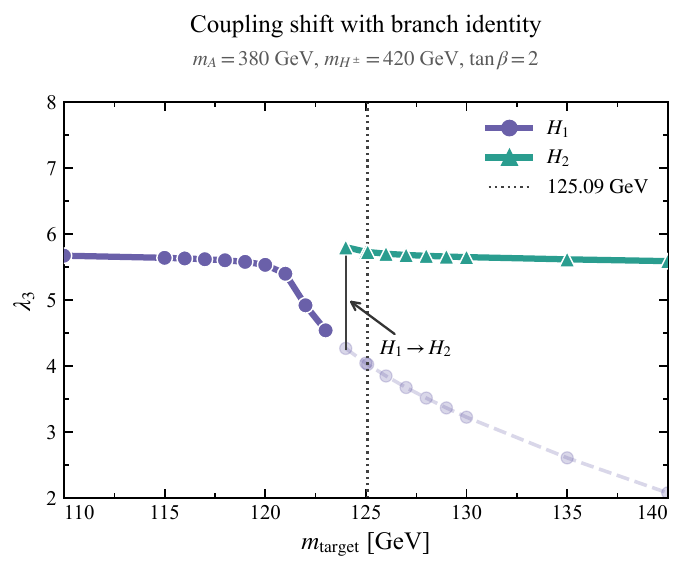}
\caption{Illustration of the CP-even branch selection in the conformal scan for \(m_{{\mathrm A}^0}=380~\mathrm{GeV}\), \(m_{{\mathrm H}^\pm}=420~\mathrm{GeV}\) and \(\tan\beta=2\). The vertical line marks \(m_{\mathrm h}=125.09~\mathrm{GeV}\). The two marker styles indicate whether the state whose mass is fixed by the matching condition is identified with \(H_1\) or \(H_2\). The physical branch is selected by requiring the 125.09 GeV state to be sufficiently SM-like, \(g_{125,VV}^2>0.8\).}
\label{fig:c2hdm_branch_strategy}
\end{figure}

Figure~\ref{fig:c2hdm_branch_strategy} illustrates this branch selection for a representative parameter point. The vertical line marks the physical Higgs mass, near which the state selected by the matching condition can be either $H_1$ or $H_2$.  

\section{Finite Temperature Effective Potential}
\label{sec:thermal-potential}

With the zero-temperature inputs fixed, we now construct the three finite-temperature potentials. For PD, we solve the 2HDM scalar gap equations, including field-dependent mixing, and integrate the PD tadpole along a straight path to obtain the potential. For Parwani and Arnold--Espinosa, we specify the truncated thermal masses of the scalars and of the longitudinal gauge bosons; the latter also enter the gauge-boson part of the PD potential. The same construction applies to both models, which differ only in their zero-temperature potentials and matching conditions, so we use a common notation unless stated otherwise.

\subsection{2HDM Implementation of Partial Dressing}
\label{sec:finitetemperature}
The starting point of our finite temperature analysis is the perturbative effective potential. At one loop, it is written as the sum of the tree-level potential, the finite counterterms, the zero temperature Coleman--Weinberg correction, and the finite temperature correction,
\begin{equation}
V
=
V_0+V_{\rm CT}+V_{\rm CW}+V_T .
\label{eq:c2hdm_finiteT_eff_potential}
\end{equation}
For the \(i\)-th degree of freedom, the thermal contribution is
\begin{equation}
V_{T,i}\big(m_i^2(\varphi)\big)
=
(-1)^{2s_i} n_i
\frac{T^4}{2\pi^2}
J_{B,F}\left(\frac{m_i^2(\varphi)}{T^2}\right),
\label{eq:2HDM_vT}
\end{equation}
where \(s_i\), \(n_i\), and \(m_i^2(\varphi)\) follows the same conventions as in Eq.~\eqref{eq:c2hdm_vcw}. The thermal functions are defined by
\begin{equation}
J_{B,F}(y)
=
\int_0^\infty dx\,x^2
\log\left(1\mp e^{-\sqrt{x^2+y}}\right),
\end{equation}
with the upper (lower) sign corresponding to bosonic (fermionic) degrees of freedom. In our numerical analysis, both these thermal functions and their derivatives are evaluated using precomputed integral tables, without any high temperature expansion.

For the discussion of PD prescription, it is useful to combine the zero temperature Coleman--Weinberg contribution Eq.~\eqref{eq:c2hdm_vcw} and the thermal contribution Eq.~\eqref{eq:2HDM_vT} of the \(i\)-th mode into
\begin{equation}
\begin{aligned}
V_i\big(m_i^2\big)
&\equiv
V_{{\rm CW},i}\big(m_i^2\big)
+
V_{T,i}\big(m_i^2\big)
\\&=(-1)^{2s_i} n_i
\left[\frac{m_i^4}{64\pi^2}
\left(
\ln\frac{m_i^2}{\bar\mu^2}
-c_i
\right)+
\frac{T^4}{2\pi^2}
J_{B,F}\left(\frac{m_i^2}{T^2}\right)\right],
\label{eq:2HDM_loop_function}
\end{aligned}
\end{equation}
where the explicit temperature dependence is suppressed. The one loop finite temperature potential can then be written as
\begin{equation}
V(\varphi,T)
=
V_0(\varphi)
+
V_{\rm CT}(\varphi)
+
\sum_i
V_i\big(m_i^2(\varphi)\big).
\label{eq:2HDM_pert_eff_potential}
\end{equation}
The PD prescription reorganizes only the scalar part of this fixed-order potential at the level of tadpoles. The dressed scalar spectrum is not inserted directly into the one-loop potential. Instead, the dressed masses are inserted into the scalar tadpole, and the corresponding contribution to the potential is reconstructed by integrating these tadpoles along a prescribed path in field space. For the straight line path used in our implementation, this gives
\begin{equation}
V_{\rm PD}(\varphi,T)
=
V_0(\varphi)
+
V_{\rm CT}(\varphi)
+
V_{\rm NS}(\varphi,T)
+
\int_0^1 dt\,
\varphi_a\,
\mathcal T_a^{\mathrm{PD}}(\varphi t,T),
\label{eq:VPD_1d}
\end{equation}
where summation over \(a\) is understood. Here
\begin{equation}
V_{\rm NS}(\varphi,T)
=
\sum_{i\in V,F}
V_i\big(m_i^2(\varphi)\big)
\label{eq:2HDMVF}
\end{equation}
denotes the non-scalar contribution, while the last term in Eq.~\eqref{eq:VPD_1d} represents the partially dressed scalar contribution. The scalar loop tadpole is
\begin{equation}
\mathcal T_a^{\mathrm{PD}}
=
\sum_{A\in S}
V_A'\!\left(M_A^2\right)
\left(
\mathbf U^T
\partial_a \mathbf m_{\rm S}^2
\mathbf U
\right)_{AA},
\label{eq:VPD_2HDM_force}
\end{equation}
where the sum runs over the scalar sector  \(A\in S=\{\mathrm h,\mathrm H,\mathrm G^0,\mathrm A^0,\mathrm G^\pm,\mathrm H^\pm\}\). The prime denotes a derivative with respect to the squared mass. The vertex uses the tree-level scalar mass matrix $\mathbf m_{\rm S}^2(\varphi)$ defined in Eq.~\eqref{eq:c2hdm_field_mass_general}, whereas the loop masses $M_A^2(\varphi,T)$ and rotation $\mathbf U(\varphi,T)$ are obtained from the 2HDM gap equation at each $(\varphi,T)$. We describe the solution of these coupled equations below.

Before closing this section, we comment on \(V_{\rm NS}(\varphi,T)\), the part of the effective potential that is not treated by the scalar PD procedure. In principle, one could also formulate gap equations for the gauge bosons by using their finite temperature self energies, thereby dressing the gauge sector together with the scalar sector~\cite{Espinosa:1993}. However, unlike the scalar case, the finite temperature gauge boson self-energy has a nontrivial tensor structure, and the corresponding gauge boson gap equations cannot be derived from the effective potential in the same direct manner as the scalar gap equations~\cite{Curtin:2016}.
Since the phase-transition dynamics considered in this work is primarily driven by the extended scalar sector, we retain a Parwani type resummation prescription for the gauge boson contribution to \(V_{\rm NS}(\varphi,T)\). We defer the detailed discussion of this prescription until after we present the scalar gap equations.

\subsection{Gap Equation Solver}
To evaluate the PD potential in Eq.~\eqref{eq:VPD_1d}, we now determine $M_A^2(\varphi,T)$ and $\mathbf U(\varphi,T)$ from the 2HDM gap equation. For a generic scalar background, denoted collectively by
\begin{equation}
\varphi
=
\left(
h_1,h_2,a_1,a_2,\phi_1^+,\phi_2^+,\phi_1^-,\phi_2^-
\right),
\end{equation}
the extended scalar sector contains non-trivial field-dependent mixing. A direct solution of the gap equation would then require repeated diagonalization of a general \(8\times8\) matrix, which is the main technical complication in the 2HDM implementation. We reduce this calculation by evaluating the potential on the neutral CP-even background subspace~\cite{Bittar:2025lcr}
\begin{equation}
\varphi_{H}(h_1,h_2)
=
\left(h_1,h_2,0,0,0,0,0,0\right).
\end{equation}
This is the field subspace relevant for the charge- and CP-preserving electroweak phase transition considered in this work. More precisely, our analysis is restricted to thermal histories whose relevant extrema and transition paths lie within the neutral CP-even field space. Parameter points for which a charged or CP-odd minimum becomes thermodynamically relevant require a separate treatment and are therefore outside the scope of the present construction.

This restriction should not be understood as removing the CP-odd and charged fields from the scalar spectrum. Rather, it only specifies the background subspace on which the effective potential is evaluated. The quantities entering the PD tadpoles in Eq.~\eqref{eq:VPD_2HDM_force} and the gap equation are still constructed in the full scalar field space before the restriction to \(\varphi_H\) is imposed. In particular, the tree-level mass matrix \(\mathbf m_{\rm S}^2(\varphi)\), the counterterm Hessian \(\bigl[\mathbf m_{\rm CT}^2(\varphi)\bigr]_{ab}\equiv \partial_a\partial_bV_{\rm CT}\), and the derivative tensors \(\partial_a\mathbf m_{\rm S}^2(\varphi)\) and \(\partial_a\partial_b\mathbf m_{\rm S}^2(\varphi)\) are all computed before setting the CP-odd and charged background fields to zero. This order of operations is essential, because taking derivatives in the full scalar field space does not in general commute with restricting the potential to a lower-dimensional background subspace. These quantities are also the only model-dependent ingredients entering the gap equations.

The task now is to solve the coupled matrix gap equation along the neutral CP-even background directions. On the background \(\varphi_H\), the tree-level matrix decomposes into one CP-even block, one CP-odd block, and two identical charged blocks, each of dimension \(2\times2\),
\begin{equation}
\mathbf m_{\rm S}^2(\varphi_H)
=
\operatorname{diag}
\left(
\mathbf m_{{\mathcal H}}^2,\,
\mathbf m_{{\mathcal A}}^2,\,
\mathbf m_{{\mathcal H}^\pm}^2,\,
\mathbf m_{{\mathcal H}^\pm}^2
\right),
\label{eq:2HDM_scalar_hessian_blocks}
\end{equation}
where mixing inside each block is retained. The same block structure is preserved by the dressed solution of the gap equation, since the neutral CP-even background preserves the residual charge and CP quantum numbers that distinguish the CP-even, CP-odd and charged sectors. Accordingly, the dressed matrix is parametrized as
\begin{align}
\mathbf M_{\rm S}^2(\varphi_H,T)
&=
\mathbf U_{\rm S}(\varphi_H,T)\,{\rm diag}(M_A^2)\,\mathbf U_{\rm S}^T(\varphi_H,T),
\label{eq:2HDM_dress_blocks}
\\
\mathbf U_{\rm S}(\varphi_H,T)
&=
\operatorname{diag}
\left(
\mathbf U_{\mathcal H},\mathbf U_{\mathcal A},\mathbf U_{{\mathcal H}^\pm},\mathbf U_{{\mathcal H}^\pm}
\right).
\label{eq:2HDM_rotation_blocks}
\end{align}
The rotation matrix is therefore specified by only three independent mixing angles, one for each independent 2×2 block. This substantially simplifies the numerical solution compared with a direct treatment of a general \(8\times8\) matrix.

At each point \((\varphi_H,T)\), the 2HDM gap equation follows directly from Eq.~\eqref{eq:gap-compact} and Eq.~\eqref{eq:2HDM_pert_eff_potential}, and takes the form
\begin{equation}
\mathbf M_{\rm S}^2
=
\mathbf m_{\rm S}^2
+\mathbf m_{\rm CT}^2
+\boldsymbol{\Pi}_{\rm NS}
+\boldsymbol{\Pi}_{\rm S}(\mathbf M_{\rm S}^2).
\label{eq:2HDM_gap_rhs}
\end{equation}
We denote the right-hand side of this equation by \(\mathbf R(\mathbf M_{\rm S}^2)\). The term \(\left[\boldsymbol{\Pi}_{\rm NS}\right]_{ab}\equiv\partial_a\partial_bV_{\rm NS}\) contains the non-scalar contributions from gauge bosons and fermions in Eq~\eqref{eq:2HDMVF}, evaluated with their ordinary field-dependent masses. Gauge boson and fermion loops therefore enter the scalar gap equation, although we do not solve separate gap equations for their masses. 

The scalar contribution \(\boldsymbol{\Pi}_{\rm S}\), by contrast, is evaluated with the dressed eigenvalues and rotations. In terms of the loop function defined in Eq.~\eqref{eq:2HDM_loop_function}, the scalar contribution entering the right-hand side of the gap equation is 
\begin{equation}
\left[\boldsymbol{\Pi}_{\rm S}\right]_{ab}
=
\sum_{A\in S}
\left[
V_A'\!\left(M_A^2\right)D_{ab}^{(A)}
+
V_A''\!\left(M_A^2\right)D_a^{(A)}D_b^{(A)}
\right],
\label{eq:2HDM_gap_scalar_self_energy}
\end{equation}
where
\begin{equation}
D_a^{(A)}
=
\left.
\partial_a m_A^2
\right|_{(m_A^2,\mathbf u)\to(M_A^2,\mathbf U)},
\qquad
D_{ab}^{(A)}
=
\left.
\partial_a\partial_b m_A^2
\right|_{(m_A^2,\mathbf u)\to(M_A^2,\mathbf U)} .
\label{eq:2hdm_gap_derivative_def}
\end{equation}
We evaluate the Feynman--Hellmann expressions in Eqs.~\eqref{eq:feynhell} and \eqref{eq:feynhell-second} with $m_A^2\to M_A^2$ and $\mathbf u\to\mathbf U$ to obtain $D_a^{(A)}$ and $D_{ab}^{(A)}$. These substitutions make $\boldsymbol{\Pi}_{\rm S}$ and hence $\mathbf R$ functions of the dressed mass matrix.

The matrix equation is solved by fixed-point iteration. Throughout the iterative procedure, the combination \(\mathbf m_{\rm S}^2+\mathbf m_{\rm CT}^2+\boldsymbol{\Pi}_{\rm NS}\) appearing in \(\mathbf R(\mathbf M_{\rm S}^2)\) is kept fixed. The initial matrix is chosen to be \(\mathbf M_{{\rm S},1}^2=\mathbf R(\mathbf m_{{\rm S}}^2)\) evaluated at \((\varphi_H,T)\). At iteration \(n\), the current dressed matrix is written as
\begin{equation}
\mathbf M_{{\rm S},n}^2
=
\mathbf U_n\,{\rm diag}(x_n)\,\mathbf U_n^T,
\qquad
x_{n,A}\equiv M_{n,A}^2 ,
\label{eq:2hdm_gap_state}
\end{equation}
The block structure in Eqs.~\eqref{eq:2HDM_dress_blocks} and~\eqref{eq:2HDM_rotation_blocks} allows us to diagonalize the three independent $2\times2$ blocks separately at each iteration. We use the resulting $x_{n,A}$ and $\mathbf U_n$ to evaluate $\boldsymbol{\Pi}_{\rm S}$ and $\mathbf R(\mathbf M_{{\rm S},n}^2)$, then damp the update at the matrix level,
\begin{equation}
\mathbf M_{{\rm S},n+1}^2
=
(1-\omega)\mathbf M_{{\rm S},n}^2
+\omega\,\mathbf R(\mathbf M_{{\rm S},n}^2),
\label{eq:2hdm_gap_iteration}
\end{equation}
and the updated matrix is diagonalized again block by block. Thus the damping is applied to the matrix itself, rather than separately to the eigenvalues and mixing angles. This avoids ambiguities associated with eigenvalue ordering and with the periodic parametrization of the mixing angles.

The iteration is repeated until convergence. In the numerical scan we start from \(\omega=1\), and reduce this factor adaptively if the fixed-point iteration fails to converge. We monitor the four step moving average
\begin{equation}
\bar x_n=\frac{1}{4}\sum_{r=0}^{3}x_{n-r},
\qquad
\epsilon_n
=
\max_A
\frac{|\bar x_{n,A}-\bar x_{n-1,A}|}
{1+|\bar x_{n-1,A}|}.
\label{eq:2hdm_gap_convergence}
\end{equation}
The average is understood componentwise in the scalar eigenvalue index \(A\). The moving average suppresses small residual oscillations of the fixed-point iteration, which otherwise make a pointwise convergence test overly sensitive near regions with rapidly varying scalar mixing. Points that still fail the convergence test after the adaptive damping retries are marked as failed gap-equation solutions and are not used in the construction of the PD effective potential. In the scan we use \(N_{\max}=80\) and \(\epsilon_{\rm tol}=10^{-8}\).

\subsection{Conventional Resummation}
We now specify how the Parwani and Arnold--Espinosa prescriptions are implemented in the 2HDM finite temperature effective potential.

We first discuss the non scalar sector, which consists of the electroweak gauge bosons and the top quark. At tree level, the transverse and longitudinal vector-boson modes have the same field dependent masses. In terms of the two Higgs doublets, these masses are given by
\begin{equation}
m_{W_{T,L}}^2(\varphi)
=
\frac{g^2}{4}\rho^2(\varphi)
,
\qquad
m_{Z_{T,L}}^2(\varphi)
=
\frac{g^2+g'^2}{4}\rho^2(\varphi),
\label{eq:WZmass}
\end{equation}
where
\begin{equation}
\rho^2(\varphi)=2\big(\Phi_1^\dagger\Phi_1+\Phi_2^\dagger\Phi_2\big).
\end{equation}
For the fermion sector, we keep only the top quark contribution, whose field-dependent mass is
\begin{equation}
m_t^2(\varphi)
=
y_t^2\Phi_2^\dagger\Phi_2.
\label{eq:tmass}
\end{equation}
The electroweak and top-Yukawa couplings are fixed from the input parameters as
\begin{equation}
g=\frac{2m_W}{v},
\qquad
g'=\sqrt{\frac{4m_Z^2}{v^2}-g^2},
\qquad
y_t=\frac{\sqrt{2}m_t}{v_2}.
\end{equation}

Thermal resummation is applied only to the bosonic modes that contain Matsubara zero modes. Accordingly, the transverse vector modes are kept unresummed, whereas the longitudinal modes receive the leading Debye corrections~\cite{Basler_2017}. For the charged gauge bosons, this gives
\begin{equation}
\bar m_{W_L}^2(\varphi,T)
=
m_{W_L}^2(\varphi,T)
+
2g^2T^2 .
\label{eq:mWL_bar}
\end{equation}
The neutral longitudinal gauge bosons are obtained by diagonalizing the truncated thermal mass matrix in the \((W_L^3,B_L)\) basis,
\begin{equation}
\bar {\mathbf m}_N^2(\varphi,T)
=
{\mathbf m}_N^2(\varphi)
+
\begin{pmatrix}
2g^2T^2
&
0
\\
0
&
2g'^2T^2
\end{pmatrix},
\label{eq:neutral_longitudinal_gauge_matrix}
\end{equation}
where
\begin{equation}
\mathbf m_N^2(\varphi)
=
\frac{\rho^2(\varphi)}{4}
\begin{pmatrix}
g^2 & -gg'
\\
-gg' & g'^2
\end{pmatrix}.
\label{eq:neutral_gauge_mass_matrix_zeroT}
\end{equation}
Its eigenvalues define the truncated thermal masses for longitudinal photon and \(Z\)-boson:
\begin{subequations}
\label{eq:neutral_longitudinal_gauge_eigenvalues}
\begin{align}
\bar m_{\gamma_L}^2(\varphi,T)
&=
(g^2+g'^2)
\left(
T^2+\frac{\rho^2(\varphi)}{8}
\right)
-
\frac{1}{8}
\Delta_Z(\varphi,T),
\label{eq:mgammaL_bar}
\\[0.5em]
\bar m_{Z_L}^2(\varphi,T)
&=
(g^2+g'^2)
\left(
T^2+\frac{\rho^2(\varphi)}{8}
\right)
+
\frac{1}{8}
\Delta_Z(\varphi,T),
\label{eq:mZL_bar}
\end{align}
where
\begin{equation}
\Delta_Z(\varphi,T)
\equiv
\sqrt{
(g^2-g'^2)^2
\left[
64T^4+16T^2\rho^2(\varphi)
\right]
+
(g^2+g'^2)^2\rho^4(\varphi)
}\, .
\label{eq:DeltaZ_def}
\end{equation}
\end{subequations}
By contrast, fermions do not possess bosonic Matsubara zero modes and therefore do not require thermal resummation.

We next turn to the scalar sector. In order to define the truncated thermal masses, we restrict the background field configuration to the CP-even subspace \(\varphi_H\). With this restriction, the field dependent scalar mass matrix decomposes into the CP-even, CP-odd, and charged blocks, as in Eq.~\eqref{eq:2HDM_scalar_hessian_blocks}.
Let
\begin{equation}
 S=\{\mathrm h,\mathrm H,\mathrm G^0,\mathrm A^0,\mathrm G^\pm,\mathrm H^\pm\}
\end{equation}
denote the scalar modes. The corresponding truncated thermal masses are obtained by diagonalizing the finite temperature mass matrices
\begin{equation}
    \bar {\mathbf m}_X^2(\varphi_H,T)
    =
    \mathbf m_X^2(\varphi_H)
    +
    \begin{pmatrix}
        c_{\Phi_1}T^2 & 0\\
        0 & c_{\Phi_2}T^2
    \end{pmatrix},
    \qquad X={\mathcal H},{\mathcal A},{\mathcal H}^\pm ,
\end{equation}
where
\begin{align}
    c_{\Phi_1}
    &=
    \frac{1}{48}
    \left[
        12\lambda_1+8\lambda_3+4\lambda_4
        +3(3g^2+g'^2)
    \right],
    \\
    c_{\Phi_2}
    &=
    \frac{1}{48}
    \left[
        12\lambda_2+8\lambda_3+4\lambda_4
        +3(3g^2+g'^2)+12y_t^2
    \right].
\end{align}

The eigenvalues of the matrices $\bar {\mathbf m}_X^2$ are denoted by \(\bar m_A^2\), with \(A\in S\). Together with the gauge boson and top quark masses defined above, they form the set of truncated thermal masses used in this work: \begin{equation}
\bar m_i^2
\in
\left\{
m_{W_T}^2,
\bar m_{W_L}^2,
m_{Z_T}^2,
\bar m_{\gamma_L}^2,
\bar m_{Z_L}^2,
m_t^2,
\bar m_A^2
\right\}.
\end{equation}

With these conventions, the Parwani prescription amounts to inserting the truncated thermal masses directly into the full one-loop functions. The effective potential is 
\begin{equation}
    V_{\rm Parwani}(\varphi,T)
    =
    V_0(\varphi)
    +V_{\rm CT}(\varphi)
    +
    \sum_{i}
    V_i\!\left(\bar m_i^2(\varphi,T)\right).
\end{equation}

In the Arnold--Espinosa prescription, the truncated thermal mass replacement is applied only to the daisy contribution.  The corresponding potential is
\begin{equation}
    V_{\rm AE}(\varphi,T)
    =
    V_0(\varphi)
    +V_{\rm CT}(\varphi)
    +
    \sum_{i}
    V_i\!\left(m_i^2(\varphi)\right)
    -
    \frac{T}{12\pi}
    \sum_{i}
    n_i
    \left[
        \left(\bar m_i^2(\varphi,T)\right)^{3/2}
        -
        \left(m_i^2(\varphi)\right)^{3/2}
    \right].
\end{equation}

Having specified how the Parwani and Arnold--Espinosa prescriptions are implemented in 2HDM, we now return to the treatment of the non scalar contribution \(V_{\rm NS}(\varphi,T)\) within the partial dressing prescription. As mentioned at the end of section~\ref{sec:finitetemperature}, we adopt a Parwani-type prescription for the gauge-boson contribution. Explicitly, we take
\begin{equation}
V_{\rm NS}^R(\varphi,T)
=
\sum_{i\in V,F}
V_i\big(\bar m_i^2(\varphi,T)\big)
\end{equation}
Thus, in the final PD effective potential, the effects of the Parwani resummed longitudinal gauge bosons are incorporated by replacing \(V_{\rm NS}\) with \(V_{\rm NS}^R\) in Eq.~\eqref{eq:VPD_1d}.

\section{Phase Transition Dynamics}
\label{sec:phase_transition_dynamics}
Using the potentials defined above, we compare the phase classifications, transition strengths, and supercooling predicted by PD, Arnold--Espinosa (AE), and Parwani at the same zero-temperature inputs. For PD, we use the straight integration path and assess its path dependence separately in Appendix~\ref{app:path-dependence}. The standard definitions of bubble nucleation, percolation, and the gravitational wave observables follow our previous analysis~\cite{Benincasa:2026c2hdm}; here we briefly recall only the ingredients required for the present implementation. The comparison shows that completed first-order transitions dominate the conformal scan in all three prescriptions, with PD giving the strongest and most supercooled transitions, whereas in the non-conformal scan changing the prescription can change the phase classification itself and no prescription is uniformly the strongest.

\subsection{One Dimensional Reduction and Transition Criteria}
\label{sec:oned_reduction_transition_setup}

\begin{figure}[t!]
    \centering
        \includegraphics[
        width=0.65\textwidth,
        height=0.45\textheight
    ]{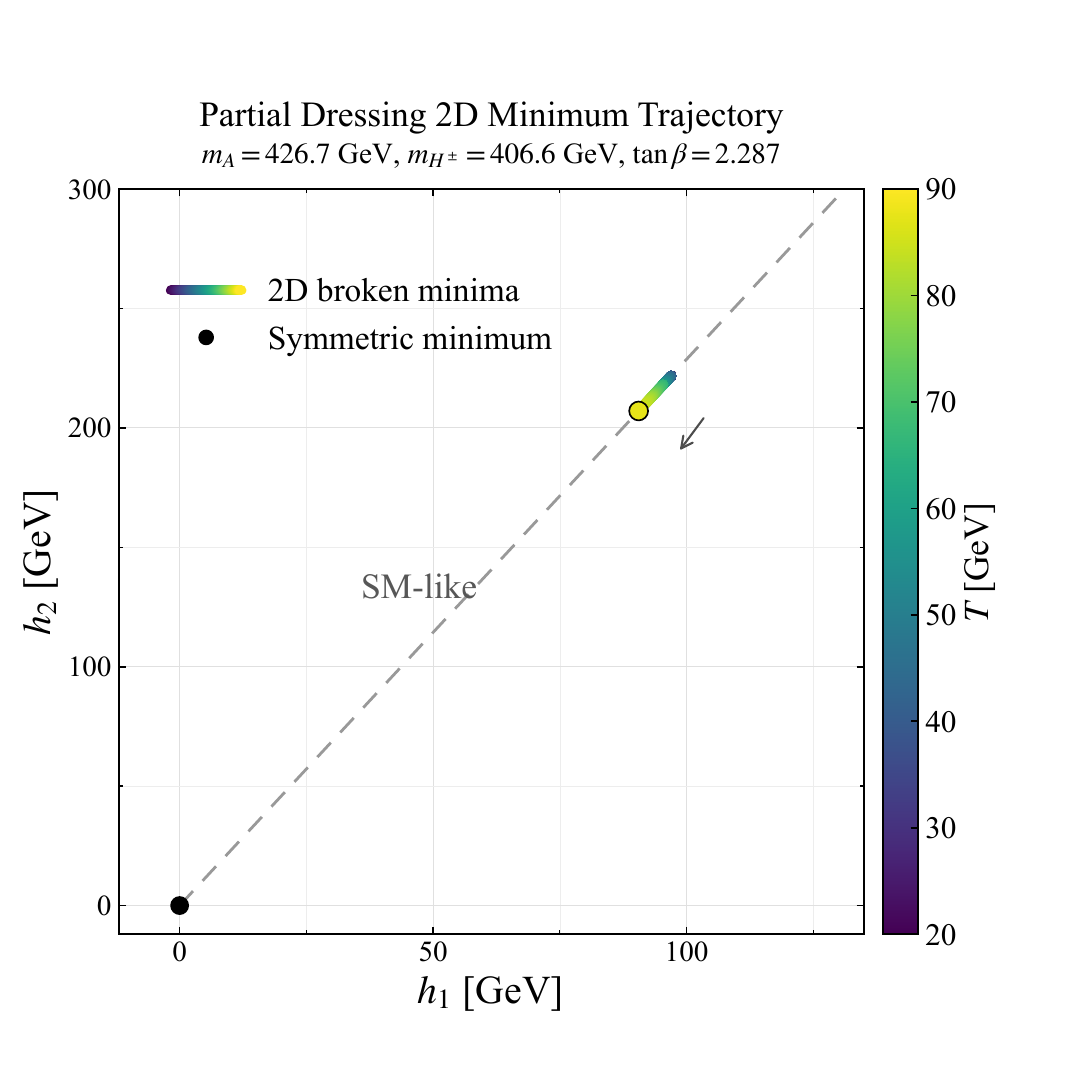}
    \caption{
    Two dimensional check of the neutral PD effective potential for one representative parameter point in the scan. The input parameters of this illustrative point are indicated in the legend. The colored points show the temperature evolution of the minimum in the \((h_1,h_2)\) plane, while the small arrows indicate the direction of the tracked minimum as the temperature is lowered. The dashed line denotes the SM-like direction \(H=0\). The absence
    of a visible displacement in the orthogonal direction illustrates the one-dimensional behavior assumed in the bounce calculation.
    }
    
    \label{fig:pd_flat_direction}
\end{figure}

The critical temperature \(T_c\) is defined as the temperature at which the symmetric and broken minima of the finite temperature effective potential are degenerate. Before computing the transition dynamics, we check the one-dimensional reduction at a representative conformal benchmark by minimizing the neutral PD effective potential in the two dimensional \((h_1,h_2)\) plane. In the rotated basis defined in Eq.~\eqref{eq:flat_dir}, the SM-like direction corresponds to \(H=0\).

As shown in figure~\ref{fig:pd_flat_direction}, the global minimum of the two-dimensional neutral PD effective potential follows the SM-like direction during the transition. At high temperature the symmetric minimum is located at the origin, while below \(T_c\) it moves to the broken branch. The displacement in the orthogonal \(H\) direction is numerically negligible. We therefore describe the transition using the reduced field configuration
\begin{equation}
    h_1=c_\beta h,\qquad
    h_2=s_\beta h,\qquad
    a_i=\phi_i^\pm=0 .
    \label{eq:one_dimensional_trajectory}
\end{equation}
This numerical check supports the one-dimensional treatment adopted below. Accordingly, we evaluate the bounce action along the SM-like trajectory, neglecting a possible bending of the tunneling path away from it.

Below \(T_c\), the symmetric phase may remain metastable and the transition proceeds through bubble nucleation. The tunneling rate per unit volume is written as \cite{Linde:1980tt,Linde:1981zj}
\begin{equation}
    \Gamma(T)
    =
    A(T)\,e^{-S(T)},
    \qquad
    S(T)=\min\left\{\frac{S_3(T)}{T},S_4(T)\right\},
    \label{eq:tunneling_rate_compact}
\end{equation}
where the bounce profiles are computed numerically with \texttt{CosmoTransitions}~\cite{Wainwright_2012}. We determine the nucleation and percolation temperatures from the spacetime integrals of the tunneling rate. The nucleation temperature \(T_n\) is obtained from the condition that the expected number of bubbles per Hubble volume becomes order one, while the percolation temperature \(T_p\) is defined by~\cite{1971AdPhy..20..325S,M_gevand_2017,Kobakhidze_2017,Cai_2017,Ellis:2018mja}
\begin{equation}
    I(T_p)=0.34,
    \label{eq:Tp_definition}
\end{equation}
with \(P(T)=e^{-I(T)}\) the false-vacuum probability. We also require the physical false vacuum volume to shrink at \(T_p\) as a criterion for completion~\cite{Turner:1992tz, Ellis:2018mja}.

The quantities characterizing the phase transition are evaluated at the percolation temperature, \(T_*=T_p\)~\cite{Athron:2023Supercool}. We use the radiation dominated Hubble rate
\begin{equation}
    H(T)
    =
    \frac{\pi T^2}{M_{\rm Pl}}
    \sqrt{\frac{g_*}{90}},
    \qquad
    g_*=110.75 .
    \label{eq:Hubble_radiation_limit}
\end{equation}
Here \(M_{\rm Pl}=(8\pi G)^{-1/2}\) is the reduced Planck mass. For the most strongly supercooled points, Eq.~\eqref{eq:Hubble_radiation_limit} neglects the vacuum-energy contribution to the expansion rate. The transition strength is characterized by
\begin{equation}
\alpha
=
\frac{1}{\frac{\pi^2}{30} g_*T_p^4}
\left[
\Delta V
-
\frac{T_p}{4}
\Delta s
\right],
\label{eq:alpha_definition}
\end{equation}
where \(\Delta V=V_{\rm false}-V_{\rm true}\) and \(\Delta s=\partial\Delta V/\partial T=s_{\rm true}-s_{\rm false}\) are evaluated at \(T_p\), with \(s=-\partial V/\partial T\) the entropy density of each phase. The inverse duration is estimated from the mean bubble separation~\cite{Turner:1992tz, Enqvist:1991xw},
\begin{equation}
    \frac{\beta}{H_*}
    =
    \frac{(8\pi)^{1/3}v_w}{H_*R_*},
    \qquad
    R_*
    =
    \left[
    T_p^3
    \int_{T_p}^{T_c} dT'\,
    \frac{\Gamma(T')}{T'^4 H(T')}
    e^{-I(T')}
    \right]^{-1/3}.
    \label{eq:beta_Rp_definition}
\end{equation}
Throughout the scan we take \(v_w\simeq1\), where \(v_w\) is the bubble-wall velocity in the plasma frame.

\subsection{Thermal History Classification in the Scan}
\label{sec:phase_diagrams}

\begin{figure}[t]
    \centering
    \includegraphics[width=\linewidth]{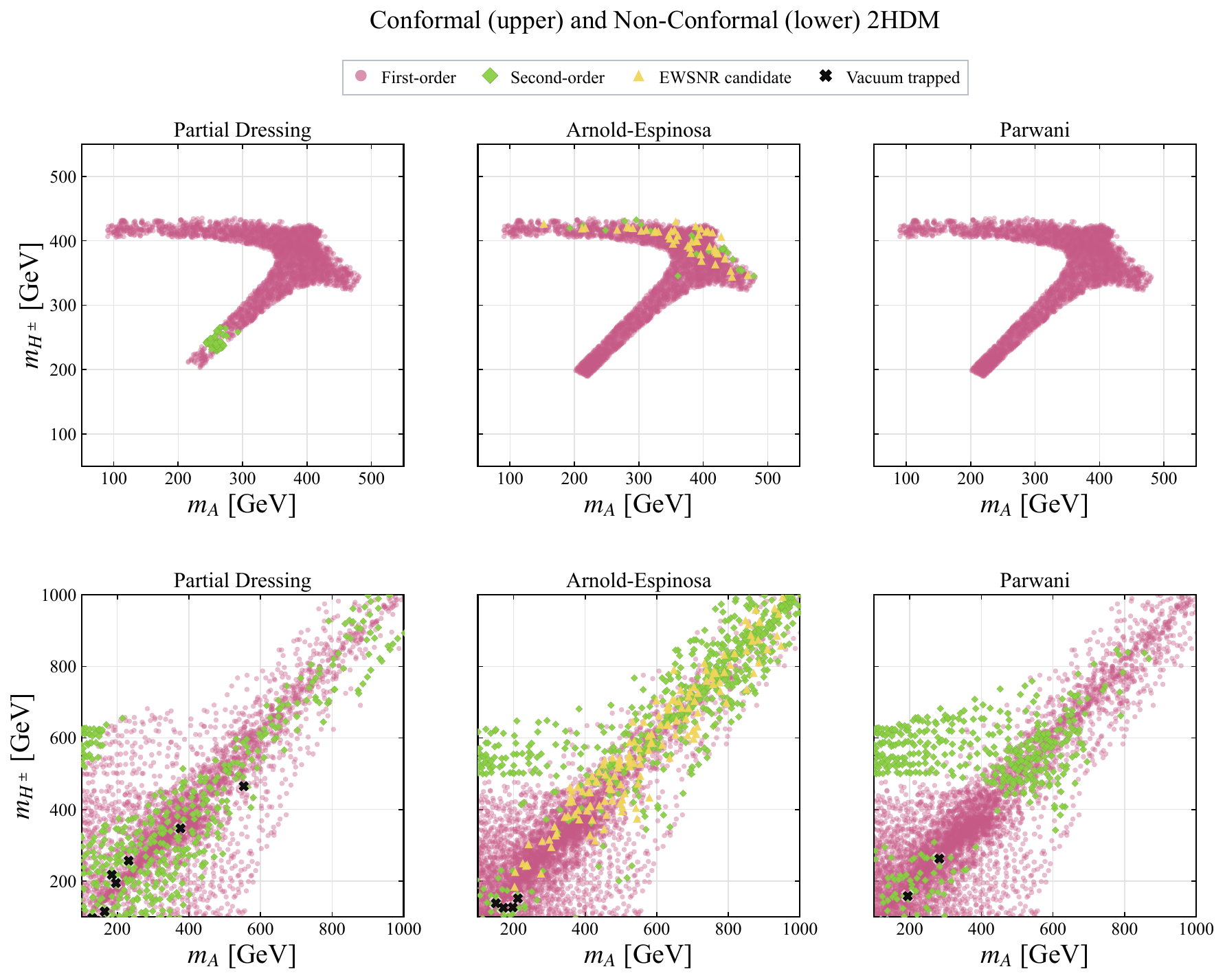}
    \caption{
    Phase classes in the \(m_A\)--\(m_{H^\pm}\) plane for the conformal (upper row) and non-conformal (lower row) scans. In each row the columns show, from left to right, the PD, AE, and Parwani thermal prescriptions. The same accepted zero temperature benchmark set is used across the three prescriptions within a given model, so changes in marker type reflect the finite temperature prescription rather than a different input sample.}
    \label{fig:three_methods_phase_diagram}
\end{figure}

The scan ranges, zero temperature constraints, and additional non-conformal inputs are specified in section~\ref{sec:constraints}. After these cuts are applied, each accepted zero temperature point is passed to the PD, AE, and Parwani finite temperature analyses. Within a given model, the same accepted zero-temperature scan sample is used for the three prescriptions, so that differences among the corresponding panels can be attributed to the finite temperature treatment rather than to a change of input sample.

We begin with the phase classifications in the \(m_A\)--\(m_{H^\pm}\) plane, shown in Fig.~\ref{fig:three_methods_phase_diagram}. For each thermal prescription, the scan points are grouped into completed first order transitions, second order transitions, electroweak symmetry non restoration (EWSNR) candidate histories, and vacuum trapped histories. The EWSNR label denotes thermal histories in which the electroweak broken minimum remains energetically favored over the symmetric configuration within the temperature range relevant for the scan \cite{Weinberg:1974hy,Meade:2018,Biekotter:2022kgf}. Vacuum-trapped histories denote cases in which a nucleation temperature \(T_n\) can be identified, but no temperature satisfies the percolation criterion. In these cases bubbles may nucleate locally, but their growth and abundance are insufficient to overcome the cosmological expansion and convert a percolating volume of the Universe. They are therefore not counted as completed first-order phase transitions. 

The conformal row of Fig.~\ref{fig:three_methods_phase_diagram} shows a clear
first-order dominance in all three prescriptions. This is the most robust
message of the conformal scan: over most of the accepted parameter space, the
existence of a first-order phase transition is not tied to a particular
resummation prescription. The prescription dependence instead appears in the remaining, non-first-order histories. PD produces a small pocket of second-order points in the relatively light part of the diagonal arm, AE gives a visible
population of EWSNR candidate points, together with a few second-order points, along the upper, nearly horizontal band and in the heavier part of the scan, whereas
Parwani classifies the conformal scan almost uniformly as first order. Thus the main first-order region is robust, but the prescription determines which points are not first order and whether they are classified as second order or as EWSNR candidates.

The non-conformal row displays a qualitatively richer pattern. Because explicit
mass scales are already present at tree level, the accepted points span a wider
region in the displayed \(m_A\)--\(m_{H^\pm}\) projection, and the thermal
histories are less uniformly first order than in the conformal case. In
particular, second-order histories appear in all three prescriptions, most prominently at light \(m_A\) with heavy \(m_{H^\pm}\) and along parts of the diagonal, AE in addition populates a broad band of EWSNR candidates along the diagonal \(m_A\simeq m_{H^\pm}\), and a few vacuum-trapped points appear, mostly at light scalar masses. This shows that the
conformal and non-conformal models should not be summarized by a single common
phase transition pattern. It also shows that the resummation ambiguity is not
purely quantitative: in parts of the scan, changing the thermal prescription
changes the qualitative thermal history assigned to the same zero temperature
point.

The prevalence of EWSNR candidate points in the AE prescription can be traced to the cubic daisy term in Eq.~\eqref{eq:AE}. Symmetry restoration is controlled by the curvature of the potential at the origin along the SM-like direction \(h\) of Eq.~\eqref{eq:one_dimensional_trajectory}. Since the first derivatives of the field-dependent masses vanish there, each bosonic mode \(A\) contributes, up to corrections of relative order \((\bar m_A/T)^2\),
\begin{equation}
\left.\frac{\partial^2V_{\mathrm{AE}}}{\partial h^2}\right|_{h=0}
\supset
\frac12\left.\frac{\partial^2\bar m_A^2}{\partial h^2}\right|_{h=0}
\left[\frac{T^2}{12}-\frac{T\,\bar m_A(0,T)}{4\pi}\right].
\label{eq:AE_curvature}
\end{equation}
The first term in the bracket is the usual thermal mass, which drives symmetry restoration. The second comes from the daisy term and contains the truncated thermal mass \(\bar m_A\); it is formally subleading, smaller by a factor of order \(\bar m_A/T\). At high temperature, however, \(\bar m_A(0,T)\simeq\sqrt{c_A}\,T\), with \(c_AT^2\) the leading thermal mass of mode \(A\) (\(c_{\Phi_1}\) or \(c_{\Phi_2}\) for the scalars), so the ratio \(\bar m_A/T\simeq\sqrt{c_A}\) does not decrease with temperature. For \(c_A>\pi^2/9\simeq1.1\), which requires large couplings, the bracket is negative and the mode lowers the curvature instead of raising it. Where such modes outweigh the positive contributions of the other fields, the curvature at the origin stays negative even at high temperature, the broken minimum remains favored, and the point is classified as an EWSNR candidate. This requires \(\bar m_A/T\) of order one, where the high-temperature expansion behind Eq.~\eqref{eq:AE_curvature} and the truncated thermal masses is no longer reliable.

The Parwani prescription does not show the same EWSNR behavior in the present scan. A natural way to understand this difference is that, unlike the AE prescription, Parwani inserts the thermal masses into the full thermal functions rather than only into the cubic daisy contribution. In the high-temperature expansion this reorganization also generates subleading positive curvature contributions, including terms proportional to \(\bar m^2\), in addition to the negative contribution associated with the cubic term. In the parameter region covered by our scan, these additional terms appear to be sufficient to prevent the AE-like negative curvature from producing a comparable population of EWSNR candidate points. PD, which, like Parwani, evaluates the full thermal functions, also produces no comparable EWSNR population in this scan.

Overall, Fig.~\ref{fig:three_methods_phase_diagram} shows that first-order phase transitions occupy a large fraction of the accepted scan in both models, especially in the conformal case. The main differences among the prescriptions appear in the non first order classifications. EWSNR candidates appear essentially only with AE, in both models, whereas second-order histories appear with all three prescriptions in the non-conformal scan and, in the conformal scan, mainly in the small PD pocket. Thus, at the level of phase classification, the comparison among PD, AE and Parwani indicates a stable first order dominated region, together with non first order sectors that depend on the thermal prescription.

\subsection{Benchmark Histories and Transition Strength}

\begin{figure}[t!]
    \centering
    \includegraphics[
        width=\textwidth
    ]{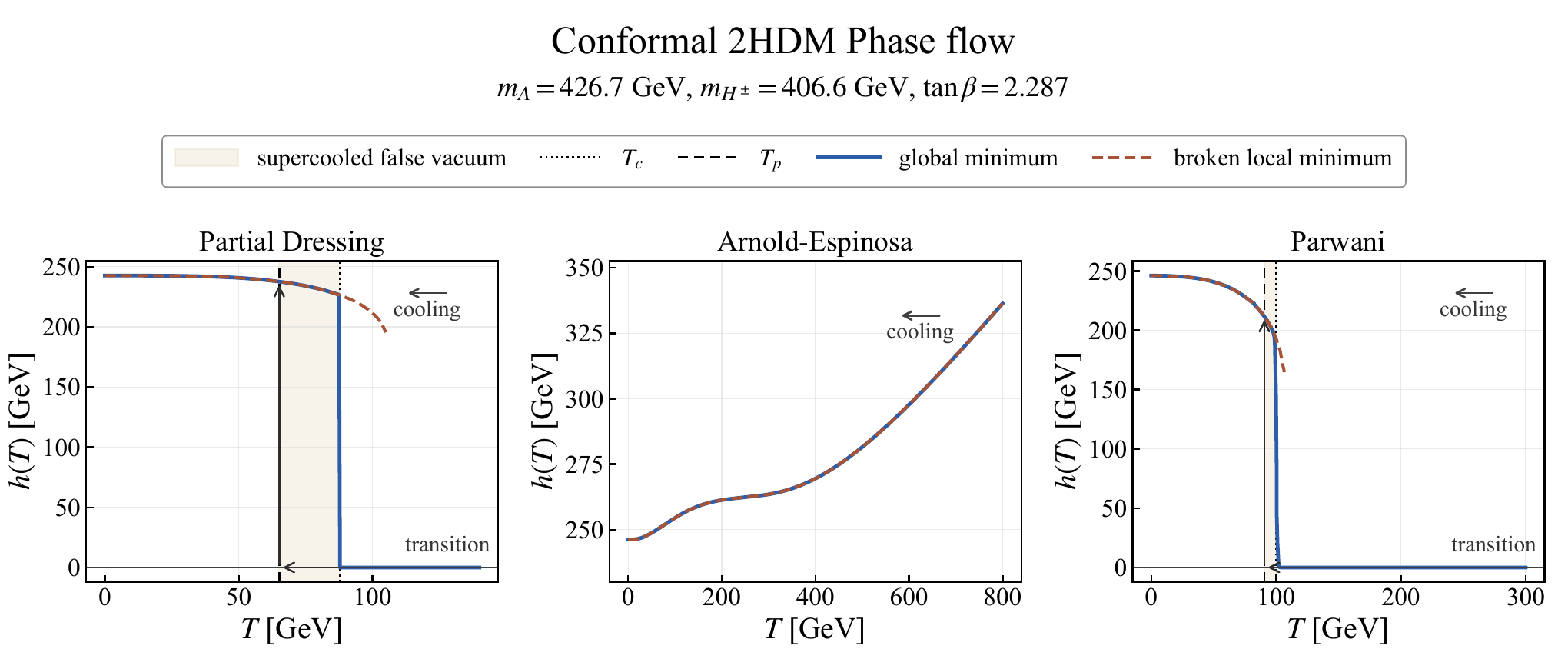}
    \caption{
    Thermal evolution of a representative benchmark selected from the viable conformal parameter space under the PD, AE and Parwani thermal prescriptions.  The panels show the field value of the relevant minimum along the SM-like direction.  The blue curve denotes the equilibrium global minimum, while the brown dashed curve denotes the broken local minimum when it is not the global one.  The vertical dotted and dashed lines mark \(T_c\) and \(T_p\), respectively.  The shaded regions indicate the supercooled interval from \(T_c\) down to \(T_p\).
    }
    \label{fig:three_modes_phase_flow}
\end{figure}

We next show the thermal evolution of a representative conformal benchmark. We visualize the evolution through \(h(T)\), the location of the thermal minimum along the SM-like field direction. Figure~\ref{fig:three_modes_phase_flow} shows that the three prescriptions can assign distinct thermal histories to the same accepted point. In the PD prescription, \(T_c\simeq87.9~\mathrm{GeV}\), while percolation occurs at \(T_p\simeq65.2~\mathrm{GeV}\), indicating a sizable period of supercooling. The Parwani prescription also gives a completed first-order transition, but with a higher critical temperature, \(T_c\simeq100.2~\mathrm{GeV}\), and a smaller separation from \(T_p\simeq90.6~\mathrm{GeV}\).  In contrast, with the AE prescription the broken minimum remains energetically favored over the symmetric configuration throughout the temperature range shown, up to \(T\simeq800~\mathrm{GeV}\). This corresponds to an EWSNR type thermal history for this benchmark. This example shows, at the level of the temperature-dependent minimum, that a change of resummation prescription can alter not only the strength of the transition but also its very nature.

\begin{figure}[t!]
    \centering
    \includegraphics[width=\linewidth]{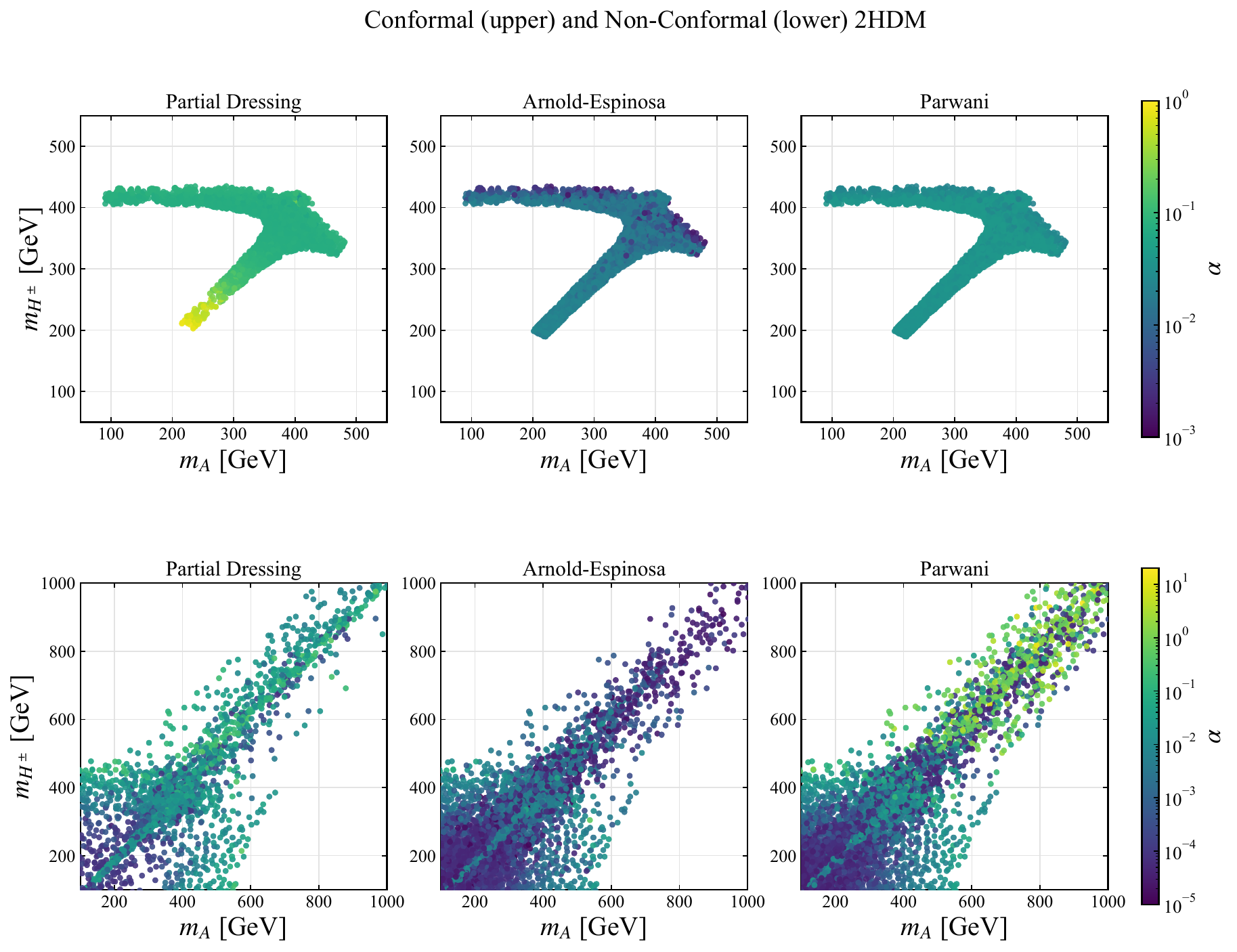}
    \caption{First-order electroweak phase transition points in the $m_A$--$m_{H^\pm}$ plane for the conformal and non-conformal 2HDM. The upper and lower rows show the conformal and non-conformal scenarios, respectively. From left to right, the columns correspond to the partial dressing, Arnold--Espinosa, and Parwani prescriptions. The color coding indicates the transition strength $\alpha$ at the percolation temperature. The figure shows that both the location of the first-order points and the corresponding transition strength depend sensitively on the thermal resummation prescription.}
    \label{fig:mass_planes_alpha}
\end{figure}

We now examine the thermodynamic strength of completed transitions across the scanned parameter space. We first focus on \(\alpha\), which controls the released energy available to the plasma and hence the overall size of the GW signal.

Figure~\ref{fig:mass_planes_alpha} shows the distribution of \(\alpha\) in the \(m_A\)--\(m_{H^\pm}\) plane for the completed first order transitions. In the conformal case, shown in the upper row, the three resummation prescriptions lead to broadly similar completed transition regions, but assign different thermodynamic strengths to these points. In the PD prescription, the largest values of \(\alpha\) are localized near the light scalar mass end of the diagonal arm. Toward the main completed transition band, the values of \(\alpha\) generally become more moderate. The AE prescription gives visibly weaker thermodynamic strength over a large part of the same projection, with many points lying at smaller \(\alpha\), especially along the nearly horizontal band and the heavier mass side of the completed region. The Parwani prescription gives an intermediate pattern, with values of \(\alpha\) that are mostly moderate and less strongly enhanced than in the strongest PD region. Across most of the conformal completed region, the strengths are therefore ordered as \(\alpha_{\rm PD}>\alpha_{\rm Parwani}>\alpha_{\rm AE}\).

The non-conformal case, shown in the lower row, exhibits a broader and more diffuse distribution in the \(m_A\)--\(m_{H^\pm}\) plane, reflecting the larger freedom associated with the additional mass parameters. Across all three prescriptions, we find a pronounced enhancement of the transition strength along the approximately mass-degenerate diagonal, \(m_A \simeq m_{H^\pm}\), compared with the surrounding regions; this feature is particularly visible in the lower-mass range. The PD prescription mainly populates a relatively narrow part of the diagonal band, with \(\alpha\) values that are mostly moderate but can become sizable in selected regions. The AE prescription gives many completed transitions with smaller \(\alpha\), showing that a first order classification does not necessarily imply a large released energy. By contrast, the Parwani prescription covers a broader range of thermodynamic strengths in this projection, including points with comparatively large \(\alpha\) at heavier scalar masses. These differences show that the thermal resummation prescription affects not only where completed first order transitions appear, but also the released energy encoded in \(\alpha\).

\begin{figure}
    \centering
         \includegraphics[width=\textwidth]{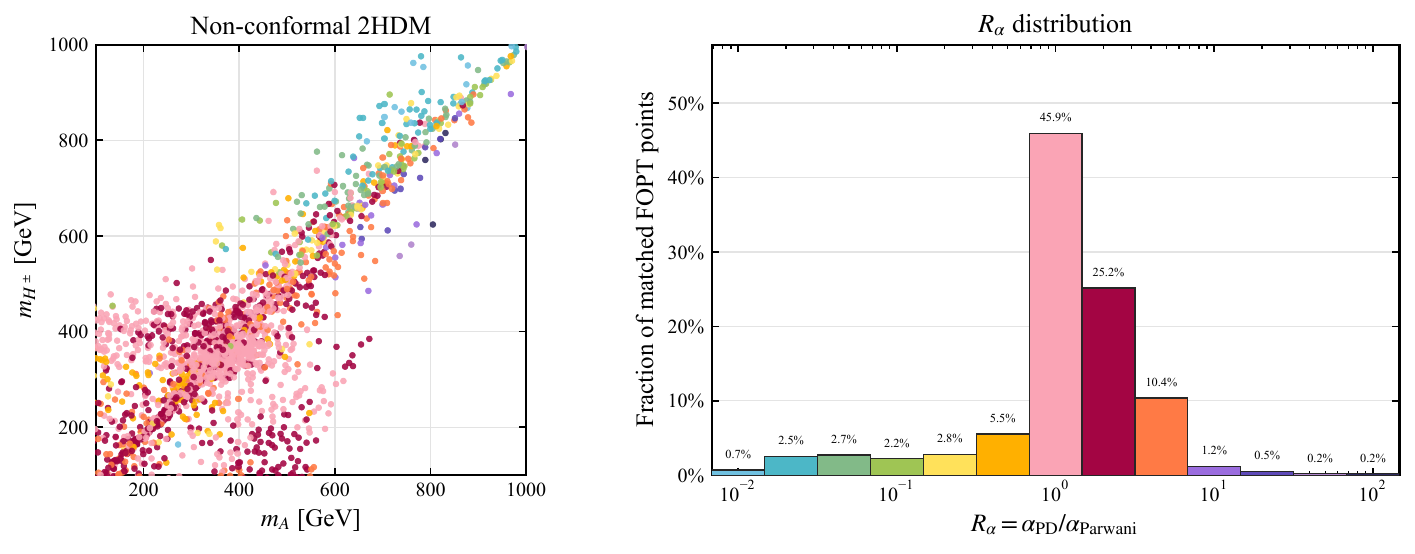}
\caption{
Comparison of the transition strength in the partial dressing and Parwani prescriptions for the NC2HDM. Only matched first-order transitions completed in both schemes are shown. The left panel shows $R_\alpha=\alpha_{\rm PD}/\alpha_{\rm Parwani}$ in the $m_A$--$m_{H^\pm}$ plane, while the right panel shows its logarithmically binned distribution. We use equal width bins in $\log_{10}R_\alpha$ with three bins per decade, and assign the same bin colours in both panels.
}
    \label{fig:alpha_ratio}
\end{figure}

We observe that, in the NC2HDM, the PD and Parwani prescriptions lead to qualitatively similar patterns, while still exhibiting quantitative differences. To make the prescription dependence more explicit, we further compare the thermodynamic strength obtained in the PD and Parwani schemes in Fig.~\ref{fig:alpha_ratio}. The color scale shows the ratio $R_{\alpha}=\alpha_{\rm PD}/\alpha_{\rm Parwani}$ for the non-conformal 2HDM. The distribution is sharply peaked at \(R_\alpha\simeq1\): almost half of the matched points (\(46\%\)) agree within a factor of \(1.5\), and about three quarters within a factor of three. The remaining deviations are asymmetric, with PD giving the larger \(\alpha\) by more than a factor of \(1.5\) for about \(38\%\) of the points and Parwani for about \(16\%\), and the tails extend over two decades on either side of unity. Thus the two prescriptions often predict comparable energy release for completed first-order transitions, but neither dominates universally. Nevertheless, sizable deviations are visible in specific regions of the $m_A$--$m_{H^\pm}$ plane.

\subsection{Supercooling and Inverse Duration}
\label{sec:resummation_schemes}

\begin{figure}[t!]
    \centering
    \includegraphics[width=\linewidth]{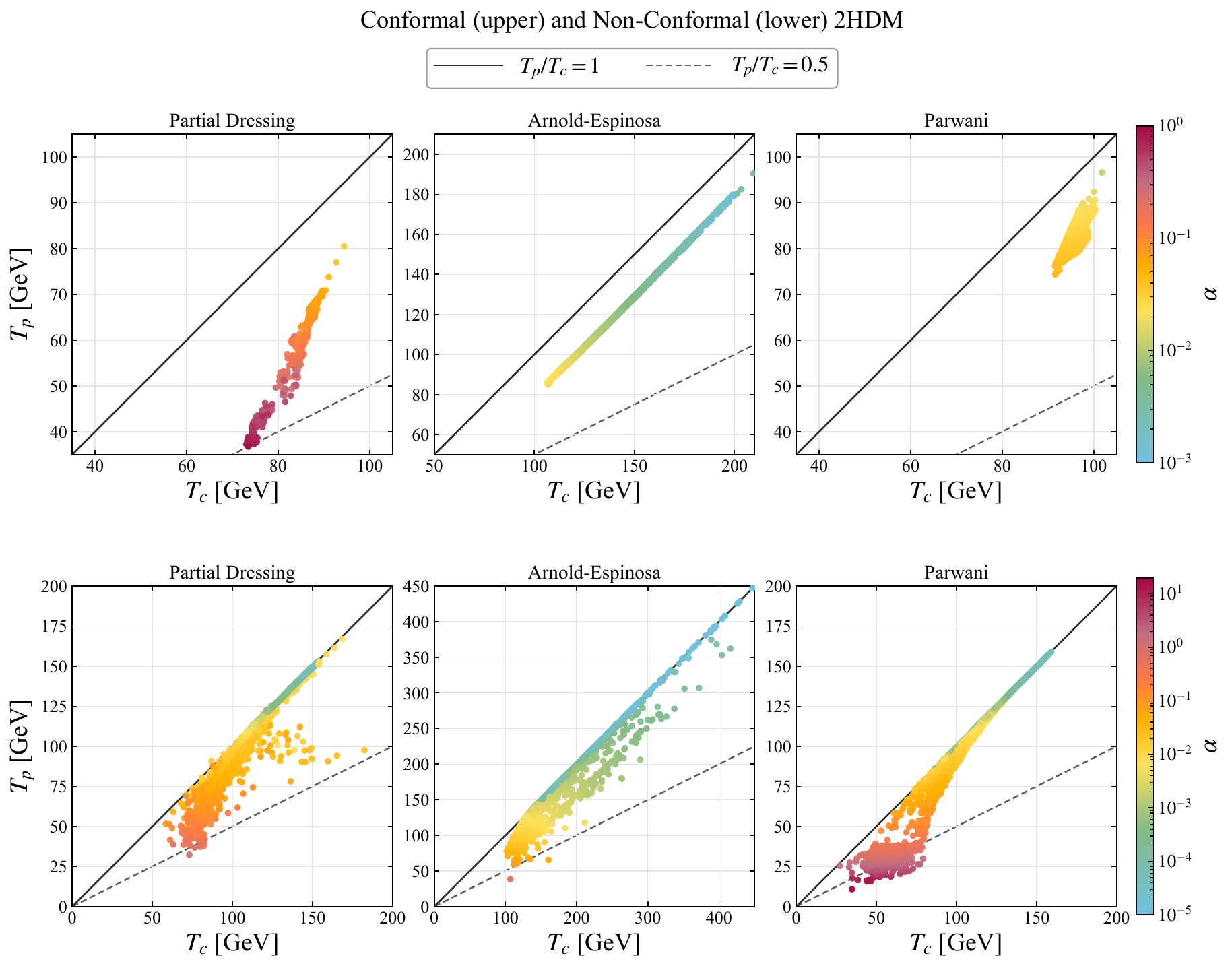}
    \caption{Completed transitions projected onto the \((T_c,T_p)\) plane for the conformal (upper row) and non-conformal (lower row) scans.  In each row the columns show the PD, AE and Parwani prescriptions. The reference lines show fixed ratios \(T_p/T_c=1\) and \(T_p/T_c=0.5\), and the color scale denotes \(\alpha\).}
    \label{fig:three_methods_tp_tc}
\end{figure}

The amount of supercooling is displayed directly in the \((T_c,T_p)\) projection in figure~\ref{fig:three_methods_tp_tc}.  Points close to the diagonal \(T_p=T_c\) correspond to transitions that complete soon after the two phases become degenerate, whereas points farther below the diagonal have a larger supercooled interval. The color scale gives the corresponding value of \(\alpha\), so the same figure also indicates how the released energy varies across the completed transitions.

As expected, the completed transitions in Fig.~\ref{fig:three_methods_tp_tc} lie on or below the diagonal \(T_p=T_c\).  In the conformal row, the PD prescription gives the most extended supercooled tail among the three prescriptions shown, reaching \(T_p/T_c\simeq0.50\). This corresponds to a sizable but still moderate amount of supercooling, rather than an extremely supercooled regime. The completed AE and Parwani transitions in the same row remain closer to the diagonal, with \(T_p/T_c\gtrsim0.8\). The non-conformal row extends over a broader range of temperature scales, reflecting the broader zero temperature parameter space. We again observe that the PD and Parwani prescriptions display qualitatively similar behavior in this comparison, except that Parwani also produces a cluster of strongly supercooled transitions with \(T_p\lesssim35~\mathrm{GeV}\), \(T_p/T_c\) as low as about \(0.3\) and \(\alpha\gtrsim0.5\), which has no counterpart of comparable strength in the PD panel. The figure also shows that a large separation between \(T_c\) and \(T_p\) is not the only way in which the prescriptions differ: the range of \(\alpha\) associated with the completed points also changes across columns, especially in the non-conformal scan. Here, unlike in Fig.~\ref{fig:three_methods_phase_diagram}, the prescription changes how strong and how supercooled the completed transitions are, rather than their type.

\begin{figure}[t!]
    \centering
    \includegraphics[width=\linewidth]{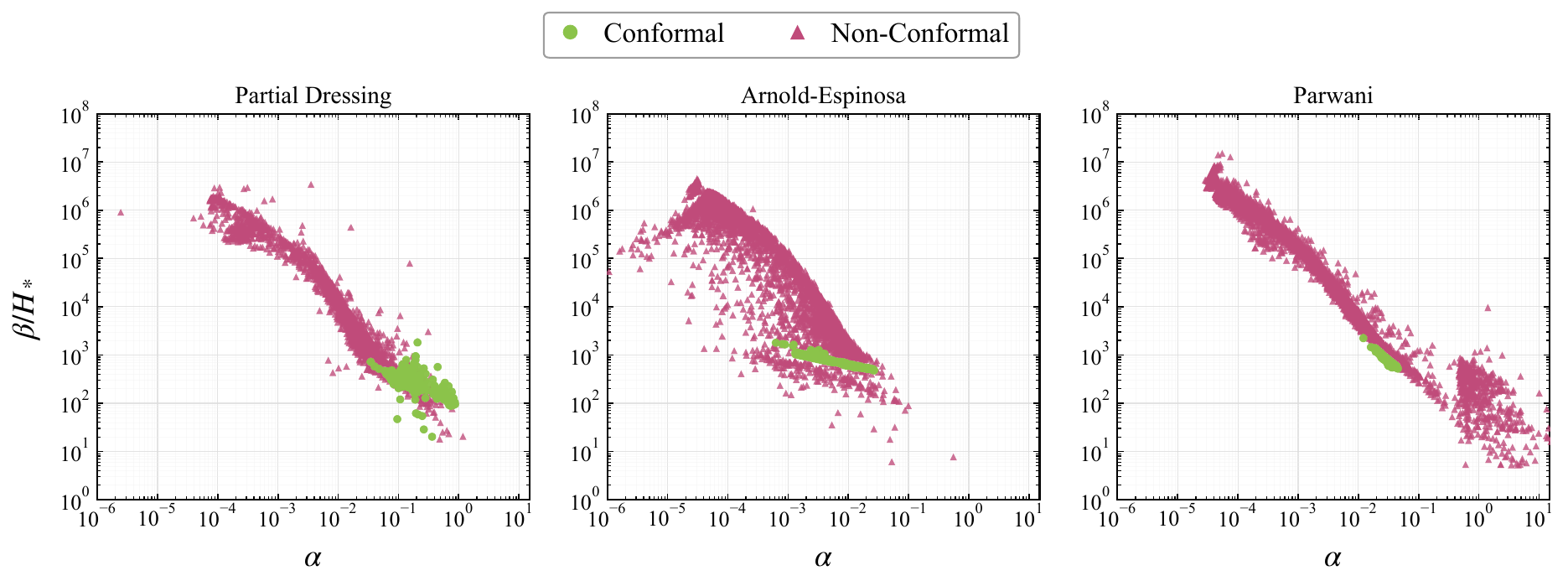}
    \caption{
    Completed transitions in the \((\alpha,\beta/H_*)\) plane for the conformal and non-conformal scans, the columns show the PD, AE and Parwani prescriptions. }
    \label{fig:pd_phase_strength}
\end{figure}

We next use the \((\alpha,\beta/H_*)\) plane to connect the completed transitions with the quantities entering the gravitational wave calculation. Figure~\ref{fig:pd_phase_strength} summarizes this projection. The parameter \(\alpha\) measures the released energy normalized to the radiation bath, while \(\beta/H_*\) characterizes the inverse duration of the transition. Thus larger \(\alpha\) implies a larger energy budget for the GW source, whereas smaller \(\beta/H_*\) corresponds to a slower transition and a larger characteristic bubble separation.

Figure~\ref{fig:pd_phase_strength} shows the completed transitions in the \((\alpha,\beta/H_*)\) plane, with the three panels corresponding to the PD, AE and Parwani prescriptions. In the conformal case, the viable completed transition points form comparatively localized branches.  In particular, under the PD prescription the conformal points extend toward relatively large \(\alpha\) and small \(\beta/H_*\), corresponding to stronger and longer lasting transitions within the completed transition sample, whereas the conformal AE and Parwani points occupy more restricted regions of the same plane. By contrast, the non-conformal points populate a much broader region, spanning several orders of magnitude in \(\alpha\) and including both rapid transitions with large \(\beta/H_*\) and slower transitions with smaller \(\beta/H_*\).  Although the main branches exhibit the expected tendency that stronger transitions have smaller \(\beta/H_*\), the sizable scatter shows that the transition duration is not determined by \(\alpha\) alone.  The broader non-conformal distribution is consistent with the additional mass scales in the non-conformal potential, while the detailed population of the plane remains prescription dependent. With Parwani, the non-conformal sample also contains a separate cluster at \(\alpha\gtrsim0.5\) and \(\beta/H_*\lesssim10^3\), consistent with the strongly supercooled transitions in Fig.~\ref{fig:three_methods_tp_tc}; it contains the strongest transitions of the entire scan.

\Needspace{7\baselineskip}
\section{Gravitational Wave Implications}
\label{sec:gw_implications}
Using the transition parameters obtained above, we calculate the gravitational-wave spectra and compare their peak amplitudes and frequencies at fixed zero-temperature inputs. At every conformal scan point where all three prescriptions predict a completed transition, PD gives the largest peak amplitude and AE the smallest, whereas the non-conformal scan has no universal ordering, although PD still gives the largest amplitude at most common points. The benchmark spectra quantify the size of this prescription dependence: the peak amplitude shifts by up to about nine orders of magnitude between PD and AE and the peak frequency by up to a factor of about fifty.

\subsection{Stochastic GW Production from an FOPT}

A first-order electroweak phase transition can source a stochastic gravitational wave background when bubbles of the broken phase nucleate, expand, and transfer energy to the surrounding plasma~\cite{Caprini_2016,Caprini:2018mtu}.
The macroscopic inputs for the spectrum are the percolation temperature and transition parameters computed in section~\ref{sec:phase_transition_dynamics}.  The distributions in figure~\ref{fig:pd_phase_strength} therefore provide the direct bridge between the phase transition dynamics and the gravitational wave spectra.

In general, the spectrum can receive contributions from scalar field bubble wall collisions, sound waves in the plasma, and magnetohydrodynamic turbulence,
\begin{equation}
    h^2\Omega_{\rm GW}(f)
    =
    h^2\Omega_{\rm col}(f)
    +
    h^2\Omega_{\rm sw}(f)
    +
    h^2\Omega_{\rm turb}(f).
    \label{eq:gw_total_general}
\end{equation}
The efficiency factor \(\kappa_\phi\) associated with
\(h^2\Omega_{\rm col}\) measures the fraction of the released vacuum energy stored in the scalar field bubble walls. In the absence of a microscopic wall friction calculation, we adopt a conservative non-runaway energy budget~\cite{Bodeker:2009qy,Espinosa:2010hh}.
In this setup, the released energy is assumed to source bulk plasma motion rather than continued wall acceleration, so that
\begin{equation}
    \kappa_\phi \simeq 0,
    \qquad
    h^2\Omega_{\rm col}(f)\simeq0 .
    \label{eq:no_collision_nonrunaway}
\end{equation}
With this choice, the GW spectrum used in the main numerical analysis is
\begin{equation}
    h^2\Omega_{\rm GW}(f)
    =
    h^2\Omega_{\rm sw}(f)
    +
    h^2\Omega_{\rm turb}(f).
    \label{eq:gw_total_default}
\end{equation}

To translate these thermodynamic quantities into a present-day stochastic gravitational wave spectrum, we use the double broken power law (DBPL) template for first-order phase transitions with moderate supercooling~\cite{Caprini:2024hue}. For the two fluid sources \(i={\rm sw},{\rm turb}\) we have
\begin{equation}
    S_i(f)=N_i
    \left(\frac{f}{f_{1,i}}\right)^{n_{1,i}}
    \left[
    1+\left(\frac{f}{f_{1,i}}\right)^{a_{1,i}}
    \right]^{\frac{n_{2,i}-n_{1,i}}{a_{1,i}}}
    \left[
    1+\left(\frac{f}{f_{2,i}}\right)^{a_{2,i}}
    \right]^{\frac{n_{3,i}-n_{2,i}}{a_{2,i}}} .
    \label{eq:dbpl_shape}
\end{equation}
where \(f_{1,i}\) and \(f_{2,i}\) are the two break frequencies. The spectrum is then determined by
\begin{equation}
    h^2\Omega_i(f)/h^2\Omega_i(f_{2,i})={S_i(f)}/{S_i(f_{2,i})},
\end{equation}
in which the constant \(N_i\) cancels.  The true peak used in Figure~\ref{fig:gw_peak_scatter} is then obtained by maximizing the total spectrum numerically.  The DBPL exponents are listed in table~\ref{tab:dbpl_slopes}.

\begin{table}[t]
\centering
\begin{tabular}{c|ccccc}
Source & \(n_1\) & \(n_2\) & \(n_3\) & \(a_1\) & \(a_2\) \\
\hline
sound waves & \(3\) & \(1\) & \(-3\) & \(2\) & \(4\) \\
MHD turbulence & \(3\) & \(1\) & \(-8/3\) & \(4\) & \(2.15\)
\end{tabular}
\caption{DBPL exponents used for the sound wave and MHD turbulence spectra.}
\label{tab:dbpl_slopes}
\end{table}

The remaining inputs specify how the phase-transition parameters are mapped to the present-day spectrum~\cite{Caprini:2019egz,Caprini:2024hue}.
The redshifted Hubble scale and the overall redshift factor are
\begin{align}
    H_{*,0}
    \simeq
    1.65\times10^{-5}\,{\rm Hz}\,g_{*,100}^{1/6}T_{*,100}
    ,\qquad
    h^2F_{\rm GW,0}
    \simeq
    1.64\times10^{-5}g_{*,100}^{-1/3}.
    \label{eq:FGW0}
\end{align}
where \(g_{*,100}=g_*/100\) and \(T_{*,100}=T_*/100\,{\rm GeV}\). The characteristic bubble separation is fixed by \(\beta/H_*\), while the kinetic energy fraction is estimated from the strength parameter \(\alpha\):
\begin{align}
    H_*R_*
    &=
    (8\pi)^{1/3}v_w
    \left(\frac{\beta}{H_*}\right)^{-1},\qquad
    K\simeq
    0.6\,\frac{\kappa(\alpha)\alpha}{1+\alpha}.
    \label{eq:K_definition}
\end{align}
Here \(\kappa(\alpha)\simeq\alpha/(0.73+0.083\sqrt{\alpha}+\alpha)\) is the single-bubble efficiency factor for a relativistic fluid~\cite{Espinosa:2010hh}.  Thus, increasing \(\alpha\) raises the kinetic-energy budget, whereas increasing \(\beta/H_*\) shortens the transition, decreases \(R_*\), and shifts the spectrum to higher frequency. 

For the sound-wave contribution, the two break frequencies are set by the inverse bubble separation and by the width of the sound shell,
\begin{equation}
    f_{1,{\rm sw}}
    \simeq
    0.2\,\frac{H_{*,0}}{H_*R_*},\qquad
    f_{2,{\rm sw}}
    \simeq
    0.5\,\frac{H_{*,0}}{\Delta_bH_*R_*} .
\label{eq:sw_breaks}
\end{equation}
Here \(\Delta_b\simeq0.42\) is the shell-width parameter. The normalization at the second break is then
\begin{equation}
 h^2\Omega_{2,{\rm sw}}(f_{2,{\rm sw}})
    =0.11\,
    \frac{1}{\pi}
    \left[
    \sqrt2+
    \frac{2(f_{2,{\rm sw}}/f_{1,{\rm sw}})}
    {1+(f_{2,{\rm sw}}/f_{1,{\rm sw}})^2}
    \right]
    h^2F_{\rm GW,0}\,
    K^2\,
    H_*R_*\,
    \Upsilon .
\label{eq:Omega2_sw}
\end{equation}
The factor \(\Upsilon\) accounts for the finite lifetime of the acoustic source~\cite{Guo:2020grp}:
\begin{equation}
\Upsilon=1-\frac{1}{\sqrt{1+2H_*R_*/\bar U_f}} .
\label{eq:sound_lifetime}
\end{equation}
The enthalpy-weighted root-mean-square of the fluid velocity is \( \bar U_f =\sqrt{K/\Gamma}\), with the mean adiabatic index \(\Gamma=4/3\) for a relativistic fluid.

For MHD turbulence, the corresponding source-specific inputs are again the two break frequencies and the normalization at the second break. The break frequencies are
\begin{equation}
f_{1,{\rm turb}}= \frac{1}{{\cal N}}\,\sqrt{\frac{\varepsilon K}{\Gamma}}\,\frac{H_{*,0}}{H_*R_*},
\qquad
f_{2,{\rm turb}} \simeq 2.2\,\frac{H_{*,0}}{H_*R_*}.
\label{eq:turb_breaks}
\end{equation}
We take \(\varepsilon=0.5\) as the template fraction of bulk kinetic energy converted into turbulence, with \({\cal N}\simeq2\)~\cite{Caprini:2024hue}.
The corresponding normalization is
\begin{equation}
h^2\Omega_{2,{\rm turb}}(f_{2,{\rm turb}})=4.37\times10^{-3}\,h^2F_{\rm GW,0}\,(\varepsilon K)^2\,(H_*R_*)^2.
\label{eq:Omega2_turb}
\end{equation}
Equations~\eqref{eq:gw_total_default}--\eqref{eq:Omega2_turb} define the numerical prescription used to obtain the spectra and peak values shown below. 

\subsection{GW Benchmark Spectra and Peak Distributions}
\label{sec:gw_benchmark_peak}

\begin{figure}[t!]
    \centering
    \includegraphics[width=\textwidth]{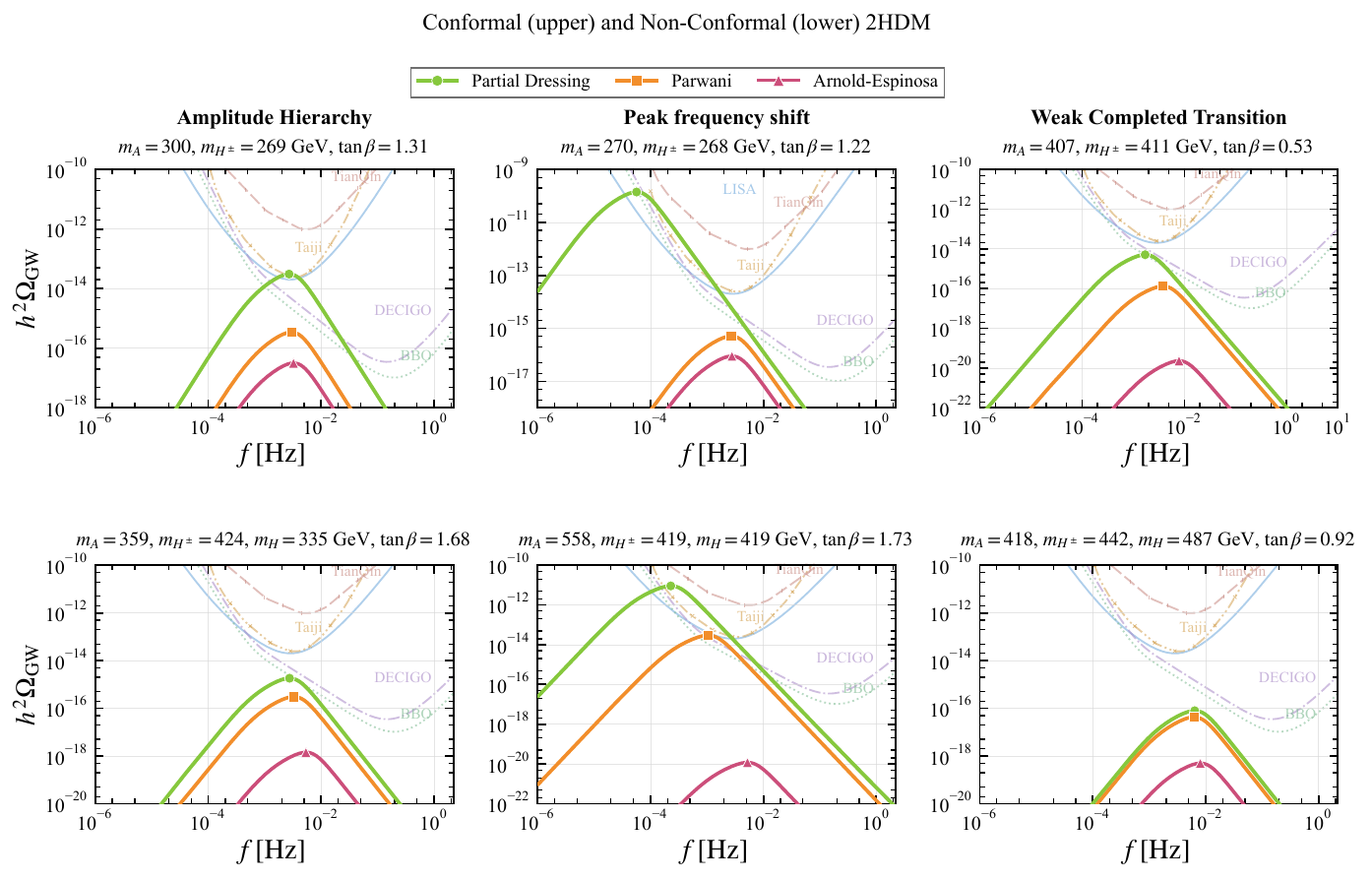}
    \caption{
GW spectra \(h^2\Omega_{\rm GW}(f)\) for six representative 2HDM benchmarks
under the PD, Parwani and AE prescriptions. The upper row shows the conformal
model and the lower row the non-conformal model. The columns illustrate,
from left to right, an amplitude hierarchy, a peak frequency shift and a weak
but completed transition. Thick curves denote the model predictions; thin
curves show the projected sensitivities of LISA, Taiji, TianQin, DECIGO and BBO.
Each panel lists the scalar masses and \(\tan\beta\) of the benchmark.
For the non-conformal benchmarks in the lower row, the soft-breaking parameter
\(m_{12}\) is \(78.11\), \(141.32\) and \(151.16\,\mathrm{GeV}\), respectively,
from left to right.
    }
    \label{fig:gw_benchmark_spectra}
\end{figure}

\begin{figure}[t!]
    \centering
    \includegraphics[width=\textwidth]{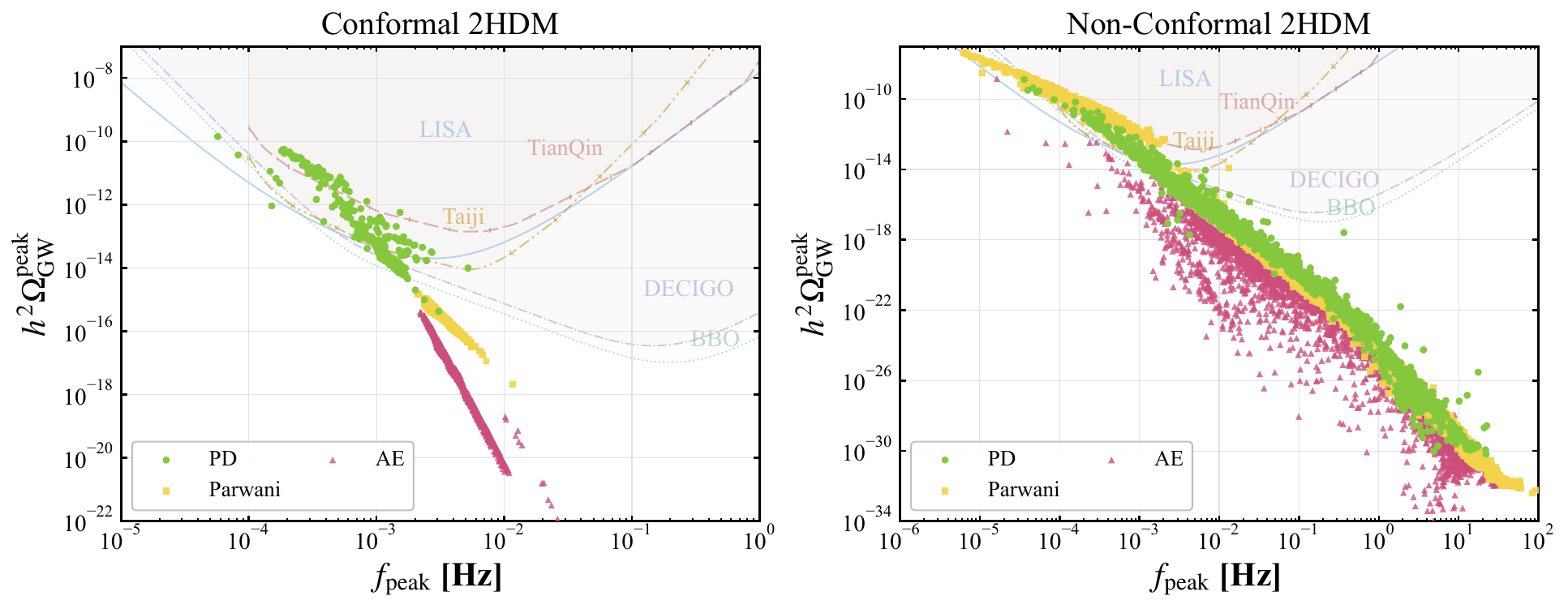}
    \caption{
    Peak amplitude \(h^{2}\Omega_{\rm GW}^{\rm peak}\) versus peak frequency
    \(f^{\rm peak}\) for scan points that give a completed FOEWPT under each prescription. The overlaid detector curves are used only as visual
    sensitivity references under the convention specified in the text.
    }
    \label{fig:gw_peak_scatter}
\end{figure}

We place the predicted stochastic gravitational wave spectra in an observational context by overlaying the projected sensitivities of the space based interferometers LISA~\cite{2017arXiv170200786A}, BBO~\cite{Corbin:2005ny}, DECIGO~\cite{Kawamura:2011zz}, TianQin~\cite{TianQin:2015yph,Liang:2021bde} and Taiji~\cite{Hu:2017mde,Liu:2023ymk}. These experiments cover the frequency range most relevant for electroweak scale phase transitions. The curves in Figs.~\ref{fig:gw_benchmark_spectra} and~\ref{fig:gw_peak_scatter} are used only as visual sensitivity references, not as exclusion limits or detector level discovery contours. The displayed sensitivity curves follow the standard power law integrated and peak integrated sensitivity conventions of Refs.~\cite{Thrane:2013oya,Schmitz:2020syl}, supplemented by the TianQin and Taiji curves of Refs.~\cite{TianQin:2015yph,Liang:2021bde,Hu:2017mde,Liu:2023ymk}. They are shown for a signal to noise ratio (SNR) threshold of 10 and an observation time of four years, or the corresponding effective operation time. Since we do not perform SNR integrations for individual missions or model foreground subtraction, overlaps with these curves should be interpreted only as comparisons at the order of magnitude level.

\begin{figure}[h!]
    \centering
    \includegraphics[width=\linewidth]{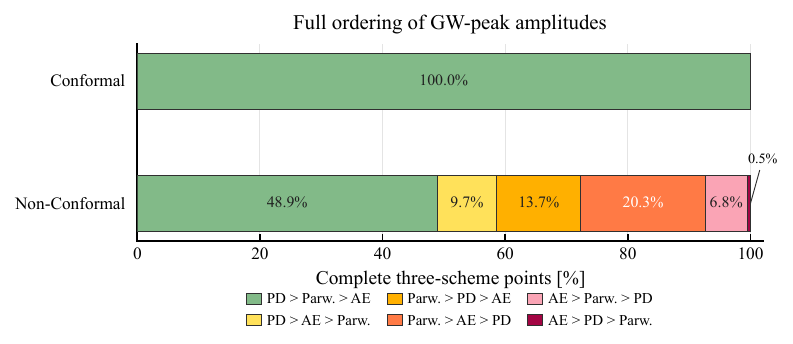}
    \caption{ Peak amplitude classification for common GW producing scan points, defined as points for which PD, Parwani and AE all predict a completed first order electroweak phase transition. In the conformal 2HDM, all common points satisfy \(\Omega_{\rm peak}^{\rm PD}> \Omega_{\rm peak}^{\rm Parwani}> \Omega_{\rm peak}^{\rm AE}\), so PD always gives the largest peak amplitude. In the non conformal 2HDM, this relation is no longer universal, and the bars show the corresponding fractions for the different peak amplitude classes. }
    \label{fig:gw_peak_ordering}
\end{figure}

Before showing representative spectra, we first compare the GW peak amplitudes obtained from the same zero temperature parameter point under the three thermal prescriptions. The comparison is made separately in the conformal and non-conformal scans, using only common completed points for which PD, Parwani and AE all yield a GW spectrum. 

Figure~\ref{fig:gw_peak_ordering} displays the corresponding peak amplitude classifications at the scan level. In the C2HDM, all common points with a GW spectrum satisfy \(\Omega_{\rm peak}^{\rm PD}>\Omega_{\rm peak}^{\rm Parwani}> \Omega_{\rm peak}^{\rm AE}\). This shows that the full hierarchy holds at every common point of the conformal scan. In the NC2HDM, this simple relation is lost: the full ordering survives at only \(48.9\%\) of the common points. PD gives the largest GW peak amplitude in \(58.6\%\) of the common points, Parwani in \(34.0\%\), and AE in \(7.3\%\). Thus, unlike in the conformal model, where \(m_H\) is not an independent input and no soft scale is present, the prescription that yields the largest GW signal in the non-conformal model depends on the values of the additional mass parameters \(m_H\) and \(M\).

Figure~\ref{fig:gw_benchmark_spectra} shows the resulting gravitational wave spectra for six representative benchmark points selected from the viable scan. They are selected as representative, phenomenologically viable examples in which the prescription dependence is clearly visible, rather than as extreme corners of parameter space or as cases driven by numerical artifacts. The three columns illustrate different aspects of this prescription dependence: an amplitude hierarchy, a peak frequency shift, and a weak but completed transition. In the amplitude hierarchy panels, PD gives the largest signal, AE the smallest, and Parwani lies between them, with the PD and AE peak amplitudes differing by about three orders of magnitude at similar peak frequencies. The peak frequency shift panels show that changing the prescription can move not only the normalization but also the frequency at which the spectrum is maximal: the PD spectrum peaks at a frequency lower than the AE one by a factor of about 20 (non-conformal) to 50 (conformal), and the PD--AE amplitude gap grows to about six (conformal) and nine (non-conformal) orders of magnitude. In the weak completed transition panels, all three spectra are suppressed relative to the stronger benchmark cases, with AE again giving the smallest amplitude in the examples shown. These benchmark spectra make the prescription dependence visible at the level of individual points, while the broader scan level pattern is summarized next.

Figure~\ref{fig:gw_peak_scatter} summarizes the scan level predictions in the \((f^{\rm peak},h^{2}\Omega^{\rm peak}_{\rm GW})\) plane. For each thermal prescription, the plotted points are those for which that prescription predicts a completed FOEWPT and hence a GW spectrum. In the conformal scan, the points occupy a comparatively narrow frequency interval, and the three prescriptions separate mainly in amplitude. The PD points form the upper envelope, Parwani lies below them over much of the same frequency range, and AE is concentrated toward the lower-amplitude tail. This scatter level hierarchy is consistent with the complete conformal ordering found in Fig.~\ref{fig:gw_peak_ordering}. The non conformal scan spans a broader region and extends to higher peak frequencies, reflecting the wider range of transition scales and inverse durations. PD and Parwani populate similar descending bands over much of the $(f_{\rm peak},h^2\Omega_{\rm GW}^{\rm peak})$ plane, whereas AE typically gives smaller amplitudes at comparable $f_{\rm peak}$. This behavior is consistent with Eq.~\eqref{eq:K_definition}: increasing $\beta/H_*$ reduces the characteristic bubble separation $H_*R_*$, shifts the spectral features to higher frequencies, and suppresses the overall normalization, while increasing $\alpha$ raises the kinetic-energy budget $K$. The finite width of the bands reflects the combined variation of $\alpha$, $\beta/H_*$, and $T_p$ across the non-conformal parameter space. Some of the strongest points lie close to the displayed detector sensitivity curves.

\section{Conclusion}
\label{sec:conclusion}

By comparing multi-field PD, Parwani, and Arnold--Espinosa resummation at the same zero-temperature inputs in the conformal and non-conformal 2HDM, we isolate the effect of the thermal prescription on electroweak phase transitions and their gravitational-wave signals. We find that the prescription can change the type of thermal history at a given input point: the same point can show a completed first-order transition under one prescription but a second-order transition or electroweak symmetry non-restoration (EWSNR) under another (Figs.~\ref{fig:three_methods_phase_diagram} and~\ref{fig:three_modes_phase_flow}). Beyond such qualitative changes, the prescription can shift the predicted gravitational-wave signal by orders of magnitude.

On the technical side, we formulate the multi-field scalar gap equations with field-dependent mixing and implement them for both 2HDMs. Feynman--Hellmann relations write the mass derivatives in these equations as tree-level couplings projected onto mass eigenstates. PD keeps the tree-level couplings but projects them onto the dressed eigenstates, so that only the propagators are dressed. For the conformal model, finite counterterms preserve the electroweak vacuum and the CP-odd and charged mass blocks, while the SM-like Higgs mass is generated radiatively along the Gildener--Weinberg direction. This provides a consistent matching prescription for the conformal scan and allows a controlled comparison with the non-conformal 2HDM.

Applying the multi-field PD prescription also reveals a theoretical inconsistency: contrary to the path-independence argument of Ref.~\cite{Bahl2024Improved}, the dressed tadpole is not a gradient, so its line integral is path dependent and the direct multi-field extension fails to define a unique effective potential. For the numerical predictions, we use the straight-path definition of the PD potential, which is only one of many inequivalent choices. To estimate the path ambiguity, we compare the straight path with two alternative contours for five benchmarks in each model (Appendix~\ref{app:path-dependence}), finding that the transition temperatures change by at most about \(7\%\) and the gravitational-wave peak amplitude by up to a factor of about three, well below the orders-of-magnitude prescription differences in Fig.~\ref{fig:gw_benchmark_spectra}. The path dependence is therefore a smaller, but intrinsic, ambiguity of multi-field PD that calls for a consistent multi-field formulation. Our subsequent work~\cite{LiWangZhao:MultifieldCJT} constructs such a formulation, to which we return below.

With this PD prescription, completed first-order transitions dominate the conformal scan, as they do under Parwani and Arnold--Espinosa (Fig.~\ref{fig:three_methods_phase_diagram}). The prescription dependence appears mainly in the subdominant non first-order classes. In the non conformal 2HDM, the pattern is less uniform: second-order histories appear under all three prescriptions, while EWSNR-like histories appear only with Arnold--Espinosa, in a broad band along \(m_A\simeq m_{H^\pm}\). We trace these AE EWSNR candidate histories to the cubic daisy term, whose negative contribution to the thermal curvature can overcome the positive thermal mass and keep the broken minimum favored over the temperature range examined.

In the conformal scan, PD typically gives the strongest and most supercooled completed transitions, reaching \(\alpha=\mathcal O(1)\) and \(T_p/T_c\simeq0.5\), Arnold--Espinosa gives the weakest, and Parwani lies between them (Figs.~\ref{fig:mass_planes_alpha} and~\ref{fig:three_methods_tp_tc}). The same tendency is also visible in the $(\alpha,\beta/H_*)$ plane in Fig.~\ref{fig:pd_phase_strength}, where the PD points extend toward larger $\alpha$ and smaller $\beta/H_*$. Supercooling alone, however, does not fix the released energy: a smaller $T_p/T_c$ need not give a larger $\alpha$.

In the non-conformal scan there is no such fixed hierarchy: which prescription gives the strongest transition changes across parameter space. All three prescriptions nevertheless share one feature: below about \(500~\mathrm{GeV}\), the transition strength is enhanced in a narrow band around \(m_A\simeq m_{H^\pm}\) compared with neighbouring points. With PD this band is broader, and large values of \(\alpha\) are also more common away from the diagonal than with AE or Parwani (Fig.~\ref{fig:mass_planes_alpha}). PD and Parwani often give comparable transition strengths and form similar broad bands in the \((\alpha,\beta/H_*)\) plane (Figs.~\ref{fig:alpha_ratio} and~\ref{fig:pd_phase_strength}). PD is more often the stronger of the two, but Parwani reaches the largest transition strengths and the strongest supercooling in the non-conformal scan, whereas PD gives the strongest conformal transitions.

Even where all three prescriptions predict a completed transition, these differences in \(\alpha\) and \(\beta/H_*\) carry over to the gravitational-wave signal: in the conformal scan the peak amplitudes always follow PD \(>\) Parwani \(>\) AE, whereas in the non-conformal scan PD gives the largest signal at about \(60\%\) of the common points and Parwani at about one third, with AE mostly at the low-amplitude end (Figs.~\ref{fig:gw_peak_ordering} and~\ref{fig:gw_peak_scatter}). For individual benchmarks, the PD and AE peak amplitudes differ by up to about nine orders of magnitude, and the peak frequencies by up to a factor of about fifty (Fig.~\ref{fig:gw_benchmark_spectra}). The overlaid LISA, Taiji, TianQin, DECIGO and BBO curves serve only as order-of-magnitude references, not as discovery contours; within this caveat, the larger PD amplitudes suggest comparatively favorable observational prospects.

Our results identify the thermal resummation prescription as a major source of theoretical uncertainty in gravitational-wave predictions from extended Higgs sectors. Even the minimal PD used here omits two subleading effects: the momentum-overlap correction to the sunset contribution in its gap equation and the lollipop diagram (Section~\ref{sec:PD}). Hybrid partial dressing (HPD) captures both by adding the full two-loop vacuum sunset to the potential, but was formulated without scalar mixing~\cite{Berthiaume2026Hybrid}. Our subsequent work~\cite{LiWangZhao:MultifieldCJT} derives both corrections from one two-loop Cornwall--Jackiw--Tomboulis functional, which yields path-independent potentials for any number of mixing scalars, including HPD. It remains to extend this framework beyond the scalar sector, notably to the gauge bosons, which here receive only leading Debye masses, and to apply it to the 2HDM.

\acknowledgments
We thank Michael Ramsey-Musolf, Xiao-Gang He, Pedro Bittar, Philipp Schicho, Yun Jiang and Jin-Wei Wang for helpful discussions. Zhi-Wei Wang, Ji-Wei Li and Yu-Bo Zhao are
partially supported by the National Natural Science Foundation of China (Grant No. 12475105, Grant No. 12622508).

\appendix

\section{Useful Identities}
\label{app:identities}
We give the scalar one-loop functions used in the gap equation, separating the vacuum and thermal terms and isolating the Matsubara zero mode resummed by Arnold--Espinosa. For a bosonic scalar degree of freedom, the one-loop contribution to the finite-temperature effective potential can be written in terms of the loop integral \cite{Kapusta:2006,Laine:2016}
\begin{equation}
\mathcal J(m^2)
\equiv
\int_K \ln(K^2+m^2)
=
T
\sum_{n=-\infty}^{\infty}
\int\frac{d^3k}{(2\pi)^3}
\ln\left(k^2+\omega_n^2+m^2\right),
\end{equation}
where \(\omega_n=2\pi nT\) are the bosonic Matsubara frequencies. Here $\int_K\equiv T\sum_n\int d^3k/(2\pi)^3$ and $K=(\omega_n,\mathbf k)$ is the Euclidean momentum. With this normalization, the one-loop potential per real scalar is $\mathcal J(m^2)/2$. Performing the Matsubara sum separates the result into a zero-temperature Coleman--Weinberg contribution and a finite-temperature correction,
\begin{equation}
\mathcal J(m^2)
=
\underbrace{
\int\frac{d^3k}{(2\pi)^3}\,E_k
}_{\mathcal J_0(m^2)}
+
\underbrace{
2T
\int\frac{d^3k}{(2\pi)^3}
\ln\left(1-e^{-E_k/T}\right)
}_{\mathcal J_T(m^2)},
\end{equation}
where \(E_k^2=k^2+m^2\). After renormalization in the \(\overline{\mathrm{MS}}\) scheme, the vacuum part of the determinant is 
\begin{equation}
\mathcal J_0(m^2)
=
\frac{m^4}{32\pi^2}
\left(
\ln\frac{m^2}{\bar\mu^2}
-\frac{3}{2}
\right),
\end{equation}
while the finite-temperature part can be written as
\begin{equation}
\mathcal J_T(m^2)=\frac{T^4}{\pi^2}J_B\left(\frac{m^2}{T^2}\right)=\frac{T^4}{\pi^2}\int_0^\infty dx\,x^2\ln\left(1-\mathrm{e}^{-\sqrt{x^2+m^2/T^2}}\right).
\end{equation}

For the Arnold--Espinosa prescription, it is more convenient to isolate the bosonic zero mode from the Matsubara sum. This defines the soft contribution, while the remaining non-zero Matsubara modes are collected into the hard part:
\begin{equation}
\mathcal J(m^2)
=
\underbrace{
T
\int\frac{d^3k}{(2\pi)^3}
\ln(k^2+m^2)
}_{\mathcal J_{\mathrm{soft}}(m^2)}
+
\underbrace{
T
\sum_{n\neq 0}
\int\frac{d^3k}{(2\pi)^3}
\ln(k^2+\omega_n^2+m^2)
}_{\mathcal J_{\mathrm{hard}}(m^2)} .
\end{equation}
In dimensional regularization, the soft contribution evaluates to
\begin{equation}
\mathcal J_{\mathrm{soft}}(m^2)
=
-\frac{T}{6\pi}
\left(m^2\right)^{3/2}.
\end{equation}
Notably, this term is non-analytic in \(m^2\) and precisely corresponds to the infrared-sensitive contribution targeted by the Arnold--Espinosa prescription.
\section{Counterterm Potential}
\label{app:counterterm}
We give the finite counterterms and matching conditions for both models. In the C2HDM, the conditions preserve the vacuum and the CP-odd and charged mass blocks while retaining the CP-even loop correction that generates the SM-like Higgs mass. In the NC2HDM, by contrast, they preserve the vacuum and the full scalar mass matrix.

\subsection{Counterterm Potential in C2HDM} 
In order to preserve the tree-level vacuum structure and scalar mass-matrix elements after including the one-loop Coleman--Weinberg correction, we introduce a UV-finite counterterm potential for the C2HDM.

Classical scale invariance restricts the structure of the counterterm potential. Since the theory contains no dimensionful parameters at tree level, the counterterm potential must likewise be built entirely from quartic operators as \cite{Gildener:1976ih,Lee:2012jn}
\begin{align}
V_{\rm CT} ={}& \frac{\delta\lambda_1}{2}\left(\Phi_1^\dagger\Phi_1\right)^2
+ \frac{\delta\lambda_2}{2}\left(\Phi_2^\dagger\Phi_2\right)^2
+ \delta\lambda_3\left(\Phi_1^\dagger\Phi_1\right)\left(\Phi_2^\dagger\Phi_2\right)
\nonumber\\
&+ \delta\lambda_4\left(\Phi_1^\dagger\Phi_2\right)\left(\Phi_2^\dagger\Phi_1\right)
+ \frac{\delta\lambda_5}{2}\left[\left(\Phi_1^\dagger\Phi_2\right)^2+\text{h.c.}\right].
\label{eq:VCT_C2HDM}
\end{align}
There are precisely five independent counterterm parameters \(\delta\lambda_{1,\ldots,5}\); in particular there are no bilinear mass counterterms \(\delta m_{11}^2\), \(\delta m_{22}^2\), or \(\delta m_{12}^2\).

The renormalization conditions read
\begin{alignat}{2}
\text{(Tadpoles):}\quad &\left.\frac{\partial\bigl(V_{\rm CW}+V_{\rm CT}\bigr)}{\partial h_i}\right|_{\rm vac} &\,{}= 0,
\label{eq:RC_tadpole}\\[2pt]
\text{(CP-odd):}\quad &\left.\frac{\partial^2\bigl(V_{\rm CW}+V_{\rm CT}\bigr)}{\partial a_i\,\partial a_j}\right|_{\rm vac} &\,{}= 0,
\label{eq:RC_Ablock}\\[2pt]
\text{(Charged):}\quad &\left.\frac{\partial^2\bigl(V_{\rm CW}+V_{\rm CT}\bigr)}{\partial\phi_i^-\,\partial\phi_j^+}\right|_{\rm vac} &\,{}= 0.
\label{eq:RC_Cblock}
\end{alignat}
Conditions~\eqref{eq:RC_tadpole} fix the electroweak vacuum to \((v_1,v_2)\) at one loop, ensuring that neither VEV is shifted by the Coleman--Weinberg correction. Conditions~\eqref{eq:RC_Ablock} and~\eqref{eq:RC_Cblock} restore the CP-odd and charged mass matrices to their tree-level forms; as discussed in the main text, this elevates \(m_A\) and \(m_{H^\pm}\) to loop-independent input parameters, defined as zero-momentum curvature masses. Crucially, no analogous condition is placed on the CP-even (\(H\)) block: it is allowed to retain the full one-loop Coleman--Weinberg correction, which is the dynamical origin of the SM-like Higgs mass through radiative symmetry breaking.

Among these five parameters, the renormalization conditions do not fix each \(\delta\lambda_{3,4,5}\) independently but only certain combinations, leaving a one-parameter freedom. We resolve it with the prescription
\begin{equation}
\delta\lambda_{345} \equiv \delta\lambda_3+\delta\lambda_4+\delta\lambda_5 = 0.
\label{eq:dl345_zero}
\end{equation}
Solving the system formed by Eqs.~\eqref{eq:RC_tadpole}--\eqref{eq:RC_Cblock} subject to Eq.~\eqref{eq:dl345_zero} yields closed-form expressions for all five counterterms as follows
\begin{align}
\delta\lambda_1 &= -\frac{2}{v_1^3}\,DV[h_1],
\label{eq:ct_dl1_C}\\[3pt]
\delta\lambda_2 &= -\frac{2}{v_2^3}\,DV[h_2],
\label{eq:ct_dl2_C}\\[3pt]
\delta\lambda_5 &= -\frac{1}{v_1 v_2}\,DV[a_1 a_2],
\label{eq:ct_dl5_C}\\[3pt]
\delta\lambda_4 &= -\frac{2}{v_1 v_2}\,DV[\phi_1^+\phi_2^-] + \frac{1}{v_1 v_2}\,DV[a_1 a_2],
\label{eq:ct_dl4_C}\\[3pt]
\delta\lambda_3 &= \frac{2}{v_1 v_2}\,DV[\phi_1^+\phi_2^-].
\label{eq:ct_dl3_C}
\end{align}
where we use the shorthand 
\begin{equation}
DV[\varphi_a] \equiv \left.\frac{\partial V_{\rm CW}}{\partial\varphi_a}\right|_{\rm vac},
\qquad
DV[\varphi_a\varphi_b] \equiv \left.\frac{\partial^2 V_{\rm CW}}{\partial\varphi_a\,\partial\varphi_b}\right|_{\rm vac},
\label{eq:DVdef_C2HDM}
\end{equation}
with all derivatives evaluated at the electroweak minimum \(h_i = v_i\), \(a_i = \phi_i^\pm = 0\).
 
Now we discuss the Gildener--Weinberg approximation in the finite-counterterm prescription \cite{Gildener:1976ih}. Along the flat direction, the tree-level potential vanishes. Therefore, the leading contribution to the effective potential along this direction is generated by the one-loop Coleman--Weinberg correction together with the finite counterterms. Since all field-dependent masses scale as
\begin{equation}
m_i^2(h)=m_i^2(v)\frac{h^2}{v^2}
\end{equation}
along the flat direction, the Coleman--Weinberg potential can be written as
\begin{equation}
V_{\rm CW}^{\rm FD}(h)
=
A h^4
+
B h^4\ln\frac{h^2}{\bar\mu^2},
\label{eq:CWfd}
\end{equation}
where \(A\) is scheme dependent, while
\begin{equation}
\begin{aligned}
B
&=
\frac{1}{64\pi^2 v^4}
\sum_i (-1)^{2s_i} n_i m_i^4(v)
\\
&=
\frac{1}{64\pi^2 v^4}
\bigl(
m_{{\mathrm H},0}^4
+m_{{\mathrm A}^0}^4
+2m_{\mathrm H^\pm}^4
+6m_W^4
+3m_Z^4
-12m_t^4
\bigr)
\end{aligned}
\end{equation}
is the coefficient of the logarithmic term. Using the conformal counterterm conditions in Eqs.~\eqref{eq:ct_dl1_C}--\eqref{eq:ct_dl3_C}, the counterterm potential restricted to the flat direction becomes
\begin{equation}
V_{\rm CT}^{\rm FD}(h)
=
\frac{h^4}{8}
\left(
\delta\lambda_1c_\beta^4
+
\delta\lambda_2s_\beta^4
\right)
=
-\frac{h^4}{4v^3}
\left.
\frac{dV_{\rm CW}^{\rm FD}}{dh}
\right|_{h=v}.
\end{equation}
Substituting Eq.~\eqref{eq:CWfd} into this expression gives
\begin{equation}
V_{\rm CT}^{\rm FD}(h)
=
-
\left(
A
+
B\ln\frac{v^2}{\bar\mu^2}
+
\frac{B}{2}
\right)h^4 .
\end{equation}
Hence the total zero-temperature effective potential along the flat direction is
\begin{equation}
V^{\rm FD}(h)
=
V_{\rm CW}^{\rm FD}(h)
+
V_{\rm CT}^{\rm FD}(h)
=
B h^4
\left(
\ln\frac{h^2}{v^2}
-
\frac{1}{2}
\right).
\end{equation}
This is the Gildener--Weinberg potential along the flat direction in the finite-counterterm prescription. The finite counterterms have absorbed the dependence on $\bar\mu$, and differentiating this potential twice at $h=v$ gives the Gildener--Weinberg squared mass,
\begin{equation}
m_{\rm GW}^2
=
\left.
\frac{d^2 V^{\rm FD}(h)}{dh^2}
\right|_{h=v}
=
8Bv^2 .
\label{eq:GWmass}
\end{equation}
A convenient approximation for \(\lambda_{345}\) is obtained by identifying this Gildener--Weinberg mass with the physical Higgs mass. This gives
\begin{equation}
m_{\mathrm h}^2 \simeq
\frac{1}{8\pi^2 v^2}
\bigl(
m_{{\mathrm H},0}^4
+m_{{\mathrm A}^0}^4
+2m_{\mathrm H^\pm}^4
+6m_W^4+3m_Z^4-12m_t^4
\bigr) .
\label{eq:GW_mh}
\end{equation}
Solving this relation for the tree-level heavy CP-even mass then yields
\begin{equation}
m_{{\mathrm H},0}^2\equiv-\lambda_{345}v^2
\simeq
\sqrt{
8\pi^2 v^2 m_{\mathrm h}^2
-m_{{\mathrm A}^0}^4
-2m_{\mathrm H^\pm}^4
-6m_W^4
-3m_Z^4
+12m_t^4
}.
\label{eq:M0_est}
\end{equation}

The approximation above can be understood more directly from the full CP-even mass matrix \(\mathcal{M}_{\mathcal H}^2=\mathcal{M}_{{\mathcal H},0}^2+\Delta\) defined in Eq.~\eqref{eq:MH2}. Treating \(\Delta\) as a perturbation, the eigenvalues of \(\mathcal{M}_{\mathcal H}^2\) can be evaluated by ordinary quantum mechanical perturbation theory. To first order, the light mass eigenvalue is
\begin{equation}
m_{\mathrm h}^2
\simeq 
e_h^T\Delta e_h
=
\left.
\frac{d^2}{dh^2}
\left[
V_{\rm CW}^{\rm FD}(h)+V_{\rm CT}^{\rm FD}(h)
\right]
\right|_{h=v}=m_{\rm GW}^2 .
\label{eq:mh_delta_first_order}
\end{equation} 
Here \(e_h=(c_\beta,s_\beta)^T\) is the eigenvector along the flat direction. Therefore, the curvature mass of the Gildener--Weinberg potential in Eq.~\eqref{eq:GWmass} is precisely the first-order perturbative eigenvalue of the full CP-even mass matrix \(\mathcal{M}_{\mathcal H}^2\).

\subsection{Counterterm Potential in NC2HDM} 
Under the symmetry constraints, the most general UV-finite counterterm potential takes the form
\begin{align}
V_{\rm CT}={}&
\delta m_{11}^2\,\Phi_1^\dagger\Phi_1
+\delta m_{22}^2\,\Phi_2^\dagger\Phi_2
-\delta m_{12}^2\left(\Phi_1^\dagger\Phi_2+\text{h.c.}\right)
+\frac{\delta\lambda_1}{2}\left(\Phi_1^\dagger\Phi_1\right)^2
+\frac{\delta\lambda_2}{2}\left(\Phi_2^\dagger\Phi_2\right)^2
\nonumber\\
&+\delta\lambda_3\left(\Phi_1^\dagger\Phi_1\right)\left(\Phi_2^\dagger\Phi_2\right)
+\delta\lambda_4\left(\Phi_1^\dagger\Phi_2\right)\left(\Phi_2^\dagger\Phi_1\right)
+\frac{\delta\lambda_5}{2}\left[\left(\Phi_1^\dagger\Phi_2\right)^2+\text{h.c.}\right].
\label{eq:VCT_2HDM}
\end{align}

The renormalization conditions are chosen such that the tadpoles and the mass matrix remain in their tree-level forms after including the CW correction:
\begin{alignat}{2}
\text{(Tadpoles):}\quad &\left.\frac{\partial\bigl(V_{\rm CW}+V_{\rm CT}\bigr)}{\partial \varphi_a}\right|_{\rm vac} &\,{}= 0,
\\[2pt]
\text{(Mass Matrix):}\quad &\left.\frac{\partial^2\bigl(V_{\rm CW}+V_{\rm CT}\bigr)}{\partial \varphi_a\,\partial \varphi_b}\right|_{\rm vac} &\,{}= 0.
\end{alignat}

Solving these conditions in the interaction basis gives the following counterterms, 
\begin{align}
\delta m_{11}^2 &=
-\left(
\frac{3}{2}v_1^2\delta\lambda_1
+\frac{v_2^2}{2}\left(\delta\lambda_3+\delta\lambda_5\right)
+DV[h_1h_1]
\right),
\label{eq:ct_dm11}
\\[3pt]
\delta m_{22}^2 &=
-\left(
\frac{3}{2}v_2^2\delta\lambda_2
+\frac{v_1^2}{2}\left(\delta\lambda_3+\delta\lambda_5\right)
+DV[h_2h_2]
\right),
\label{eq:ct_dm22}
\\[3pt]
\delta m_{12}^2 &=
2\,DV[\phi_1^+\phi_2^-]-DV[a_1a_2],
\label{eq:ct_dm12}
\\[6pt]
\delta\lambda_1 &=
\frac{DV[h_1]-v_1\,DV[h_1h_1]-\delta m_{12}^2\,v_2}{v_1^3},
\label{eq:ct_dl1}
\\[3pt]
\delta\lambda_2 &=
\frac{DV[h_2]-v_2\,DV[h_2h_2]-\delta m_{12}^2\,v_1}{v_2^3},
\label{eq:ct_dl2}
\\[3pt]
\delta\lambda_3 &=
\frac{DV[a_1a_2]-DV[h_1h_2]}{v_1v_2},
\label{eq:ct_dl3}
\\[3pt]
\delta\lambda_5 &=
\frac{2\left(DV[\phi_1^+\phi_2^-]-DV[a_1a_2]\right)}{v_1v_2},
\label{eq:ct_dl5}
\end{align}
We have chosen the remaining freedom as
\begin{equation}
\delta\lambda_4=0.
\label{eq:dl4zero_choice}
\end{equation}
These expressions were checked explicitly by differentiating the counterterm potential in Eq.~\eqref{eq:VCT_2HDM} and solving the resulting linear system for the seven counterterms. In particular, the off-diagonal conditions in the CP-even, CP-odd and charged sectors uniquely determine \(\delta\lambda_3\), \(\delta\lambda_5\) and \(\delta m_{12}^2\), while the tadpole and diagonal CP-even conditions fix \(\delta\lambda_{1,2}\) and \(\delta m_{11,22}^2\).

\section{Path Dependence of multifield PD}
\label{app:path-dependence}

To assess the numerical effect of the path dependence identified in Section~\ref{sec:PD}, we compare three integration paths for five benchmarks in each model. We find that changing the path shifts the transition temperatures by at most about \(7\%\), the quantities \(\alpha\), \(\beta/H_p\) and \(f_{\rm peak}\) by at most about \(30\%\), and the GW peak amplitude by up to a factor of \(2.7\). To relate this variation to the PD tadpole, recall that path independence requires
\begin{equation}
  \partial_b\mathcal T_a^{\mathrm{PD}}=\partial_a\mathcal T_b^{\mathrm{PD}}
  \label{eq:pd-integrability}
\end{equation}
When this condition is not satisfied, the reconstructed potential depends on the contour $\gamma$ connecting the origin to the same endpoint. Omitting the contour-independent counterterm and non-scalar contributions for brevity, we write
\begin{equation}
  V_{\rm PD}^{(\gamma)}(\varphi,T)-V_{\rm PD}^{(\gamma)}(0,T)
  =V_0(\varphi)-V_0(0)+\int_\gamma d\varphi_a\,\mathcal T_a^{\mathrm{PD}}(\varphi,T).
  \label{eq:pd-contour-potential-app}
\end{equation}
Although omitted here for brevity, $V_{\rm CT}$ and $V_{\rm NS}^R$ are included in the numerical potentials as specified below Eq.~\eqref{eq:VPD_1d}. Since they depend only on the endpoints, they cancel in comparisons between paths.

\begin{figure}[t]
	\centering
	\includegraphics[width=\textwidth]{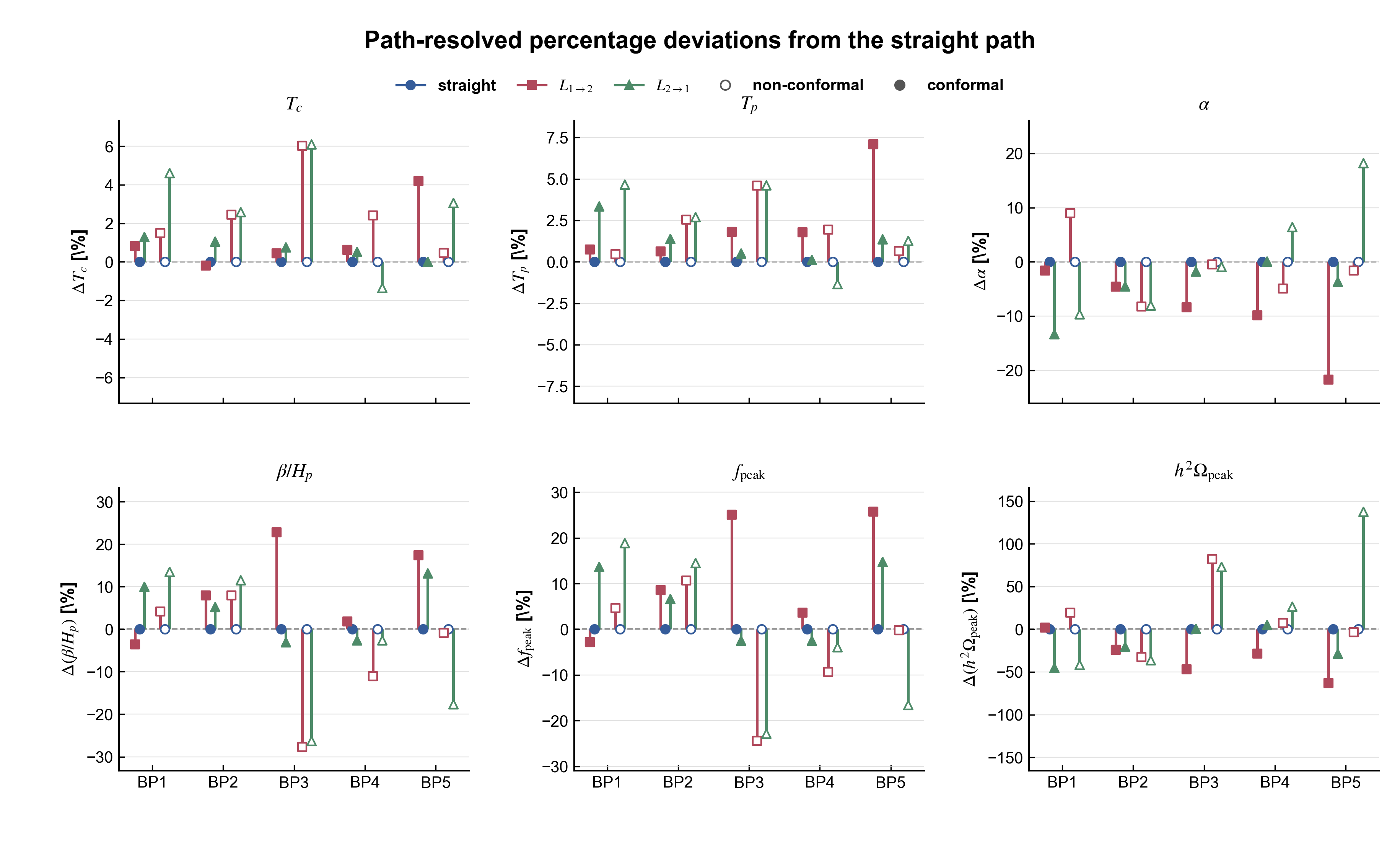}
	\caption{Path-resolved deviations of the phase-transition and gravitational-wave observables from the straight contour. The vertical segments show $\Delta X \equiv 100\left(\frac{X}{X_{\rm straight}}-1\right)\%$\protect\del{.} for the two $L$-shaped contours; colors identify the contour and filled (open) markers denote the conformal (non-conformal) model. The horizontal positions are offset only for readability.}
	\label{fig:path-spread-log}
\end{figure}

We evaluate Eq.~\eqref{eq:pd-contour-potential-app} along the straight path $\gamma_a(t)=t\varphi_a$, $t\in[0,1]$, used in the main text. To estimate the path dependence of this reference result, we compare it with two L-shaped contours in the $(\phi_1,\phi_2)$ background plane:
\begin{align}
  L_{1\to2}:&\quad (0,0)\to(\phi_1,0)\to(\phi_1,\phi_2),
  \label{eq:pd-L12-contour-app}\\
  L_{2\to1}:&\quad (0,0)\to(0,\phi_2)\to(\phi_1,\phi_2).
  \label{eq:pd-L21-contour-app}
\end{align}
Table~\ref{tab:path-benchmarks} lists five benchmarks for each model, labelled independently in increasing $\tan\beta$. Tables~\ref{tab:path-observables-conformal} and~\ref{tab:path-observables-nonconformal} report the phase-transition and GW observables for all three paths. Here $\beta/H_p$ is obtained from the mean bubble separation, and the GW peak refers to the combined sound-wave and turbulence spectrum. Figure~\ref{fig:path-spread-log} summarizes the percentage deviations from the straight-path results.

\begin{table}[htbp]
	\centering
	\scriptsize\setlength{\tabcolsep}{4pt}
	\begin{tabular}{lcrrrrr}
		\hline
		Model & benchmark & $\tan\beta$ & $m_H$ [GeV] & $m_A$ [GeV] & $m_{H^\pm}$ [GeV] & $m_{12}$ [GeV] \\
		\hline
		\multirow{5}{*}{conformal} & BP1 & 0.5989 & 411.71 & 395.05 & 379.54 & -- \\
		& BP2 & 1.1529 & 363.16 & 396.52 & 386.85 & -- \\
		& BP3 & 1.5014 & 329.76 & 412.96 & 396.39 & -- \\
		& BP4 & 1.5381 & 316.85 & 414.98 & 398.48 & -- \\
		& BP5 & 2.5894 & 183.27 & 407.27 & 420.42 & -- \\
		\hline
		\multirow{5}{*}{non-conformal} & BP1 & 0.5993 & 457.02 & 387.60 & 469.50 & 136.30 \\
		& BP2 & 1.0104 & 265.73 & 590.84 & 249.84 & 128.10 \\
		& BP3 & 1.2034 & 441.57 & 577.67 & 431.57 & 207.70 \\
		& BP4 & 1.7012 & 378.95 & 441.49 & 336.60 & 69.83 \\
		& BP5 & 2.4763 & 200.68 & 546.09 & 190.85 & 99.30 \\
		\hline
	\end{tabular}
	\caption{Benchmark points for the path comparison in the conformal and non-conformal models, ordered separately by increasing $\tan\beta$. All mass units are shown in the column headings. The soft mass parameter $m_{12}$ is absent in the conformal model and is denoted by --; the conformal $m_H$ is derived rather than an independent input.}
	\label{tab:path-benchmarks}
\end{table}

\begin{table}[p]
	\centering
	\scriptsize\setlength{\tabcolsep}{3pt}\renewcommand{\arraystretch}{1.12}
	\textbf{Conformal Benchmark Points}\par\vspace{3pt}
	\begin{tabular}{ccccccccc}
		\hline
		benchmark & Path & $T_c$ [GeV] & $T_n$ [GeV] & $T_p$ [GeV] & $\alpha$ & $\beta/H_p$ & $f_{\rm peak}$ [mHz] & $h^2\Omega_{\rm peak}$ \\
		\hline
		BP1 & Straight & 88.16 & 70.00 & 68.46 & 0.0655 & 408.7 & 1.484 & $8.44\times10^{-15}$ \\
		& $L_{1\to2}$ & 88.89 & 70.54 & 68.96 & 0.0644 & 394.1 & 1.441 & $8.60\times10^{-15}$ \\
		& $L_{2\to1}$ & 89.30 & 72.23 & 70.75 & 0.0568 & 449.8 & 1.687 & $4.65\times10^{-15}$ \\
		\hline
		BP2 & Straight & 86.78 & 68.52 & 66.63 & 0.0735 & 376.7 & 1.331 & $1.37\times10^{-14}$ \\
		& $L_{1\to2}$ & 86.62 & 68.56 & 67.04 & 0.0701 & 406.6 & 1.445 & $1.04\times10^{-14}$ \\
		& $L_{2\to1}$ & 87.70 & 69.11 & 67.55 & 0.0701 & 396.5 & 1.420 & $1.09\times10^{-14}$ \\
		\hline
		BP3 & Straight & 87.09 & 67.59 & 66.02 & 0.0790 & 382.5 & 1.339 & $1.65\times10^{-14}$ \\
		& $L_{1\to2}$ & 87.47 & 68.60 & 67.21 & 0.0724 & 469.9 & 1.675 & $8.76\times10^{-15}$ \\
		& $L_{2\to1}$ & 87.76 & 67.97 & 66.35 & 0.0777 & 371.1 & 1.306 & $1.65\times10^{-14}$ \\
		\hline
		BP4 & Straight & 87.08 & 67.40 & 65.82 & 0.0801 & 379.5 & 1.325 & $1.74\times10^{-14}$ \\
		& $L_{1\to2}$ & 87.62 & 68.57 & 67.00 & 0.0722 & 386.4 & 1.373 & $1.24\times10^{-14}$ \\
		& $L_{2\to1}$ & 87.53 & 67.53 & 65.91 & 0.0802 & 369.6 & 1.292 & $1.83\times10^{-14}$ \\
		\hline
		BP5 & Straight & 86.61 & 64.66 & 62.84 & 0.0980 & 304.8 & 1.016 & $4.61\times10^{-14}$ \\
		& $L_{1\to2}$ & 90.24 & 69.00 & 67.31 & 0.0767 & 357.9 & 1.277 & $1.71\times10^{-14}$ \\
		& $L_{2\to1}$ & 86.62 & 65.35 & 63.70 & 0.0944 & 345.1 & 1.166 & $3.30\times10^{-14}$ \\
		\hline
	\end{tabular}
	\caption{Phase-transition and gravitational-wave observables for five benchmark points and the three integration paths. $\beta/H_p$ is obtained from the mean bubble separation; the gravitational-wave entries give the peak of the total sound-wave plus turbulence spectrum.}
	\label{tab:path-observables-conformal}
\end{table}

\begin{figure}[p]
	\centering
	\includegraphics[width=\textwidth]{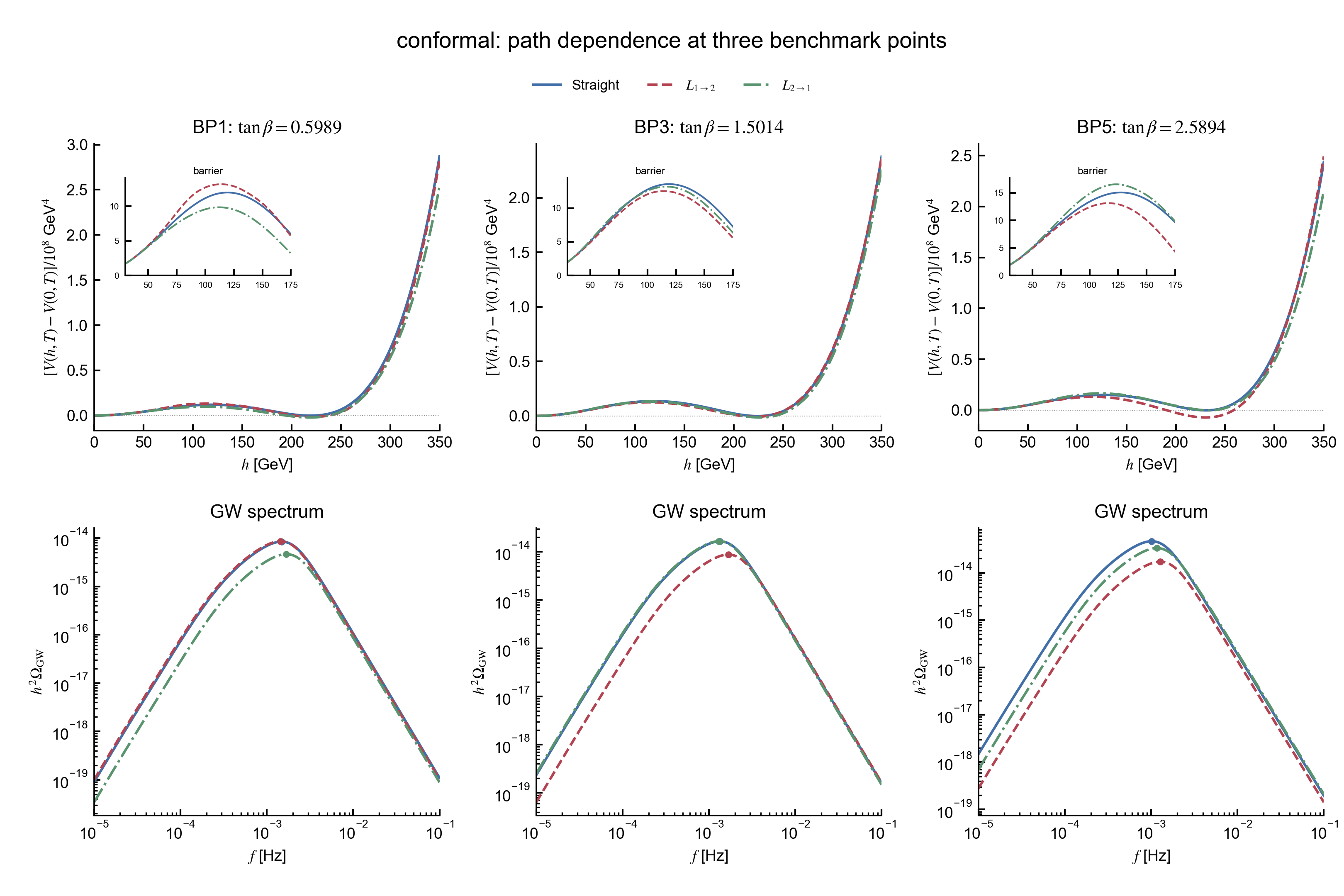}
	\caption{Path dependence in the conformal model for BP1, BP3, and BP5, ordered from small to large $\tan\beta$. The upper row shows the reconstructed potential along the straight, $L_{1\to2}$, and $L_{2\to1}$ contours; the upper-left inset in each panel enlarges the barrier region. The lower row shows the corresponding total GW spectra.}
	\label{fig:path-conformal-panel}
\end{figure}

\begin{figure}[p]
	\centering
	\includegraphics[width=\textwidth]{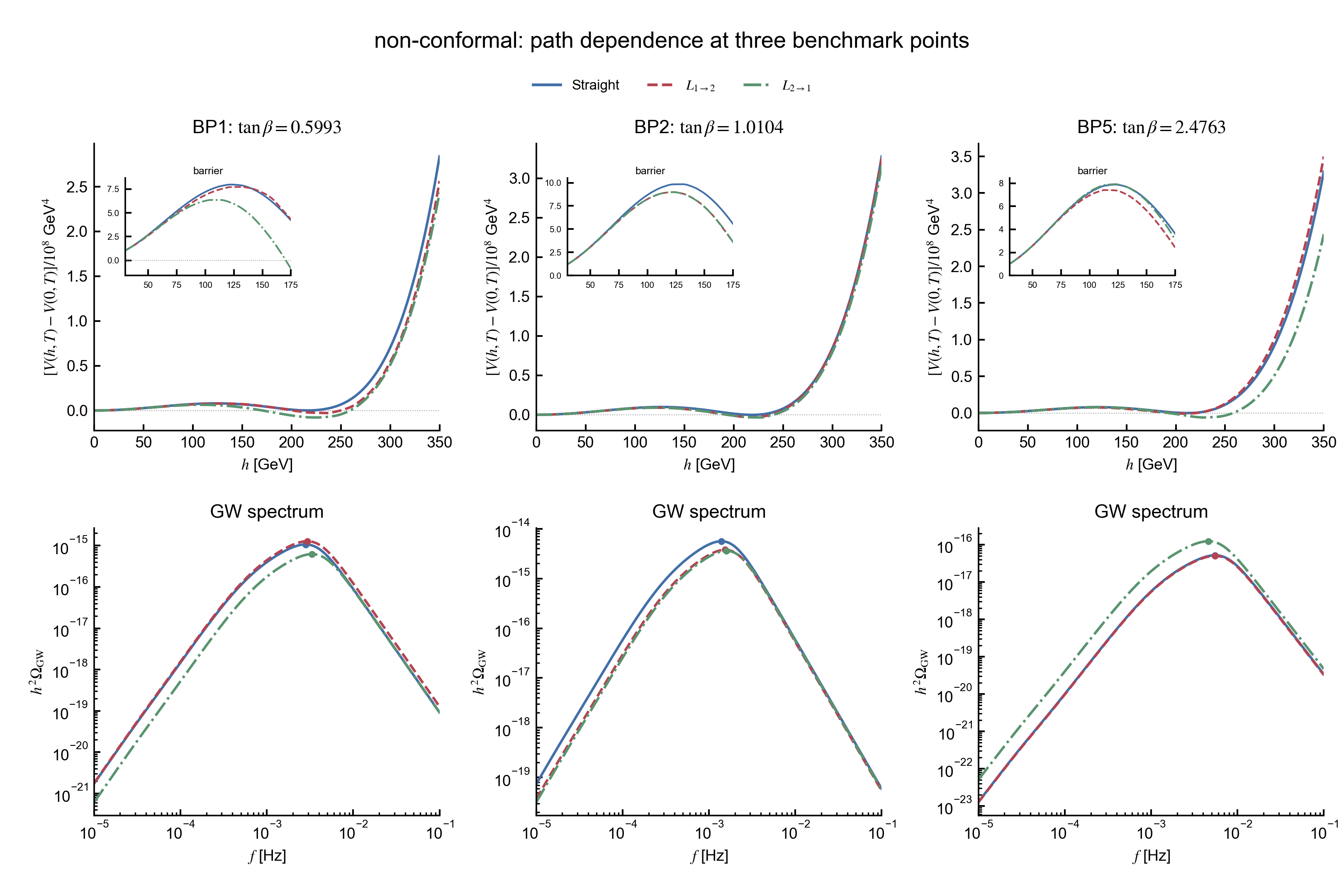}
	\caption{Path dependence in the non-conformal model for BP1, BP2, and BP5, ordered from small to large $\tan\beta$. The upper row shows the reconstructed potential and barrier zooms; the lower row shows the corresponding total GW spectra.}
	\label{fig:path-nonconformal-panel}
\end{figure}

\begin{table}[p]
	\centering
	\scriptsize\setlength{\tabcolsep}{3pt}\renewcommand{\arraystretch}{1.12}
	\textbf{Non-conformal Benchmark Points}\par\vspace{3pt}
	\begin{tabular}{ccccccccc}
		\hline
		benchmark & Path & $T_c$ [GeV] & $T_n$ [GeV] & $T_p$ [GeV] & $\alpha$ & $\beta/H_p$ & $f_{\rm peak}$ [mHz] & $h^2\Omega_{\rm peak}$ \\
		\hline
		BP1 & Straight & 87.76 & 75.89 & 74.90 & 0.0457 & 707.4 & 2.809 & $1.07\times10^{-15}$ \\
		& $L_{1\to2}$ & 89.08 & 76.37 & 75.25 & 0.0498 & 736.7 & 2.939 & $1.28\times10^{-15}$ \\
		& $L_{2\to1}$ & 91.82 & 79.52 & 78.40 & 0.0413 & 803.5 & 3.340 & $6.23\times10^{-16}$ \\
		\hline
		BP2 & Straight & 87.85 & 69.12 & 67.45 & 0.0553 & 388.6 & 1.390 & $5.61\times10^{-15}$ \\
		& $L_{1\to2}$ & 90.00 & 70.70 & 69.16 & 0.0507 & 419.5 & 1.538 & $3.77\times10^{-15}$ \\
		& $L_{2\to1}$ & 90.11 & 70.87 & 69.27 & 0.0509 & 433.5 & 1.592 & $3.59\times10^{-15}$ \\
		\hline
		BP3 & Straight & 89.20 & 85.09 & 84.61 & 0.0196 & 2835.0 & 12.717 & $6.15\times10^{-18}$ \\
		& $L_{1\to2}$ & 94.58 & 89.06 & 88.49 & 0.0195 & 2049.2 & 9.614 & $1.12\times10^{-17}$ \\
		& $L_{2\to1}$ & 94.65 & 89.06 & 88.51 & 0.0194 & 2091.4 & 9.814 & $1.06\times10^{-17}$ \\
		\hline
		BP4 & Straight & 91.09 & 85.66 & 85.14 & 0.0264 & 2185.8 & 9.867 & $2.48\times10^{-17}$ \\
		& $L_{1\to2}$ & 93.28 & 87.45 & 86.80 & 0.0251 & 1943.9 & 8.946 & $2.66\times10^{-17}$ \\
		& $L_{2\to1}$ & 89.86 & 84.53 & 84.00 & 0.0281 & 2128.9 & 9.481 & $3.14\times10^{-17}$ \\
		\hline
		BP5 & Straight & 99.61 & 89.53 & 88.65 & 0.0229 & 1172.0 & 5.508 & $5.21\times10^{-17}$ \\
		& $L_{1\to2}$ & 100.07 & 90.08 & 89.22 & 0.0226 & 1161.5 & 5.495 & $5.03\times10^{-17}$ \\
		& $L_{2\to1}$ & 102.66 & 90.76 & 89.78 & 0.0271 & 965.9 & 4.598 & $1.24\times10^{-16}$ \\
		\hline
	\end{tabular}
	\caption{Phase-transition and gravitational-wave observables for five benchmark points and the three integration paths. $\beta/H_p$ is obtained from the mean bubble separation; the gravitational-wave entries give the peak of the total sound-wave plus turbulence spectrum.}
	\label{tab:path-observables-nonconformal}
\end{table}
Figures~\ref{fig:path-conformal-panel} and~\ref{fig:path-nonconformal-panel} illustrate the reconstructed potentials and GW spectra for selected benchmarks spanning small to large $\tan\beta$. The upper panels show the potentials and barrier regions, while the lower panels show the corresponding spectra.

The transition temperatures depend only modestly on the integration path, by at most about \(7\%\), whereas the GW peak amplitudes, which depend steeply on \(\alpha\) and \(\beta/H_p\), can differ by up to a factor of \(2.7\). For these benchmarks, however, the path dependence does not produce order-of-magnitude shifts in the peak amplitude. This variation is much smaller than the differences between PD, Arnold--Espinosa, and Parwani illustrated in Fig.~\ref{fig:gw_benchmark_spectra}, which reach up to about nine orders of magnitude between PD and Arnold--Espinosa. We use the straight-path definition for the prescription comparison and treat the smaller path variation as a separate uncertainty intrinsic to multi-field PD. This numerical estimate is limited to the five benchmarks per model and three paths tested here; a more detailed theoretical treatment is given in our subsequent work~\cite{LiWangZhao:MultifieldCJT}.

\clearpage
\bibliographystyle{JHEP}
\bibliography{bib/biblio_resummation_effective_potential,bib/biblio_phenomenology,bib/biblio_phase_transition,bib/biblio_gravitational_wave}

\end{document}